\documentclass[aps,prapplied,reprint,showpacs,amsmath,amssymb,floatfix,superscriptaddress,noeprint]{revtex4-2}
\usepackage{bm}
\usepackage{color}
\usepackage[colorlinks,linkcolor=blue,urlcolor=blue,citecolor=blue,anchorcolor=blue]{hyperref}
\usepackage{graphicx}
\usepackage[capitalise]{cleveref}
\usepackage{physics}
\usepackage[detect-weight=true, separate-uncertainty=true]{siunitx}
\usepackage{multirow}
\usepackage{circledsteps}
\usepackage{placeins}

\begin{document}

\title{Number-resolved photocounter for propagating microwave mode}

\author{R. Dassonneville}
\affiliation{Univ Lyon, ENS de Lyon, Univ Claude Bernard, CNRS, Laboratoire de Physique,F-69342 Lyon, France}
\author{R. Assouly}
\affiliation{Univ Lyon, ENS de Lyon, Univ Claude Bernard, CNRS, Laboratoire de Physique,F-69342 Lyon, France}
\author{T. Peronnin}
\affiliation{Univ Lyon, ENS de Lyon, Univ Claude Bernard, CNRS, Laboratoire de Physique,F-69342 Lyon, France}
\author{P. Rouchon}
\affiliation{Centre Automatique et Syst\`emes, Mines-ParisTech, PSL Research University, 60 bd Saint-Michel, 75006 Paris, France. \\
QUANTIC team, INRIA de Paris, 2 rue Simone Iff, 75012 Paris, France.}
\author{B. Huard}
\email{benjamin.huard@ens-lyon.fr}
\affiliation{Univ Lyon, ENS de Lyon, Univ Claude Bernard, CNRS, Laboratoire de Physique,F-69342 Lyon, France}

\date{\today}

\begin{abstract}
Detectors of propagating microwave photons have recently been realized using superconducting circuits. However a number-resolved photocounter is still missing. In this article, we demonstrate a single-shot counter for propagating microwave photons that can resolve up to $3$ photons. It is based on a pumped Josephson Ring Modulator that can catch an arbitrary propagating mode by frequency conversion and store its quantum state in a stationary memory mode. A transmon qubit then counts the number of photons in the memory mode using a series of binary questions. Using measurement based feedback, the number of questions is minimal and scales logarithmically with the maximal number of photons. The detector features a detection efficiency of $0.96 \pm 0.04$, and a dark count probability of $0.030 \pm 0.002$ for an average dead time of \SI{4.5}{\micro s}. To maximize its performance, the device is first used as an \emph{in situ} waveform detector from which an optimal pump is computed and applied. Depending on the number of incoming photons, the detector succeeds with a probability that ranges from \SI{54 \pm 2}{\percent} to $99\%$.
\end{abstract}

\maketitle

\section{INTRODUCTION}
Photon detectors are an important element in the quantum optics toolbox. At optical frequencies, detectors such as single-photon avalanche photodiodes or superconducting nanowire single-photon detectors are readily available~\cite{Hadfield2009}. 
In contrast, at GHz frequencies, these kinds of absorptive detectors are harder to realize due to the low energy of the microwave photons, roughly 5 orders of magnitude lower compared to their optical counterparts. Detecting and counting the microwave photons of a stationary mode is nowadays routinely performed using the dispersive interaction with a qubit~\cite{Gleyzes2007,Guerlin2007,Johnson2010,Leek2010,Sun2014}. These operations remain challenging for propagating photons because the light-matter interaction time is smaller. Yet some photon detectors for propagating modes have been proposed~\cite{Romero2009,Helmer2009,Koshino2013,Sathyamoorthy2014,Fan2014,Kyriienko2016,Sathyamoorthy2016,Gu2017,Wong2017,Leppakangas2018,Royer2018} and developed based on various approaches: direct absorption~\cite{Chen2011,Inomata2016}, encoding parity in the phase of a qubit~\cite{Besse2018,Kono2018}, encoding the probability to have a single photon in a qubit excitation~\cite{Narla2016} or reservoir engineering~\cite{Lescanne2020}. Several implementations of a photocounter -- a microwave photodetector able to resolve the photon number -- for a propagating mode have been proposed \cite{Romero2009,Royer2018,Kono2018,Sokolov2020,grimsmo2020quantum}. However, such a device has yet to be demonstrated. Indeed, Refs.~\cite{Besse2018,Kono2018} only distinguish the parity of the photon number. References \cite{Narla2016,Lescanne2020} only distinguish Fock state $\ket{1}$ from the rest while Refs.~\cite{Chen2011,Inomata2016} distinguish $0$ photon from at least $1$. 
\begin{figure}[h!]
    \centering
     \includegraphics[width=8.6cm]{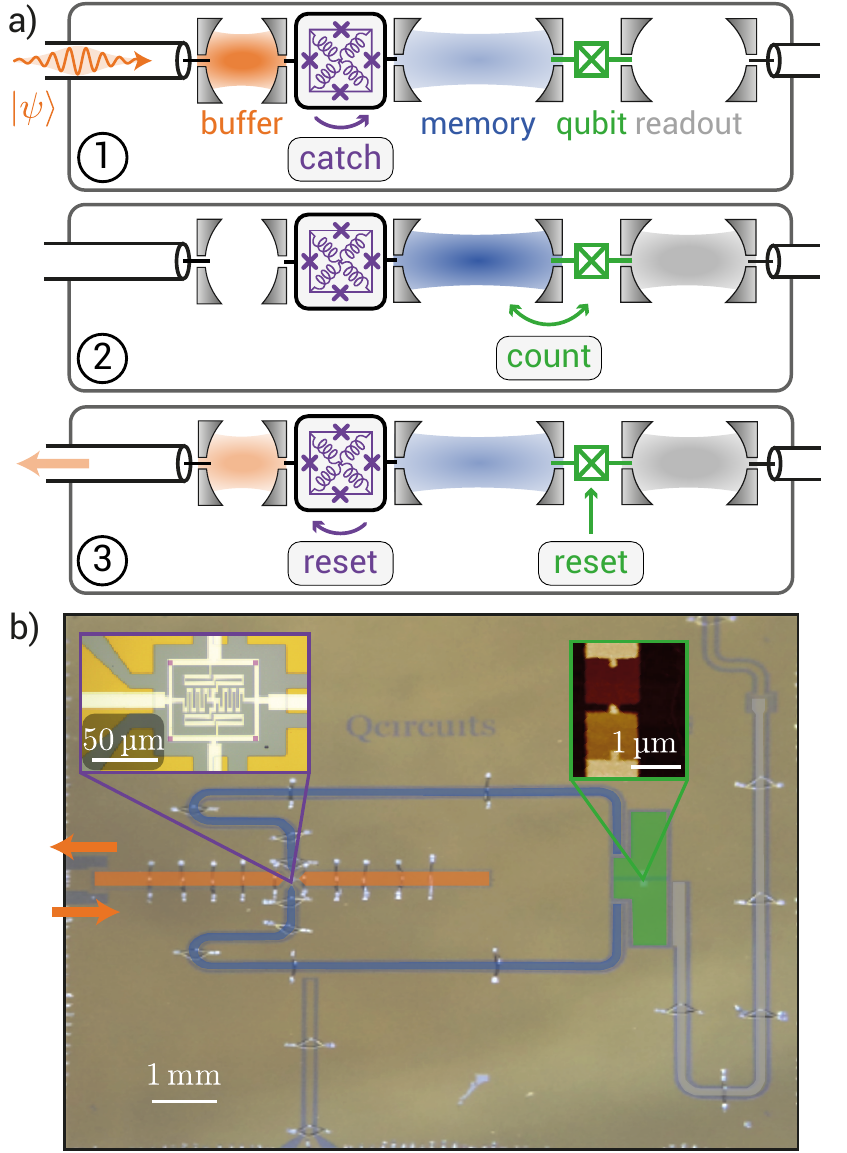}
    \caption{Principle of operation and device. a) A propagating microwave mode in state $\ket{\psi}$ is sent to the device via a buffer resonator. It is caught~\Circled{1} into the memory  by pumping the Josephson ring modulator (JRM). The qubit then counts~\Circled{2} the photon number in the memory. The device is finally reset~\Circled{3}. Pumping the JRM empties the memory by releasing its photons into an arbitrary outgoing mode. The qubit is put into its ground state by measurement-based feedback. b) False color image of the device where a JRM (left inset) is located at the crossing between buffer and memory $\lambda/2$ resonators. A transmon qubit (right inset) is coupled both to the memory and readout resonators.
    }
    \label{fig:schematic}
\end{figure}

Here, we demonstrate a photocounter that resolves the number of photons in a given propagating mode. To optimize the efficiency of our counter, we devise a way to calibrate \emph{in situ} the arrival time and envelope of the propagating mode. The device can distinguish between 0, 1, 2, and 3 photons in a 20-\SI{}{\mega\hertz} band around \SI{10.220}{\giga\hertz} using measurement-based feedback. Finally, we propose a parameter-free model that accurately predicts the behavior of the counter, as demonstrated by coherent-state photocounting and Wigner tomography.

\section{DEVICE AND OPERATION}
The purpose of a photocounter is to count the photon number in a propagating mode with state $\ket{\psi}$ by providing an integer outcome $n$ with probability $\abs{\braket{\psi}{n}}^2$. Our photocounter proceeds in three steps (\cref{fig:schematic}.a). In step \Circled{1}, it catches the incoming wavepacket and converts it into a high-Q stationary mode (memory). Then, in step \Circled{2}, it counts the number of photons in the memory using an ancillary qubit. Finally (step \Circled{3}), it resets the memory and qubit in their ground state. The catch and memory-reset operations (\Circled{1}, \Circled{3}) are performed by frequency conversion using a Josephson ring modulator (JRM)~\cite{Bergeal2010,Roch2012}. The input transmission line is coupled to a buffer mode at frequency $\omega_{\rm b}/2\pi = \SI{10.220}{\giga\hertz}$, which sets the operating bandwidth of the counter to $\kappa_{\rm b} = 2\pi\times \SI{20}{\mega\hertz} = (\SI{8.0}{ns})^{-1}$. When pumped by a coherent tone of amplitude $p(t)$ at $ \omega_{\rm b} - \omega_{\rm m}$, the JRM introduces a frequency-conversion term $\hat{\mathrm{H}}_{\rm JRM} = g_3 p(t) \hat{b} \hat{m}^{\dag} + h.c. $ between the buffer $\hat{b}$ and the memory $\hat{m}$. The memory resonates at $\omega_{\rm m}/2\pi=$ \SI{3.74527}{\giga\hertz} with a relaxation time $T_{1,\rm{ m}}= \SI{4}{\micro\second}$. When the memory is initially empty, this term enables us to catch the incoming wavepacket onto the buffer by storing its quantum state in the memory. Conversely, when the counting operation is over, we use it to release the photons from the memory into an arbitrary outgoing wavepacket.

From the point of view of the memory, the pumped JRM induces a tunable coupling to a transmission line~\cite{Peronnin2019}. It is thus possible to catch or release an arbitrary wavepacket into and from the memory~\cite{Yin2013,Wenner2014,Flurin2014,Axline2018,Zhong2019,Campagne-Ibarcq2018,Kurpiers2018}. Besides, the parasitic nonlinearities induced by the Josephson junctions of the JRM can be canceled by setting the flux through the JRM optimally, which we did (\cref{sec:flux_dep}).
Using input-output formalism in the rotating frame, and neglecting the relaxation of the memory, the dynamics is captured by 
\begin{align}
    \dv{\hat{b}}{t} =& - \frac{\kappa_{\rm b}}{2} \hat{b} - g_3 p^*(t) \hat{m} + \sqrt{\kappa_{\rm b}} \hat{b}_{\rm in}(t) \textrm{,  } \notag \\
    \dv{\hat{m}}{t} =& g_3^* p(t) \hat{b} .
    \label{eq:inout}
\end{align}
For any given envelope $\langle b_{\rm in}(t)\rangle$ of the incoming wavepacket that fits inside the buffer bandwidth $\kappa_\mathrm{b}$, there exists an optimal pump $p_{\mathrm{opt}}(t)$ for which the incoming quantum state is perfectly swapped into the memory ~\cite{Korotkov2011}. For instance if the incoming wavepacket is $\langle b_{\rm in}(t)\rangle \propto 1/ \cosh(\sqrt{\pi/2}~t/\sigma )$ (\cref{fig:SSphotocounting}.a), the optimal catching pump is given by $p_{\mathrm{opt}}(t) \propto \left[1+\frac{\lambda}{2} \tanh{(\lambda \kappa_{\rm b} t /4)}\right] (e^{\lambda \kappa_{\rm b}t/2}+1-\lambda/2)^{-1/2} $ where $\lambda = \sqrt{8\pi}/\kappa_{\rm b}\sigma$. Note that even at nonoptimal flux through the JRM or with finite relaxation time of the memory, an optimal pump can be found to catch the entire wavepacket (\cref{sec:opti_catch}).

\section{Built-in sample and hold power meter}
\begin{figure}
\centering
\includegraphics[width=8.6cm]{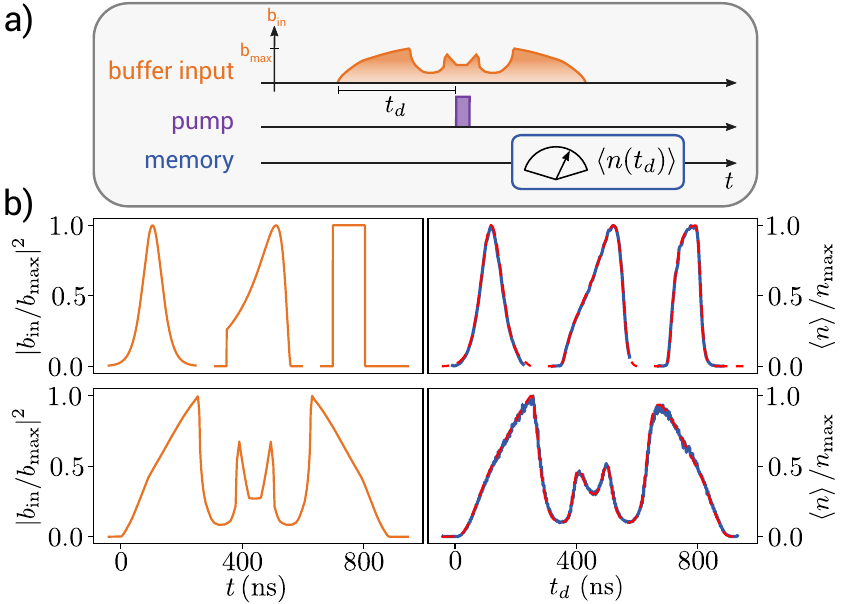}   
\caption{\emph{In situ} calibration of the incoming wavepacket envelope. a) Amplitude of an arbitrary incoming wavepacket sent onto the buffer and of the sampling pump pulse. A following measurement of the mean photon number $\expval{n(t_d)}$ in the memory is performed using the qubit. b) Left panels: various incoming waveforms. Right panels: solid blue (dashed red) lines show the measured [predicted using \cref{eq:inout}] mean photon number $\expval{n}$ normalized by its maximum $n_{\rm max}$. }
\label{fig:envelope_detection}
\end{figure}

In order to generate the optimal pump $p_{\rm opt}(t)$ for an arbitrary incoming wavepacket at $\omega_{\rm b}$, one needs to determine the envelope $\langle b_{\rm in}(t)\rangle$. Interestingly, the envelope of any incoming waveform, can be determined \emph{in situ}. The photocounter can indeed operate as a sample-and-hold power meter. Turning on the pump for a short sampling time of \SI{20}{\nano\second} after a variable delay $t_d$ and counting the mean number of photons in the memory, using the coupled transmon qubit (\cref{average_counting}), enables us to directly probe $ \expval{b_{\rm in}
^\dagger b_{\rm in}}$ up to a global prefactor [\cref{fig:envelope_detection}.a].
We demonstrate this functionality on a variety of generated waveforms displayed in \cref{fig:envelope_detection}.b (left panel). The distortion of the waveforms introduced by the finite bandwidth $\kappa_{\rm b}$ of the counter and the nonzero sampling time can be seen in the measured mean photon number $\expval{n}$ as a function of $t_d$ (right panel). The simple model \cref{eq:inout} accurately reproduces the measured envelopes, where the only free parameter is the 15-\SI{}{\nano\second} difference in propagation time between buffer and pump lines.

\section{Catch efficiency}
\label{catch_efficiency}

\begin{figure}[h]
    \centering
    \includegraphics[width=8.6cm]{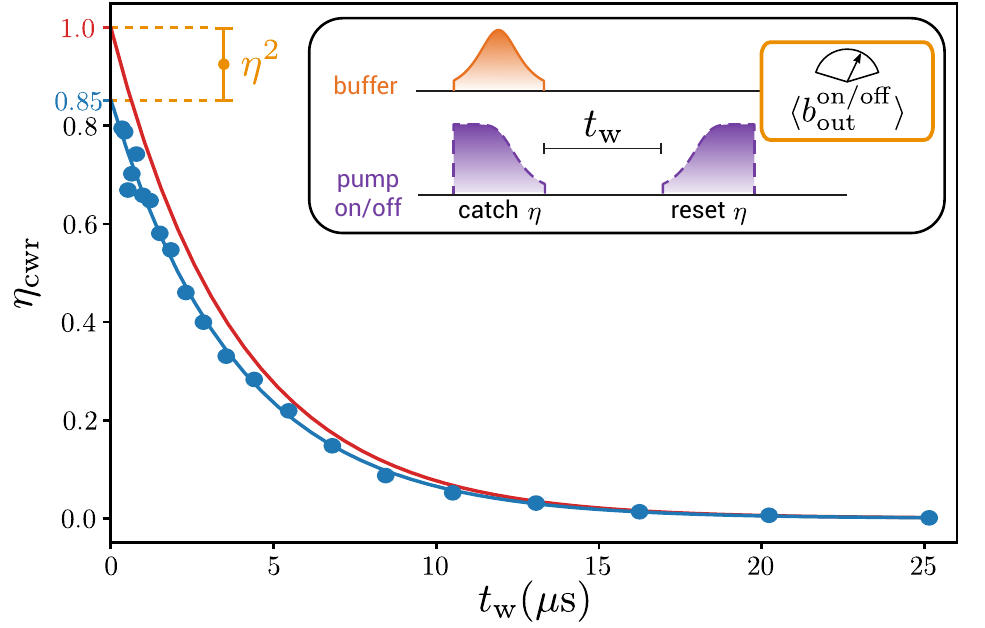}
    \caption{Red solid line: round trip efficiency $\eta_\mathrm{CWR}$ as a function of the waiting time $t_{\rm w}$ assuming that the only imperfection comes from the memory decay with a characteristic time $T_{1, \rm m}$. Blue dots: lower bound on the measured round-trip efficiency $\eta_{\rm CWR}$. Blue solid line: exponential decay with characteristic time $T_{1, \rm m}$. Orange error bar: range of possible values for $\eta^2$ that leads to a catch efficiency $\eta= 0.96 \pm 0.04$. Inset: pulse sequence of the catch-wait-release protocol. We measure the average outgoing amplitude $\expval{b_\mathrm{out}^\mathrm{on/off}}$ when the pump is on or off, from which we compute the round trip efficiency $\eta_\mathrm{CWR} = \expval{b_\mathrm{out}^\mathrm{on}}^2/\expval{b_\mathrm{out}^\mathrm{off}}^2$.}
    \label{fig:catch_eff}
\end{figure}

In order to measure the catch efficiency $\eta$, we follow a catch-wait-release protocol as in Refs.~\cite{Flurin2014,Flurin2015}. We send an input signal $\expval{b_{\rm in}(t)} \propto 1/ \cosh(\sqrt{\pi/2}~t/\sigma )$ of width $\sigma=\SI{52}{ns}$ and use the corresponding optimal pump shape, computed using \cref{eq:inout}. The calibration consists in measuring the outgoing amplitude $b_\mathrm{out}$ in various configurations. First, we measure the directly reflected amplitude  $b_\mathrm{out}^\mathrm{off}$ without pumping, which provides a reference. Then, we measure the re-emitted amplitude $b_\mathrm{out}^\mathrm{on}$ after optimally catching, waiting a time $t_{\rm w}$ and releasing. The round-trip efficiency is then given by $\eta_{\rm CWR} = \expval{b_\mathrm{out}^\mathrm{on}}^2/\expval{b_\mathrm{out}^\mathrm{off}}^2$. Besides, assuming that the catch and release operations have the same efficiency $\eta$, we get $\eta_{\rm CWR}=\eta^2 e^{-t_{\rm w}/T_{1, \rm{ m}}}$, and thus an estimation of $\eta$.

In practice, due to the finite directivity of the directional coupler used to drive the buffer (\cref{sec:setup}), there are interferences between the signal parasitically bypassing the coupler towards the output line and the desired signal coming from the buffer. This problem exclusively affects the denominator of the measured energy ratio since the parasitic signal does not spatially overlap with the signal that is released after $t_{\rm w}$. In our case, the interferences are destructive, which leads to an underestimation of the denominator. As a consequence, we obtain apparent energy ratios in excess of \SI{100}{\%}. 

It is, however, possible to get a lower bound on the actual efficiency $\eta_{\rm CWR}(t_{\rm w})$ by measuring the coupler directivity. Right after the run, we measure a 16-\SI{}{dB} directivity at room temperature using a calibrated vector network analyzer. In \cref{fig:catch_eff}, the lowest possible values of $\eta_{\rm CWR}(t_{\rm w})$ (dots) are shown assuming fully destructive interferences in the denominator (correction by a factor $0.746$ on the apparent energy ratio). Fitting these lower values by an exponential decreasing function at rate $1/T_{1, \rm m}$, we get a lower bound on the catch efficiency $\eta=\sqrt{\eta_{\rm CWR}(t_{\rm w}=0)}\geq 0.92$.

\section{Binary decomposition of the photon number}

\begin{figure}
\centering
\includegraphics[width=8.6cm]{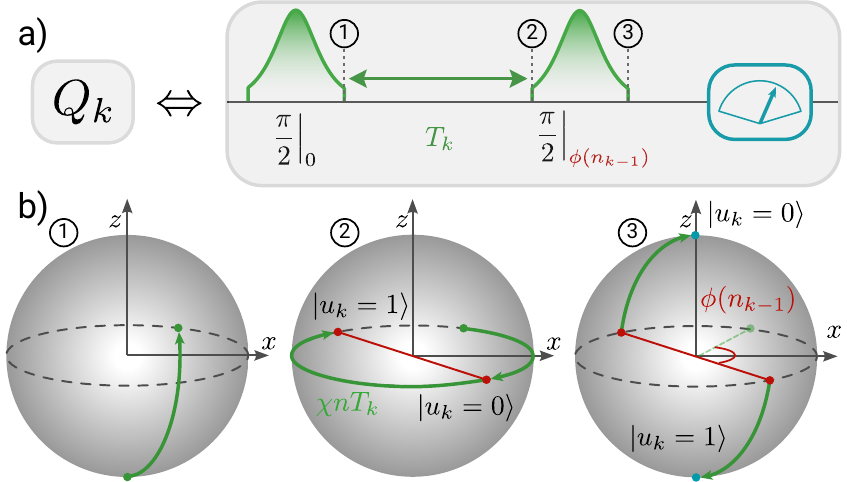}   
\caption{Binary decomposition. a) 
Pulse sequence used to extract $u_k$ experimentally. Green corresponds to taking the remainder modulo $2^k$ of the photon number, red to the subtraction of the previously found digits $n_{k-1}=[u_{k-1}\cdots u_1]_2$ and blue to the extraction of the result via a measurement of the qubit. The $\frac{\pi}{2}$ pulses consist of $\sech$ waveforms with $\sigma = \SI{4}{ns}$ truncated at $4\sigma$ further optimized to mitigate the effect of the transmon qubit's low anharmonicity of \SI{-98}{MHz}~\cite{Motzoi2009}. 
b) Trajectory of the qubit on the Bloch sphere when the cavity is in a Fock state $\ket{n}$ with yet unknown bit $u_k$. Left \Circled{1}: the qubit is prepared in $(\ket{g}+i\ket{e})/\sqrt{2}$ with an unconditional $\pi/2$ pulse around $x$. Middle \Circled{2}: trajectory of the qubit states $\ket{u_k = 0}$ and $\ket{u_k = 1}$ corresponding to the two possible values of the $k$-th bit of the photon number during the waiting time $T_k$. Right \Circled{3}: the last $\pi/2$ pulse around an axis shifted by an angle $\phi(n_{k-1})$ from the $x$ axis maps $u_k$ onto ground or excited states.} 
\label{fig:binary_decomp}
\end{figure}

Once the incoming wavepacket is characterized and efficiently caught, step \Circled{2} consists in measuring the photon number present in the memory in a single-shot manner.
To do so, we use a transmon qubit at frequency $\omega_q/2\pi= \SI{4.32731}{\giga\hertz}$ dispersively coupled to the memory such that $\hat{\mathcal{H}}_{\mathrm{qm}} = -\chi \hat{m}^{\dag}\hat{m} \ketbra{e}{e} $. Owing to a dispersive shift $\chi/2\pi = \SI{3.28}{\mega\hertz}$ much larger than the qubit decoherence rate $\Gamma_2=\left(\SI{13.6}{\micro\second}\right)^{-1}$, the device operates in the photon-number-resolved regime~\cite{Schuster2007}. 
It is thus possible to access information about the photon number by entangling the memory mode with the qubit and reading out its state. 
It is made possible by another resonator (readout), with frequency $\omega_{\rm r}/2\pi = \SI{6.293}{GHz}$, dispersively coupled to the qubit. We optimize the readout fidelity up to \SI{97}{\percent} in \SI{252}{\nano\second}, using a CLEAR-like sequence~\cite{McClure2016}, mostly limited by the finite qubit relaxation time $T_1 = \SI{7.1}{\micro\second}$ (\cref{sec:readout}). 
The actual counting uses a scheme that measures the photon number bit by bit~\cite{Haroche1992,Heeres2016}. We denote $u_k$ the $k$-th least significant bit of $n=[u_{N} u_{N-1}...u_1]_2$. Starting from $u_1$, each value of $u_k$ is encoded into the qubit state and then read out. The main difficulty in implementing this scheme comes from the need to know the value of $n_{k-1} = [u_{k-1} \cdots u_1]_2$ in order to extract $u_k$. Each step $Q_k$ (\cref{fig:binary_decomp}.a) of the recursive determination of the $u_k$'s is based on the relation
\begin{equation}
    2^k u_k = n - n_{k-1} \text{ mod }2^k.
\end{equation}
The qubit is prepared in $\frac{\ket{g} + i \ket{e}}{\sqrt 2}$ with a $\frac{\pi}{2}$ pulse (\cref{fig:binary_decomp}.b\,\Circled{1}). Then, the memory and qubit interact dispersively for a time $T_k = \frac{2\pi}{\chi 2^k}$. $T_k$ is chosen such that the qubit ends up in one of two orthogonal states $\ket{u_k=0}$ and $\ket{u_k=1}$ that only depend on the value $u_k$ (\cref{fig:binary_decomp}.b\,\Circled{2}). Precisely, the phase of the qubit states picks up an offset $\phi(n_{k-1}) = - \frac{n_{k-1} 2 \pi}{2^k}$ for $u_k=0$. Finally, using the knowledge of $n_{k-1}$, it is possible to map $\ket{u_k=0}$ and $\ket{u_k=1}$ onto $\ket{e}$ and $\ket{g}$ using a second $\pi/2$ pulse around the right axis (\cref{fig:binary_decomp}.b\,\Circled{3}). Reading out the qubit state thus provides $u_k$ directly. This scheme minimizes the number of qubit readouts required as each binary question $Q_k$ is able to extract one bit of new information about the photon number. The number of binary questions required to determine a photon number $n$ is $N=\left\lceil \log_2 n \right\rceil$, with the caveat that the necessary precision in the waiting time increases exponentially with $N$. Note that it is possible to avoid the feedback for $n_{k-1}$ using an optimal quantum-control algorithm~\cite{Wang2020}, although with the device used here, it leads to longer questioning time and thus degraded counting fidelities. 

\section{Single-shot photocounting}
\begin{figure}    
    \centering    
    \includegraphics[width=8.6cm]{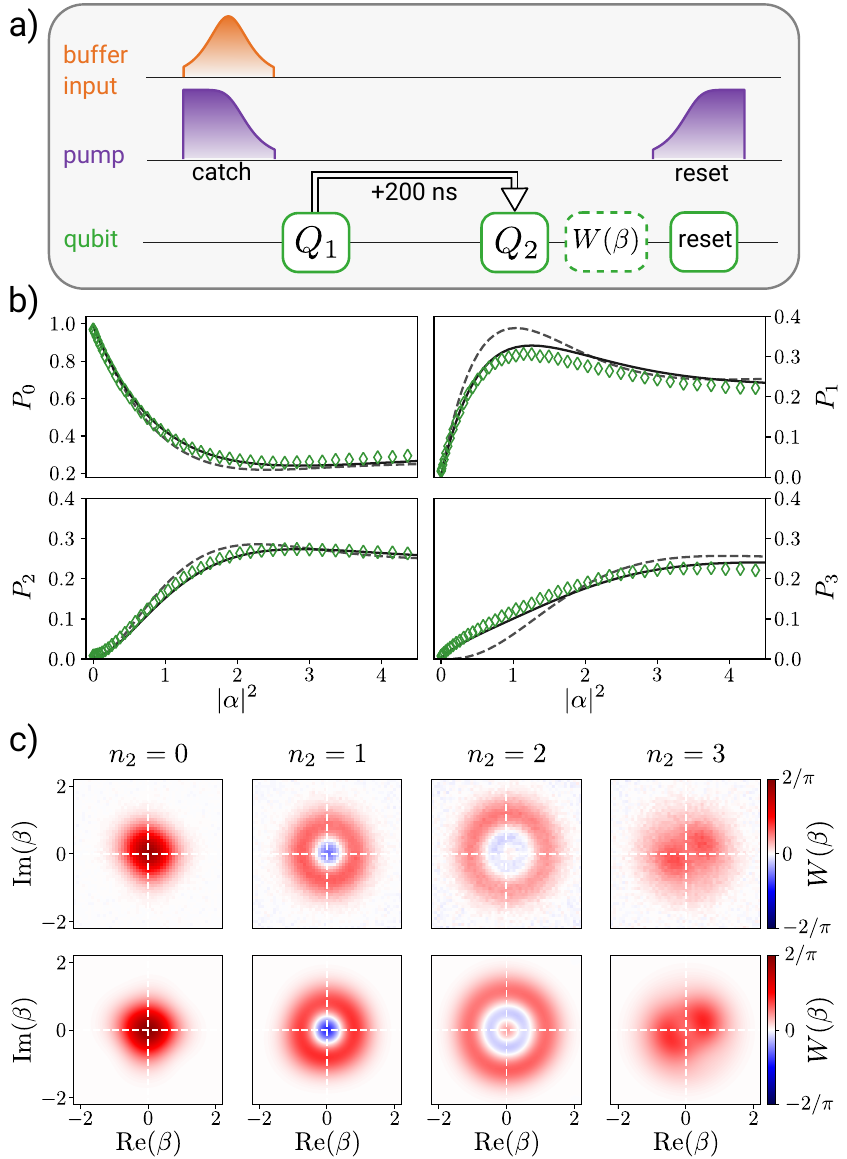}    
    \caption{Photocounting coherent states. a) Pulse sequence showing an incoming mode on the buffer with a coherent state of amplitude $\alpha$ and the optimal shape of the pump to catch the wavepacket with minimal distortion (\cref{sec:opti_catch}). The qubit performs photocounting bit by bit with pulse sequences $Q_k$'s described in \cref{fig:binary_decomp}. $Q_2$ uses the outcome of $Q_1$ in a feedback protocol that adds as little as 200-\SI{}{ns} delay. Finally, a direct Wigner tomography of the memory can be performed~\cite{Lutterbach1997,Bertet2002,Vlastakis2013} before the memory and qubit are reset. b) Green diamonds: measured probabilities $P_{n_2}$ of finding a number $n_2=n\text{ mod }4$ photons as a function of the mean photon number $\abs{\alpha}^2$ of the incoming coherent state after 200000 runs of the sequence. Dashed lines: modulo-4 Poisson distribution. Solid lines: master-equation solution without any free parameter. c) Corresponding measured (top) and simulated (bottom) Wigner functions for $\alpha=\sqrt{0.5}$ mean photons (\cref{sec:wigner,sec:num_model}). From left to right, the Wigner function is heralded on the counter outcome $n_2=0$, $1$, $2$, and $3$ out of a total of $44000$ realizations per pixel.}    
    \label{fig:SSphotocounting}
\end{figure}

We now demonstrate the number-resolved photocounting using questions $Q_1$ and $Q_2$. The device thus resolves photon numbers from $0$ to $3$. The feedback of $n_1$ is performed with minimal added latency (\SI{200}{ns}) using Quantum Machines' FPGA-based control system (OPX). To benchmark the photocounter, we send at its input a $\sech$ waveform in a coherent state of complex amplitude $\alpha$ using a microwave source (\cref{fig:SSphotocounting}.a). This state is caught in the memory using an optimal pump followed by the two binary questions $Q_1$ and $Q_2$ that reveal a number $n_2=[u_2u_1]_2$ between $0$ and $3$. 
Owing to an active reset, the counter presents a short nondeterministic average dead time of \SI{4.5}{\micro s}. The memory is reset by applying a release pump on the JRM that empties its photons into the transmission line. The qubit is reset to its ground state using a measurement-based feedback loop.

In an ideal photocounter, the distribution of $n_2$ follows a Poisson distribution modulo 4, $P_{n_2}= e^{- \abs{\alpha}^2} \sum_j [ \abs{\alpha}^{2(n_2+4j)}/(n_2+4j)!]  $ [dashed lines in \cref{fig:SSphotocounting}.(b)]. The measured probabilities $P_{n_2}$ (green diamonds) qualitatively follow the ideal Poisson distribution. However, we obtain a more quantitative agreement by solving a master equation that takes into account imperfections like the finite lifetimes of the memory and qubit, the nonzero effective temperature, and nonlinear terms~\cite{Khezri2016} $\hat{\mathcal{H}_K} = -K \hat{m}^{\dag 2} \hat{m}^2 -  K_e \ketbra{e}{e} \hat{m}^{\dag 2} \hat{m}^2 $. In the following, owing to large but uncertain value of the catch efficiency $\eta$, we set it to 1 in the simulations. The transmon qubit nonlinearity induces a self-Kerr term on the memory with rate $K/2\pi = $ \SI{27}{kHz}. When the transmon is excited in $\ket{e}$, the self-Kerr rate is offset by $K_e/2\pi = \SI{75}{kHz}$. All the above parameters are calibrated using independent measurements (\cref{sec:flux_dep}). 

A more stringent test for this model consists in predicting the measurement backaction on the quantum state of the incoming mode. Using the qubit, it is possible to perform a Wigner tomography of the collapsed quantum state of the memory conditioned on the outcome $n_2$ of the counter~\cite{Lutterbach1997,Bertet2002,Vlastakis2013} (\cref{sec:wigner}). The top panels of \cref{fig:SSphotocounting}.c show the Wigner functions for $n_2$ from $0$ to $3$ after catching a coherent state of amplitude $\abs{\alpha}=\sqrt{0.5}$. The bottom panels show the computed Wigner functions using our model above. For an outcome $n_2$, an ideal photocounter would project the incoming state $\ket{\psi}$ into $\ket{\psi_{n_2}} \propto \sum_j \ket{n_2+4j} \braket{n_2+4j}{\psi}$. Given the small mean photon number $\abs{\alpha}^2= 0.5$, the ideal state is close to Fock state $\ket{n_2}$. 
The measured Wigner functions $W(\beta)$ for $n_2\leq 2$ are indeed close to what would be obtained for pure Fock states $\ket{n_2}$. However for $n_2=3$, the relaxation of both memory and qubit induce a mixture of various Fock states, and the Wigner function does not exhibit the expected fringes. 
To quantify this agreement, we compute the fidelity $\mathcal{F}(\rho,\rho_{n_2})$ between the collapsed quantum state of the memory $\rho$ and the ideal projected quantum state $\rho_{n_2}=\dyad{\psi_{n_2}}$. Many definitions of fidelity exist for mixed states.
We chose the fidelity~\cite{Paulo2008,Miszczak2009} $\mathcal{F}(\rho,\rho')= \Tr(\rho \rho') + \sqrt{[1-\Tr(\rho^2)] [1-\Tr(\rho'^2)]} $, which can be computed in a numerically robust manner from the measured Wigner functions since $\Tr(\rho_1 \rho_2) = \pi \int W_{\rho_1}(\beta) W_{\rho_2}(\beta) \dd[2]{\beta}$. From the measured Wigner functions in \cref{fig:SSphotocounting}.c (top panels), we obtain fidelities of $86$, $52$, $32$, and $4.9\%$ for $n_2 = 0$, $1$, $2$, and $3$ respectively. This deviation from the ideal case is well captured by our model, which predicts the measured collapsed quantum states with fidelities between top and bottom panels of \cref{fig:SSphotocounting}.c above $97\%$ for the four outcomes of the counter. Simulations show that the dominant origin for the nonidealities is the qubit and memory relaxation (\cref{sec:error_budget}).

Since the model is backed up by the photon-number statistics and by the Wigner tomography, we can compute the probabilities $\mathcal{P}_{\ket{n}}(m)$ that the counter would have measured $m$ mod $4$ if a Fock state $\ket{n}$ was sent at the input (see \cref{tab:fidelities}). If the detector is giving totally random outcomes, the probabilities are equal to \SI{25}{\%} since there are four possible answers. 
Here, we obtain fidelities $\mathcal{P}_{\ket{n}}(n)$ well above \SI{25}{\%} and infidelities $\mathcal{P}_{\ket{n}}(m \neq n)$ smaller or of the same order of \SI{25}{\%}. 
Interestingly, downgraded to a photodetector that clicks when $m\neq 0$, these figures imply a detection fidelity of $1-\mathcal{P}_{\ket{1}}(0) = \SI{93 \pm 4}{\percent}$ for a single photon.
The model reveals three main sources of errors: the finite lifetimes of the memory and qubit and the rate $K_e$ (\cref{sec:error_budget}). The finite qubit lifetime affects the various $n_2$ values differently owing to the choice of encoding in the qubit state during questions $Q_k$'s. It is possible to choose which photon number to affect the least by swapping the roles of $\ket{g}$ and $\ket{e}$. The photon number corresponding to the qubit being in the excited state after each question is the one with maximum error. Here, we choose to minimize the error on $n_2=0$ and thus minimize the dark count of the counter to a measured probability of \SI{3}{\%} (measurement of $1-P_0$ at $\alpha=0$ in \cref{fig:SSphotocounting}.b). When the incoming photon number increases, the memory relaxation starts to limit the fidelity since the loss rate of the memory increases with photon number. It explains most of the decrease of fidelity with photon number from 99\% down to 56\%. Finally, because of the nonzero $K_e$, during the time $T_k$ of the question $Q_k$, the qubit acquires an additional parasitic phase that rapidly increases with the photon number resulting in larger infidelities for higher $n$.

\begin{table}[]
\centering
\begin{ruledtabular}

\begin{tabular}{c c c c c}
 $\mathcal{P}_{\ket{n}}(m)$   & $\ket{0}$ & $\ket{1}$ & $\ket{2}$ & $\ket{3}$  \\   \hline
  m = 0   & \textbf{\SI{99}{\%}} & ($7 \mp 4$)\SI{}{\percent} & ($24 \mp 3$)\SI{}{\percent} & ($9 \mp 4$)\SI{}{\percent}  \\
  
 m = 1 & $<$\SI{1}{\percent} & \textbf{\SI{76 \pm 3}{\percent}} & \SI{4.2 \pm 0.2 }{\percent} & \SI{27 \pm 1}{\percent}\\
   
   m = 2 & $<$\SI{1}{\percent} & \SI{1.03 \pm 0.01}{\percent} & \textbf{\SI{71 \pm 3}{\percent}} & \SI{9.7 \pm 0.3}{\percent} \\
    
    m = 3 & $<$\SI{1}{\percent} & \SI{16 \pm 1}{\percent} & \SI{1.5 \pm 0.1}{\percent} & \textbf{\SI{54 \pm 2}{\percent}} \\ 
\end{tabular}
 \end{ruledtabular}
 \caption{Probabilities of getting the outcome $m$ if the incoming mode is in Fock state $\ket{n}$. The probabilities are computed using the master equation validated by \cref{fig:SSphotocounting}. The uncertainties correspond to the range of possible values on the catch efficiency $\eta$. Diagonal terms are all above \SI{25}{\%}, which would correspond to a completely random counter with 4 possible outcomes.}
 \label{tab:fidelities}
\end{table}

\section{Conclusion}
We develop a photocounter using measurement-based feedback that is able to resolve the photon number from $n=0$ up to $n=3$ in a propagating microwave mode. The counter features a time-resolved power meter able to determine the envelope of the incoming waveform \emph{in situ}, which optimizes the detection efficiency up to $\eta =\SI{0.96}{} \pm \SI{0.04}{}$. Future devices with longer lifetimes would considerably improve the fidelities $\mathcal{F}$ above. The reset would then release a faithfully collapsed quantum state into the line, making the photocounter quantum nondemolition. The counter would then quickly scale up to resolve higher photon number thanks to its logarithmic complexity. 
The photocounter can also be used in a degraded mode to measure parity by asking a single question $Q_1$ as in Refs.~\cite{Kono2018,Besse2018}, and thus perform propagating Wigner tomography~\cite{Besse2019}. 
Microwave photodetection and photocounters enable quantum-optics-like experiments in the microwave range and facilitate the implementation of a quantum network. For instance, photodetection has made possible the entanglement between remote stationary qubits~\cite{Narla2016,Campagne-Ibarcq2018,Kurpiers2018}. However any protocol requiring feedback on the photon number in a propagating mode needs a single-shot photocounter. Therefore, a direct application consists in reaching the quantum limit for the discrimination between two coherent states \cite{Dolinar1973}, with obvious applications in quantum sensing.

\begin{acknowledgments}
We are grateful to Olivier Buisson, Michel Devoret, Zaki Leghtas, Danijela Markovi\'c, Mazyar Mirrahimi, Alain Sarlette for discussions.
We acknowledge IARPA and Lincoln Labs for providing a Josephson Traveling-Wave Parametric Amplifier. The device is fabricated in the cleanrooms of Coll\`ege de France, ENS Paris, CEA Saclay, and Observatoire de Paris. The feedback code is developed in collaboration with Quantum Machines~\cite{Github}. This work is part of a project that has received funding from the European Union's Horizon 2020 research and innovation program under Grant Agreement No. 820505. 
\end{acknowledgments}

\appendix
\section{Measurement setup}
\label{sec:setup}
\begin{figure*}[h]
    \centering
    \includegraphics[width=16cm]{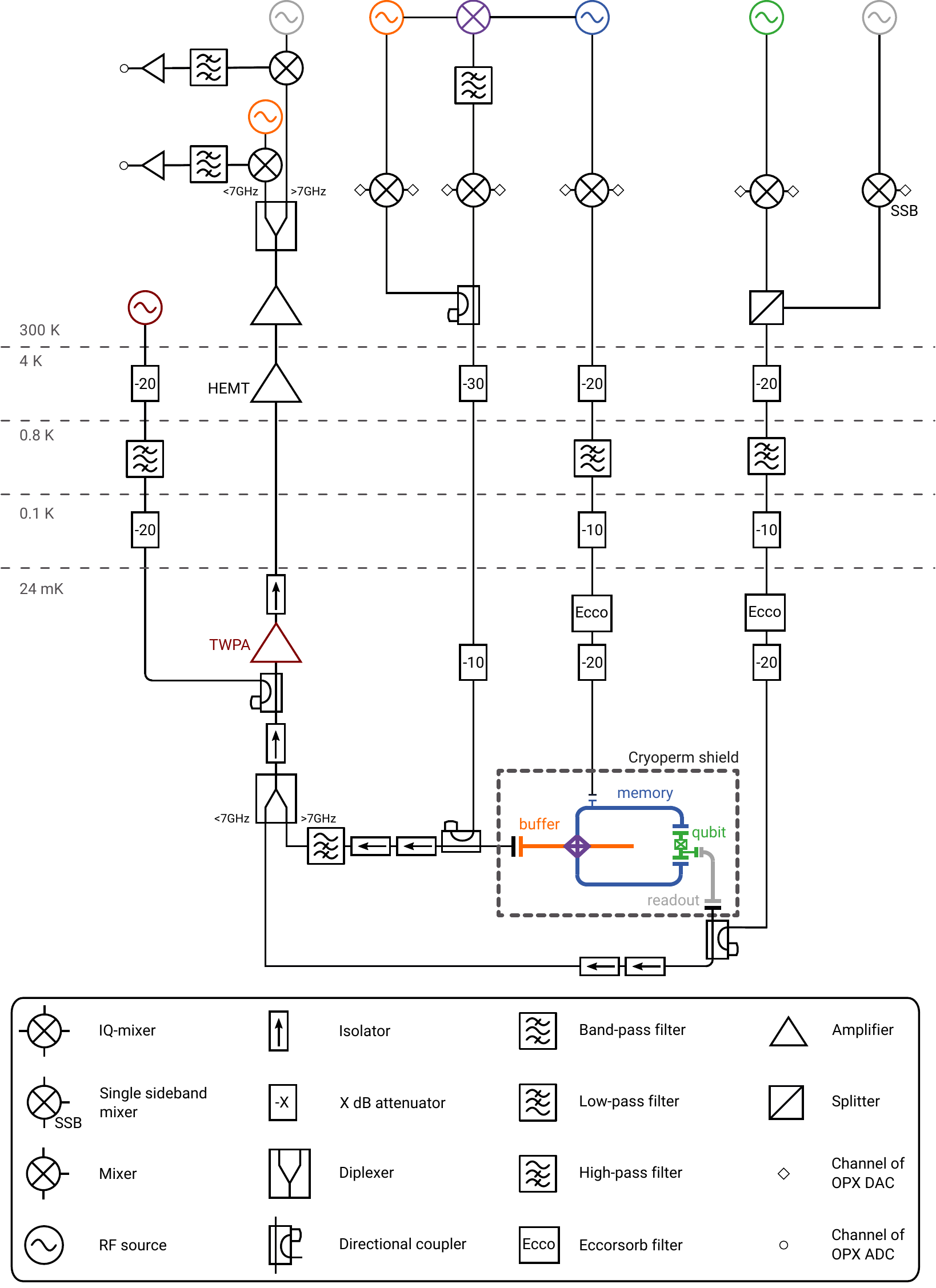}
    \caption{Scheme of the measurement setup. The rf sources color refers to the frequency of the  matching element in the device up to a modulation frequency. Identically colored sources represent a single instrument with a split output.}
    \label{fig:wiring}
\end{figure*}

The sample and its fabrication are described in Ref.~\cite{Peronnin2019}. The sample is cooled down to \SI{24}{\milli\kelvin} in a BlueFors LD250 dilution refrigerator. The diagram of the microwave wiring is given in \cref{fig:wiring}. The buffer, memory, qubit and readout pulses are generated by modulation of continuous microwave tones produced, respectively, by generators E8752D from Keysight, SGS100A from Rohde$\&$Schwarz, SGS100A from Rohde$\&$Schwarz, and  SynthHD PRO from Windfreak set, respectively, at frequencies $f_b+$\SI{50}{}, $f_m-$\SI{120}{}, $f_q+$\SI{200}{}, and $f_r+$\SI{51}{MHz}. The pump pulses are also generated by modulation of continuous microwave tone, however the local oscillator at $f_b-f_m +$\SI{170}{MHz} is produced by mixing the buffer and the memory rf sources for phase stability. The readout is modulated through a single sideband mixer while the others are modulated via IQ mixers.
The IF modulation pulses are generated by nine channels of an OPX from Quantum Machines with a sample rate of \SI{1}{GS\per \second}. The acquisition is performed, after down-conversion by their local oscillators, by digitizing a \SI{51}{MHz} (readout) or a \SI{50}{MHz} (buffer) signal with the \SI{1}{GS\per \second} analog-to-digital converter (ADC) of the OPX from Quantum Machines. The signals coming out of the buffer mode and of the readout mode are multiplexed into a single transmission line using a diplexer before getting amplified by a traveling wave parametric amplifier \cite{Macklin2015} (TWPA, provided by IARPA and the Lincoln Labs). The TWPA is pumped at a frequency $f_{TWPA}=$ \SI{7.636}{GHz} and at a power that allowed the TWPA to reach a system efficiency of \SI{18}{\%} from the buffer output to the ADC. The signal coming out of the buffer mode is filtered using a \SI{20}{cm} waveguide WR62 with a cutoff frequency at \SI{9.8}{GHz} in order to prevent the strong pump of the JRM from reaching the TWPA and reciprocally. The next stage of amplification is performed by a HEMT amplifier (from Caltech) at \SI{4}{K} and by two room-temperature amplifiers.

\section{System characterization and flux dependence}
\label{sec:flux_dep}
\begin{figure*}[h]
    \centering
    \includegraphics[width=14cm]{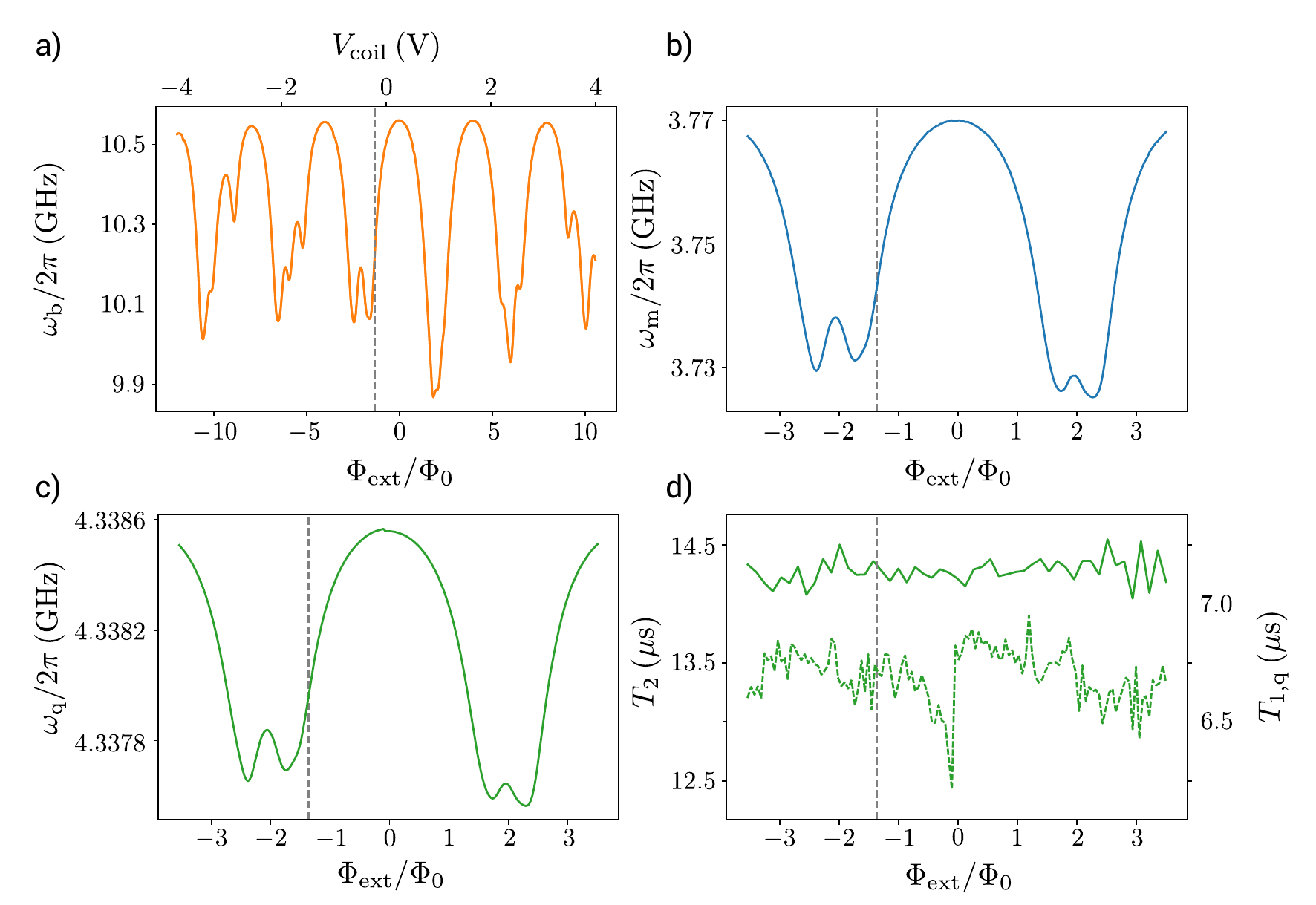}
    \caption{a) Buffer frequency, b) memory frequency, c) qubit frequency, d) qubit decoherence time $T_2$ (dashed line) and lifetime $T_{1,\rm q}$ (solid line) as a function of flux $\Phi_{\rm ext}$ in the inner loops of the JRM. Notice that the flux range is different in a) compared to b-d). Vertical dashed line: working point for the main text.}
    \label{fig:freq_and_lifetimes}
\end{figure*}

Using a vector network analyzer we measure the buffer resonance frequency as a function of the current running through a superconducting coil directly above the sample. The extracted buffer frequency $\omega_{\rm b}$ is displayed in \cref{fig:freq_and_lifetimes}.a. The current is generated by applying a voltage $V_\mathrm{coil}$ to a resistor in series with the coil. The periodicity of the buffer frequency allows us to convert the voltage $V_\mathrm{coil}$ into a flux $\Phi_\mathrm{ext}$ through the four inner loops of the JRM. 

Even though the qubit consists in a single junction transmon, its frequency $\omega_{\rm q}$ has a slight flux dependence due to its coupling with the memory. The qubit frequency, as a function of the flux, is extracted from Ramsey oscillations (\cref{fig:freq_and_lifetimes}.c). With these measurements, we are also able to extract the qubit coherence time $T_2$ as a function of flux $\Phi_\mathrm{ext}$ (solid line in \cref{fig:freq_and_lifetimes}.d).

The memory cannot be probed directly in reflection nor in transmission with the measurement setup. To measure its frequency $\omega_{\rm m}$ (\cref{fig:freq_and_lifetimes}.b), we use the qubit to determine at what excitation frequency the memory gets populated. We send a probe pulse on the memory via its weakly coupled port followed by a conditional $\pi$ pulse on the qubit at $\omega_{\rm q}$. The qubit is thus excited only if the memory has zero photons.  Measuring the qubit average excitation as a function of probe frequency leads to determining the frequency $\omega_{\rm m}$ at which the state $\ket{0}$ is most depleted. We also measure the relaxation times of the qubit  $T_{1,\rm q}$ (see  \cref{fig:freq_and_lifetimes}.d). The qubit decoherence time is limited by the relaxation since $T_{2}$ is close to $2T_{1, \rm q}$.

\begin{figure*}[h]
    \centering
    \includegraphics[width=14cm]{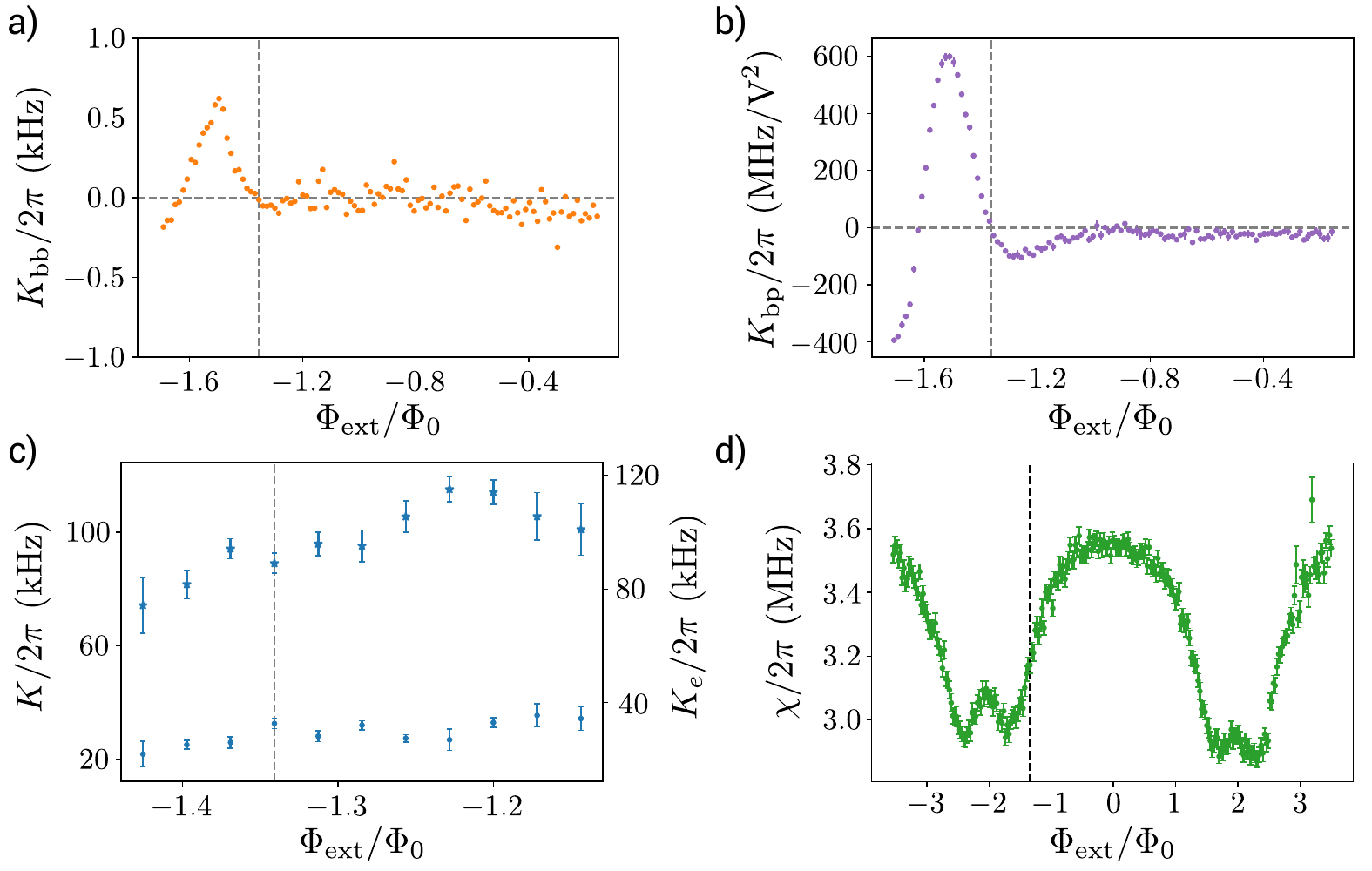}
    \caption{Rates of nonlinear terms in the device as a function of the external flux $\Phi_{\rm ext}$. Notice that the flux range is different for each panel. a) Buffer self-Kerr rate $K_{\rm bb}$, b) Pump-buffer cross-Kerr rate $K_{\rm bp}$, c) Dots, memory self-Kerr rate $K$, stars, nonlinear rate $K_e$. d) Dispersive shift $\chi$ between qubit and memory.
    }
    \label{fig:Kerr}
\end{figure*}

 We extract the buffer self-Kerr rate $K_{\rm bb}$ from the dependence of its frequency $\omega_{\rm b}$ as a function of probe power (\cref{fig:Kerr}.a). To measure the pump-buffer cross-Kerr rate $K_{\rm bp}$ (\cref{fig:Kerr}.b), we measure $\omega_{\rm b}$ while driving the pump at various powers. The pump is driven off resonance from $\omega_{\rm b}-\omega_{\rm m}$ to avoid frequency conversion. The buffer self-Kerr and buffer-pump cross-Kerr rates both vanish at the same flux point \cite{Flurin2014}, which we hence choose as our working point. A nonzero cross-Kerr rate would indeed make the pump optimization more challenging for catch and reset operations.

The measurement of the memory self-Kerr rate $K$ and the qubit-dependent nonlinear rate $K_e$ are done in a previous cool down by monitoring the average phase acquired by a coherent state in the memory mode as a function of time while varying the mean photon number and the initial qubit state. Having prepared the qubit in either $\ket{g}$ or $\ket{e}$, we load the memory with a coherent state of amplitude $\alpha=\sqrt{n}$. We then wait for a time $t_{\rm int}$.  Finally, we release the state of the memory into the transmission line and record the average phase $\phi(t_{\rm int})$ of the released pulse. The detuning $\delta \omega_{\rm m}$ between the resonant frequency of the memory $\omega_{\rm m}$ and a reference resonant frequency (when the memory is in the vacuum state and the qubit in $\ket{g}$) can be determined as $\delta \omega_{\rm m} =  \dv{\phi}{t_{int}}$. The slope of $\delta \omega_{\rm m}$ as a function of mean photon number $n$ then gives the self-Kerr rate $K$ ($K_e+K$) when the qubit is prepared in $\ket{g}$ ($\ket{e}$). The rates $K$ and $K_e$ are plotted as a function of flux in \cref{fig:Kerr}.c. 

Using a populated Ramsey protocol (see details in \cref{fig:average_counting}) as function of flux, we also extract the qubit-memory dispersive coupling $\chi$ (\cref{fig:Kerr}.d). It is also performed in a previous cool down.

\section{Readout optimization}
\label{sec:readout}
\begin{figure*}[t]
    \includegraphics[width=14cm]{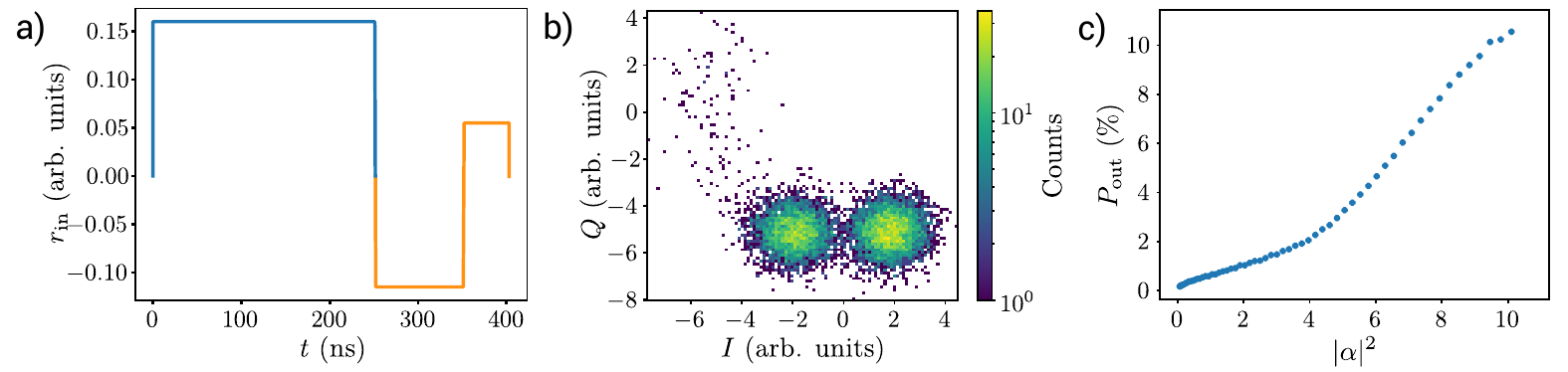}
    \caption{Readout optimization. a) CLEAR-like readout pulse sequence. Driving amplitude $r_{\rm in}$ of the readout as a function of time $t$. Blue, readout excitation; orange, readout reset. b) Histogram of the two demodulated quadratures $I$ and $Q$ of the reflected readout pulse for $10^4$ realizations after applying a $\pi/2$ pulse on the qubit. The two peaks correspond to the $\ket{g}$ and $\ket{e}$ states of the qubit. The few points in the upper-left corner correspond to the transmon in an ionized state. c) Probability to observe the transmon outside of its qubit subspace as a function of the mean number of photons inside of the memory for the readout power used in the main text.}
    \label{fig:clear}
\end{figure*}

The readout strategy is a compromise between readout speed, fidelity and QNDness. Note that the feedback protocol of the photocounter requires a QND measurement so that non-QNDness limits the counter fidelity.
In order to make fast and faithful qubit measurements, we implement a CLEAR-like sequence \cite{McClure2016} with amplitude $r_{\rm in}(t)$ shown in \cref{fig:clear}.a. The QNDness of the readout is limited by the possible ionization of the transmon out of the qubit subspace~\cite{Sank2016,Lescanne2019b}. We find that not only this constraint limits the amplitude of the readout pulse but also that the ionizaiton probability increases with the occupation of the memory mode (\cref{fig:clear}.b). In future design, the efficiency of the photocounter could be improved by using less sensitive coupling schemes~\cite{Touzard2019,Ikonen2019,dassonneville2020}.

In order to determine the state of the qubit as a function of the reflected signal with the best fidelity, we use a set of optimized demodulation weights that we compute to maximize the complex signal difference between the ground and excited states as shown in Ref.~\cite{Ryan2015}. It is convenient to quantify the readout error using the overlap $\epsilon_0$ between the two Gaussian distributions corresponding to the two qubit states~\cite{Walter2017}.

The qubit temperature is measured by repeatedly measuring the qubit, recording the demodulated signal from the readout into a complex histogram (such as the one shown in \cref{fig:clear}b) and fitting it with a set of two two-dimensionnal (2D) Gaussians of equal width. The temperature is then extracted by taking the ratio of the amplitudes of the two Gaussians. For additional precision, the center of the Gaussian corresponding to the qubit being in the excited state $\ket{e}$ is estimated by doing the same measurements after performing a $\pi$ pulse such that the final fit only had two free parameters: the center of the Gaussian corresponding to $\ket g$ and the qubit temperature. We find an effective temperature of \SI{33}{mK}. 

\section{Optimal catching pump}
\label{sec:opti_catch}
In this section, we derive the optimal pump to catch an arbitrary wavepacket with a bandwidth smaller than the bandwidth of the buffer $\kappa_{\rm b} = 2\pi\cdot\SI{20}{MHz}$. We first derive the optimal pump to catch an incoming wavepacket assuming $\kappa_{\rm m} = 0$ and we then show that a small memory relaxation rate $\kappa_{\rm m}$ and a  cross-Kerr rate $K_{\rm bp}$ do not prevent the catch from being complete. 

\subsection{Ideal case}
Let us consider the Langevin equations for the buffer $b$ and memory $m$ with a conversion pump $p$ in the frame rotating with $b_\mathrm{in}$ and $m$
\begin{align*}
    \dv{b}{t} &= -\frac{\kappa_{\rm b}}{2} b(t) - g_3 p^*(t) m(t) +\sqrt{\kappa_{\rm b}} b_{in}(t) \\
    \dv{m}{t} &= g_3 p(t) b(t),
\end{align*}
where, for simplicity, we assume that the external flux used is chosen such that all the self-Kerr and cross-Kerr terms cancel out. Note that an arbitrary choice of phase reference allows us to constrain $b$ to be a real function.

We start by parametrizing the equations with dimensionless variables using $\tau = \frac{\kappa_{\rm b}}{2}t$, $u=\frac{2g_3p}{\kappa_{\rm b}}$

\begin{align*}
    \dot{b} &= -b - u^*m + \frac{2}{\sqrt{\kappa_{\rm b}}} b_{in}\\
    \dot{m} &= ub
\end{align*}
where the dots denote the derivatives with respect to $\tau$.

Catching the incoming wavepacket $b_{\mathrm{in}}$ perfectly comes down to finding the pump $u(\tau)$ such that $b_{\rm out} = 0$ uniformly. Since $b_{\mathrm{in}} + b_{\rm out} = \sqrt{\kappa_{\rm b}} b$, $u$ is the solution of the following differential equations

\begin{align}
   u m^* &= b - \dot{b} \label{eq:catch1}\\
    \dot{m} &= u b.\label{eq:catch2}
\end{align}

For any signal with a bandwidth lower than the buffer coupling rate $\kappa_b$, these equations can be solved numerically. In the following subsection, we focus on the case of a $\sech$ input waveform, where the calculation can be carried out analytically.

\subsubsection*{Case of an incoming hyperbolic secant waveform}

In the experiment, we frequently use an incoming hyperbolic secant waveform $b(\tau) = \frac{\sqrt{\lambda/2}}{2} \sech(\lambda \tau / 2)$. To do so, we remark that 
\[
y = \abs{m}^2 + b^2
\]
is a flat output~\cite{Fliess1995}, meaning that $m$, $u$ and $b$ can be expressed as functions of $y$, $\dot{y}$ and $\ddot{y}$.  Combining \cref{eq:catch1} and \cref{eq:catch2}, we get $m^* \dot{m} = (b - \dot{b})b$. Taking the real part and using the limited bandwidth ($\dot y \leq 2y$) and the assumption that there is no loss ($0 \leq \dot y$), we get 
\begin{equation}
    \label{eq:flat}
    b^2 = \dot y/2 \qc \abs{m}^2 = y - \dot y/2.
\end{equation}

Setting $y = \frac{1}{1 + e^{-\lambda \tau}}$ with $0 \leq \lambda \leq 2$, using \cref{eq:flat}, we get $b(\tau) = \frac{\sqrt{\lambda/2}}{2} \sech(\lambda \tau / 2)$ as desired and $\abs{m} = \frac{\sqrt{e^{\lambda \tau} + 1 - \lambda/2}}{2} \sech(\lambda \tau / 2)$.
Multiplying \cref{eq:catch1} by its complex conjugate, we get $\abs{u} = \frac{b - \dot{b}}{\abs{m}}$.
From \cref{eq:catch1}, we can also see that $\arg(u) = \arg(m)$. Hence, there is a function $\theta$ such that $m = \abs{m} e^{i\theta}$ and $u = \abs{u} e^{i \theta}$. By multiplying \cref{eq:catch2} by $m^*$ and  using  \cref{eq:catch1} one gets $\dot m m^* = (b-\dot b)b$. Since  $b$ is real,   the imaginary part, yields  $\dot{\theta} = 0$. For simplicity, we choose $\theta(\tau) = 0$, which leads to
 \begin{equation}
    \label{eq:flat2}
    u = \frac{b - \dot{b}}{\abs{m}}.
\end{equation}

Finally we find
\begin{equation}
    u(\tau) = \sqrt{\frac{\lambda / 2}{e^{\lambda \tau} +1 - \lambda /2}}\left( 1 + \frac{\lambda}{2} \tanh(\lambda \tau / 2) \right).
\end{equation}

Going back to the original time variable $t$, we conclude that an incoming wavepacket with a shape $b_{\rm in}(t) = \sqrt{\frac{\lambda}{8 \kappa_{\rm b}}} \sech(\lambda \kappa_{\rm b} t / 4)$ is perfectly caught by a pump $p_{\rm opt}(t) = \frac{2g_3}{\kappa_{\rm b}} \sqrt{\frac{\lambda / 2}{e^{\lambda \kappa_{\rm b} t / 2} +1 - \lambda /2}}\left( 1 + \frac{\lambda}{2} \tanh(\lambda \kappa_{\rm b} t / 4) \right)$.

\subsection{Finite memory lifetime}
In order to account for the memory relaxation rate $\kappa_{\rm m}$, the Langevin equations become
\begin{align*}
    \dv{b}{t} &= -\frac{\kappa_{\rm b}}{2} b(t) - g_3 p^*(t) m +\sqrt{\kappa_{\rm b}} b_{in}(t) \\
    \dv{m}{t} &= -\frac{\kappa_{\rm m}}{2} m(t) + g_3 p(t) b(t).
\end{align*}
Without loss of generality, we assume that $b_{\rm in}$ and $p$ are real, hence $m$ and $b$ are also real. Using the same definition for $y$ and introducing $\varepsilon = \kappa_{\rm m}/\kappa_{\rm b}$, we get the following  modified version of \cref{eq:flat} to derive $b$ and $m$ as algebraic functions of $y$ and $\dot y$.
\begin{equation}
     (1+\varepsilon) b^2 = \dot y/2+ \varepsilon y \qc (1+\varepsilon)\abs{m}^2 = y - \dot y/2 \label{eq:flatbis}
\end{equation}
Given \cref{eq:flat2}, $u$ can be expressed as an algebraic function of $y$, $\dot y$, and $\ddot y$. 
In this case the no-loss assumption is replaced by the weaker constraint that the ratio between the outgoing power $-\dv{y}{t}$ and the total energy $y$ is smaller than $\kappa_{\rm m}$, i.e., $ \dv{y}{t} \geq -\kappa_{\rm m} y$ (i.e., $\dot y/2+ \varepsilon y\geq 0$). The bandwidth limit $\dv{y}{t} \leq \kappa_{\rm b} y$ remains valid (i.e., $ \dot y \leq 2y$).

To carry on the calculation analytically, we set $y = \frac{1}{e^{2 \varepsilon \tau} + e^{-\lambda \tau}}$ so that
\begin{align*}
    b(\tau) =\sqrt{\frac{\lambda / 2 + \varepsilon}{1+\varepsilon}}\frac{1}{e^{(\lambda / 2 + 2\varepsilon) \tau} + e^{-\lambda \tau / 2}} .
\end{align*}
We also get
\begin{align*}
    m(\tau) = \sqrt{ e^{(\lambda + 2\varepsilon) \tau} + \frac{1 - \lambda/2}{1+\varepsilon}} \frac{1}{e^{(\lambda / 2 + 2\varepsilon) \tau} + e^{-\lambda \tau / 2}}.
\end{align*}
From the above expressions for $b$ and $m$, we can then compute $u$ using \cref{eq:flat2}.
Given the small value of $\varepsilon\approx 0.002$ in the device of the main text, we choose to neglect the memory relaxation and to use the results from the ideal case above.

\subsection{Finite cross-Kerr rate}
Even in the presence of a small cross-Kerr rate $K_{\rm bp}$ between the buffer and the pump, an optimal catch pump can be found which guarantees that no signal is reflected \textit{i.e.} $b_{out} = 0$.
The modified Langevin equations are as follows
\begin{align*}
    \dv{b}{t} &= -\left( \frac{\kappa_{\rm b}}{2} + i K_{\rm bp}\abs{p(t)}^2 \right) b(t) - g_3 p^*(t) m(t) +\sqrt{\kappa_{\rm b}} b_{in}(t) \\
    \dv{m}{t} &= -\frac{\kappa_{\rm m}}{2} m(t) + g_3 p(t) b(t).
\end{align*}
Introducing the dimensionless cross-Kerr rate $k = K_{\rm bp} \kappa_{\rm b} / g_3^2$, we get a modified version of equations (\ref{eq:catch1}) and (\ref{eq:catch2})
\begin{align*}
    u m^* &= b - \dot{b} + i k \abs{u}^2b\\
    \dot{m} &= -\varepsilon m + u b.
\end{align*}
Since $b$ is real, the real quantity $y=\abs{m}^2+b^2$ can still be used to parametrize the system, despite the fact that $m$ and $u$ are now complex. The values of $b$ and $\abs{m}$ can still be expressed as functions of $y$ and $\dot y$ by \cref{eq:flatbis}. The modulus $\abs{u}$ of the pump is obtained by solving 
\begin{align*}
    \abs{u}^2 \abs{m}^2 = (b - \dot{b})^2 +  k^2 \abs{u}^4 b^2
    .
\end{align*}
The argument $\theta_m$ of $m$ results from the integration
\begin{equation*}
    \theta_m (\tau) = \theta_m(0) +  k \int_0^\tau \frac{\abs{u(s)}^2 b(s)^2}{\abs{m(s)}^2}\dd{s},
\end{equation*}
where $\abs{u}$, $\abs{m}$ and $b$ are algebraic functions of $y$, $\dot y$ and $\ddot y$.   
 The argument $\theta_u$ of $u$ is given by the argument of $m(b - \dot{b} + i k \abs{u}^2b)$ which coincides then with the argument of $(\dot m+\varepsilon m)/b$. 

Using the above derivation, one sees that finding the optimal pump in the case of cross-Kerr effect requires not only to adjust the envelope of the pump, as done in the main text, but also adjusting the phase of the pump $\theta_u$ dynamically to compensate for the time-dependent buffer frequency shift.

\section{Different methods for measuring the mean photon number}
\label{average_counting}

\begin{figure*}[t]
    \includegraphics[width=14cm]{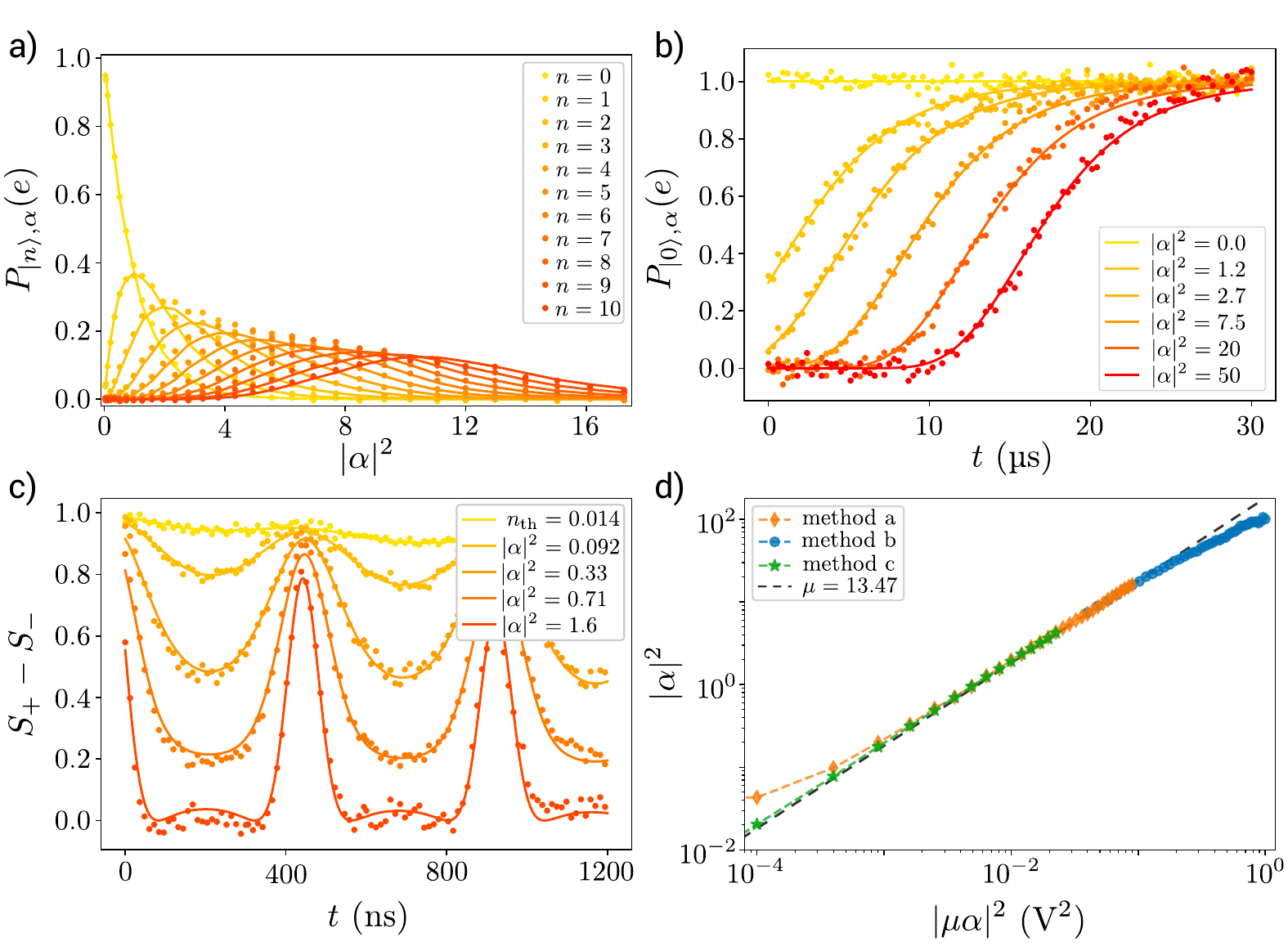}
    \caption{Three methods for calibrating the memory-displacement amplitude. The measurements are performed on a previous cool down. a) Photon-number selective $\pi$ pulse. Dots: measured probability to have $n$ photons in the memory as a function of $\abs{\alpha}^2$. Solid lines: Poisson distribution fitted to calibrate the mean photon number on the $x$ axis. b) Vacuum detector. Dots: probability $P_{\ket{0}, \alpha}(e)$ that the memory is empty as a function of waiting time for various preparation amplitudes. Solid lines: fit of the measured probabilities using the expression for memory relaxation in the text. c) Populated Ramsey. Dots: signal difference $S_+ - S_-$ between two encodings of the Ramsey-like interferences in the presence of various mean photon numbers $\expval{n}$. Solid lines: theoretical prediction allowing to calibrate the displacement amplitude and the thermal occupancy $n_\mathrm{th}$. d) Result of the calibration using the three methods: photon number selective $\pi$ pulse in orange diamonds, vacuum detector in blue dots and populated Ramsey in green stars. The black dashed line represents the overall fitted value for $\mu$.}
    \label{fig:average_counting}
\end{figure*}

We use several methods to measure the mean photon number $\expval{n}$ in the memory in order to calibrate the buffer and memory displacement pulses (\cref{fig:average_counting}). The experiment begins by a displacement pulse on the memory mode with a driving voltage $\mu\alpha$, where $\mu$ is a conversion factor between voltages and amplitudes to be determined. The following procedures then determine the mean photon number $\expval{n}=\abs{\alpha}^2+n_\mathrm{th}$ as a function of the driving voltage by different ways and thus calibrate $\mu$. $n_\mathrm{th}$ is the residual equilibrium thermal photon number in the memory.

\subsection{Photon number selective $\pi$ pulse}
The first method relies on the possibility to perform a $\pi$ pulse $\Pi_{\ket{n}}$ conditionally on the photon number $n$. It is done by driving the qubit at frequency $\omega_{\rm q}-\chi n$ with a long enough pulse so that the frequency spreading is smaller than $\chi/2$. The pulse maps the probability to have $n$ photons $P(n, \alpha)$ into the measured probabilities $P_{\ket{n}, \alpha}(e)$ for the qubit to be found in its excited state (\cref{fig:average_counting}.a). Fitting the distribution $P(n,\alpha)$ for each $\alpha$ by a Poisson distribution, we calibrate $\mu$ neglecting the thermal population. A limitation of this method occurs at high photon number. Indeed, the dispersive shift $\chi$ slightly depends on photon number $n$, so that the qubit drive frequency is off resonant. 

\subsection{Vacuum detector}
To calibrate the conversion factor $\mu$ at high photon numbers $\abs{\alpha}^2\gg 1$, we perform another method, which is to use the qubit as a vacuum detector~\cite{Peronnin2019}. Applying a $\pi$ pulse $\Pi_{\ket{0}}$ encodes the probability that the memory is empty into the probability for the qubit to be in the excited state. Now, after a waiting time $t$, the memory has relaxed and, neglecting $n_\mathrm{th}$ for the large $\abs{\alpha}^2$, the measured probability $P_{\ket{0}, \alpha}(e)$ evolves following $\exp(-\abs{\alpha}^2 e^{- t/T_{1, \rm m}} )$ (\cref{fig:average_counting}.b). Fitting the value of $\mu$ for each value of $\mu\alpha$ to match this expression with the measured $P_{\ket{0}, \alpha}(e,t)$ leads to an accurate determination of the conversion factor $\mu$ as a function of $\alpha$. This photon number calibration has a higher range than the previous one but is less sensitive for low average photon numbers.

\subsection{Populated Ramsey oscillations}
Our last method to calibrate the conversion factor $\mu$  relies on a Ramsey-like sequence \cite{Campagne-ibarcq2015} (\cref{fig:average_counting}.c). After the coherent displacement of the memory, we prepare the qubit in an equal superposition of ground and excited states by applying an unconditional $\frac{\pi}{2}$ pulse. After a waiting time $t$, the phase of the superposition increases by $\chi n t$ for each Fock state $\ket{n}$. We then apply a second unconditional $\pm \frac{\pi}{2}$ pulse giving the signal $S_{\pm}$. The signal difference is given by $S_+ - S_- = \cos(\expval{n} \sin(\chi t)) \exp(\expval{n}(\cos(\chi t)-1) -t/T_{2}) $ from which we extract the mean photon number $\expval{n}$. Without driving the memory, the measured mean number gives the thermal population of the memory $n_{\rm th} =0.014$ corresponding to an effective temperature of \SI{44}{mK}. Offsetting the measured  $\expval{n}$ by this thermal occupation leads to a calibration of $\mu$. This last method has a good sensitivity at low photon numbers, however, it cannot be used for large photon numbers where the pattern becomes insensitive to $\expval{n}$.

\subsection{Comparison}
In \cref{fig:average_counting}.d, we show the outcome of the three methods by plotting the measured $\abs{\alpha}^2$ as a function of driving power. The methods agree over their respective ranges. For large mean photon number $\abs{\alpha}^2 > 20$, due to memory self-Kerr, the mean photon number is expected to differ and be smaller than the linear behavior $\vert \mu \alpha \vert^2$.


\section{Numerical model}
\label{sec:num_model}
We simulate our system using the QuantumOptics.jl library\cite{KRAMER2018109}.

The device Hamiltonian reads~\cite{Peronnin2019}
\begin{align}
    \hat{\mathcal{H}}/\hbar =& \omega_{\rm b} \hat{b}^\dag \hat{b} + \omega_{\rm m} \hat{m}^\dag \hat{m} + \frac{\omega_{\rm q}}{2} \hat{\sigma}_z  \notag \\
    & +g_3 p \hat{m}^\dag \hat{b} + g_3^* p^* \hat{m} \hat{b}^\dag \notag \\    
    &-\chi \hat{m}^{\dag}\hat{m} \ketbra{e}{e} -K \hat{m}^{\dag 2} \hat{m}^2 -  K_e \ketbra{e}{e} \hat{m}^{\dag 2}\hat{m}^2. \notag
\end{align}
To simplify the model, we restrict the transmon to its first two levels and we do not consider the readout resonator and its dispersive coupling to the qubit. We simulate the readout of the qubit by an instantaneous projective measurement taking place at half of our experimental readout duration. During the readout time, before and after the projection, the system evolves freely. We also take into account the overlap error
$\varepsilon_o$ \cite{Walter2017} in the readout, which we measure to be below 1\%.

Moreover, we consider the catch of the wavepacket incoming onto the buffer to be optimal (\cref{sec:opti_catch}). Thus, we further reduce the numerical Hilbert space by putting aside the buffer and the pump. The catch is then simulated by an instantaneous displacement on the memory field. 

Finally, we model our system in the memory and qubit rotating frame using the following Hamiltonian.
\begin{align}
    \hat{\mathcal{H}}/\hbar = & -\chi \hat{m}^{\dag}\hat{m} \ketbra{e}{e} -K \hat{m}^{\dag 2} \hat{m}^2 -  K_e \ketbra{e}{e} \hat{m}^{\dag 2}\hat{m}^2 \notag \\
    & + \Re(f(t)) \hat{\sigma}_x + \Im(f(t)) \hat{\sigma}_y
\end{align}
with $f(t)$ the complex envelope containing all the qubit drives. Using a time-dependent Hamiltonian allows us to simulate the optimal counting with the questions $Q_0$ and $Q_1$. For instance, we can thus accurately take into account the finite duration of the $\frac{\pi}{2}$ pulses.
A Lindblad master equation enables us to take into account the qubit relaxation time $T_{1, \rm q}$ and pure dephasing time $T_{\phi}$ and the cavity lifetime $T_{1, \rm m}$ as well as temperatures of qubit and memory. We restrict the Hilbert space of the memory mode between 0 and $29$ photons.

\section{Wigner tomography}
\label{sec:wigner}
\begin{figure*}[t]
    \centering
    \includegraphics[height=22cm]{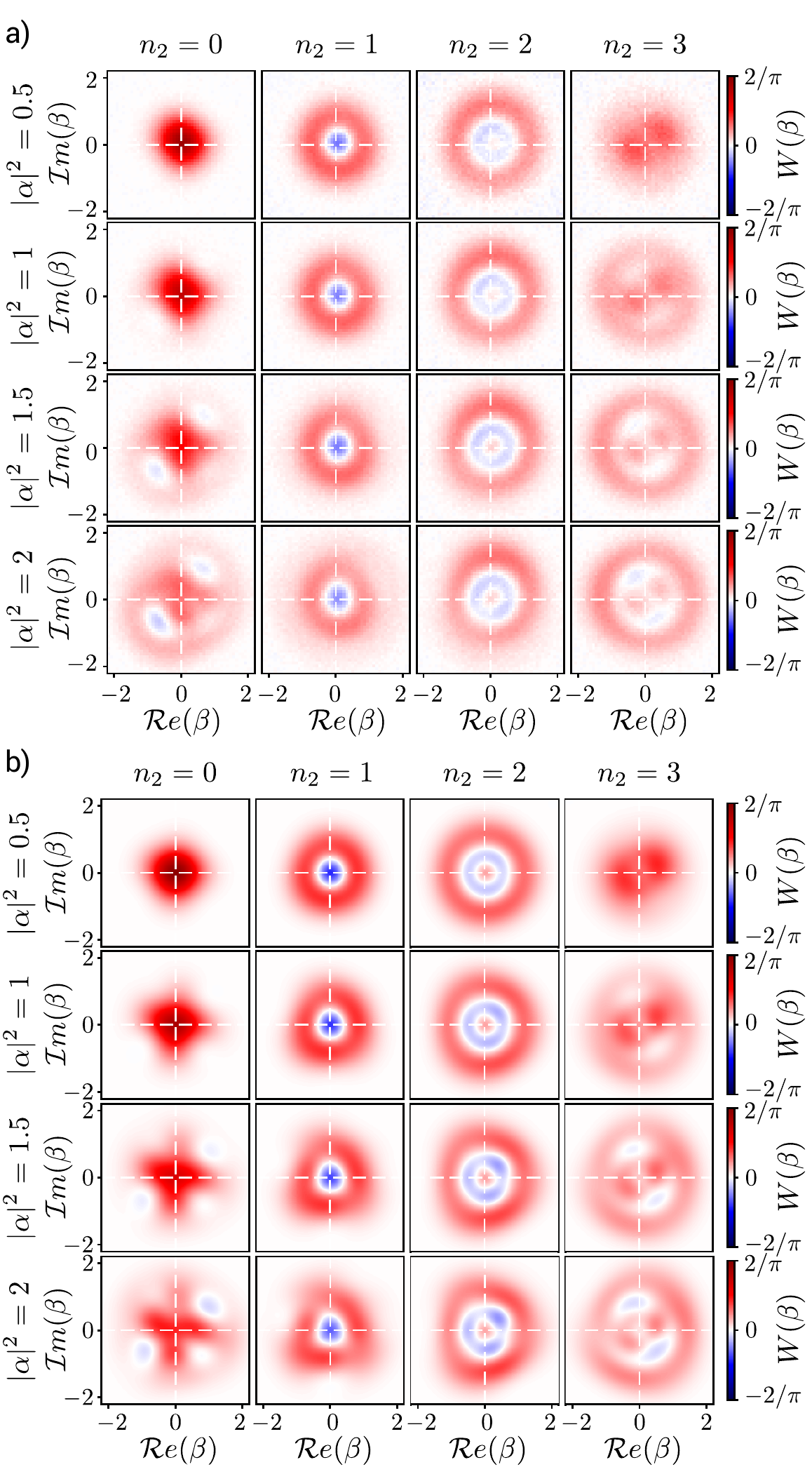}
    \caption{Measured (a) and computed (b) Wigner functions after catching a coherent state with a mean photon number $\abs{\alpha}^2= 0.5$, $1$, $1.5$ and $2$ from top to bottom respectively and heralding on a detected number $n_2=$ $0$, $1$, $2$ or $3$ from left to right respectively. For each panel, the fidelity between the measured Wigner function and the predicted one does not get below $95\%$.}
    \label{fig:Wigner}
\end{figure*}
\begin{figure*}[t]
    \centering
    \includegraphics[width=14cm]{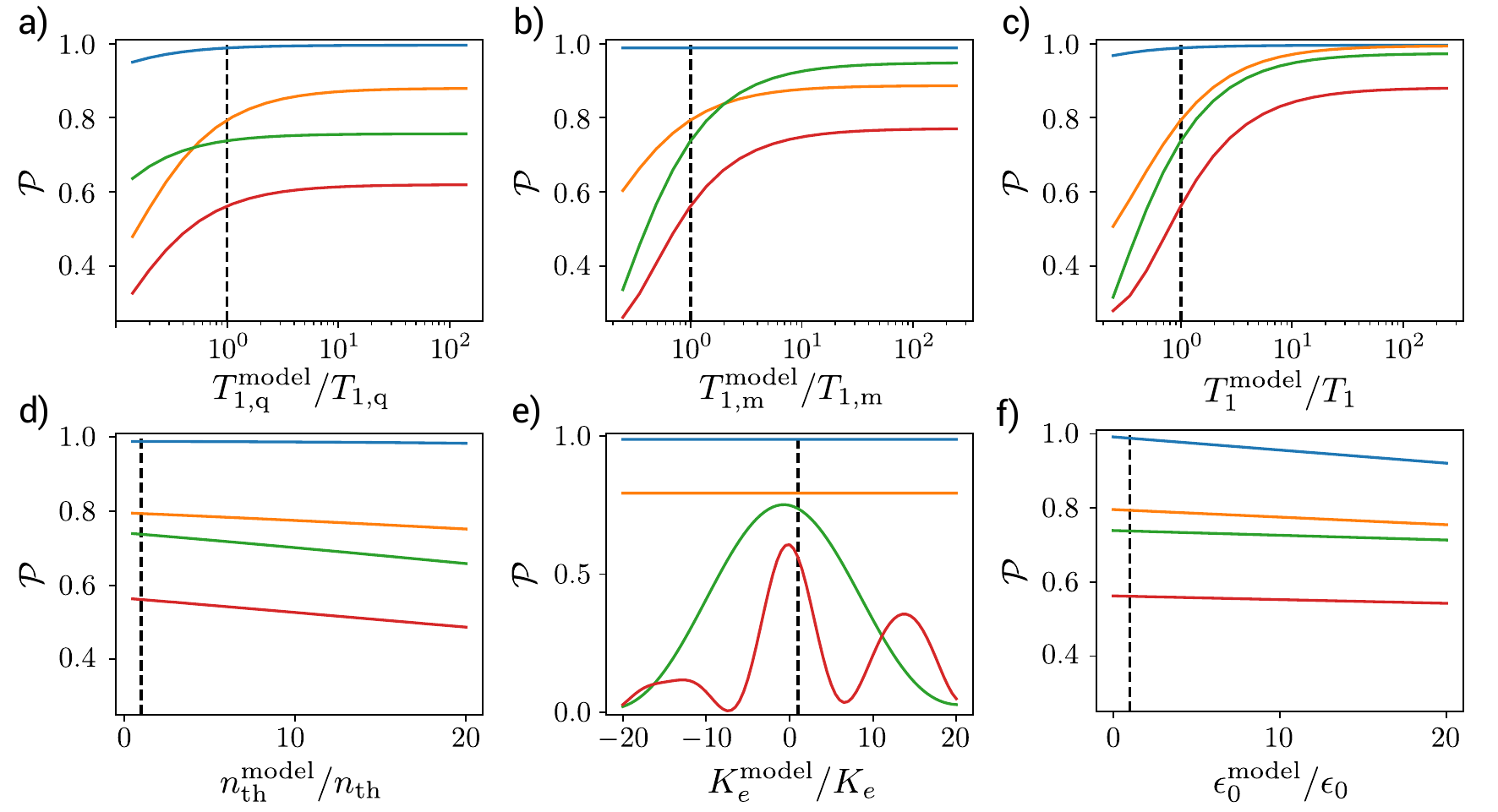}
    \caption{Success probabilities $\mathcal{P}_{\ket{0}}(0)$ (blue), $\mathcal{P}_{\ket{1}}(1)$ (orange), $\mathcal{P}_{\ket{2}}(2)$ (green) and $\mathcal{P}_{\ket{3}}(3)$ (red) as a function of the ratio between the parameter in the model and the same parameter in experiment. All curves are calculated in the case of an initial coherent state of amplitude $\abs{\alpha}= \sqrt{0.5}$. Vertical dashed lines indicate the result of the model for the actual experiment. Each panel probes the errors coming from a) the qubit relaxation time $T_{1, \rm q}$, b) the memory relaxation time $T_{1, \rm m}$, c) both qubit and memory relaxation times $T_1 = (T_{1, \rm q}$, $T_{1, \rm m})$, d) qubit and memory thermal population, e) additional Kerr rate $K_e$ when the qubit is excited, and f) readout error $\epsilon_0$.
    }
    \label{fig:error_budget}
\end{figure*}

\begin{figure*}[t]
    \centering
    \includegraphics[width=14cm]{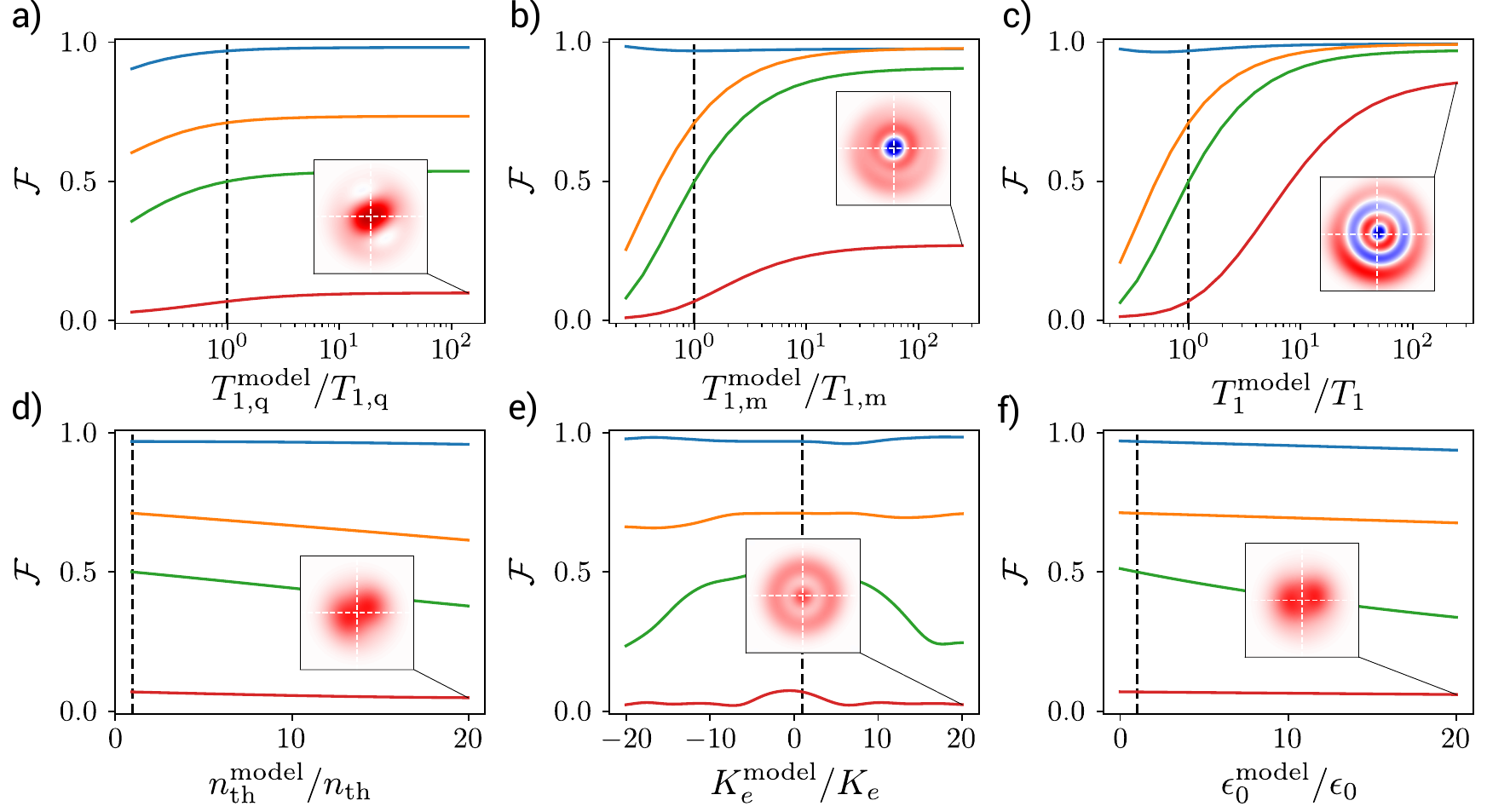}
    \caption{QNDness of the detector. Fidelity $\mathcal{F}$ between the quantum state $\rho$ predicted by our model and the ideal projected state $\rho_{n_2}$ after catching a coherent state of amplitude $\abs{\alpha}= \sqrt{0.5}$ for the outcomes $n_2=0$ (blue), $n_2=1$ (orange), $n_2=2$ (green) and $n_2=3$ (red). Each panel addresses the same parameter as in \cref{fig:error_budget}. Insets are the Wigner functions heralded on the counter outcome $n_2 =3$ for the maximal value of the model parameter. Note that on the top panels, the maximal value improves QNDness while it deteriorates it for  bottom panels.}
    \label{fig:error_budget_wigner}
\end{figure*}
We use the method of Refs.~\cite{Lutterbach1997,Bertet2002,Vlastakis2013} to directly measure the Wigner function $W(\beta) = \frac{2}{\pi}\expval{ \mathcal{D}_{\beta} \mathcal{P} \mathcal{D}_{\beta}^\dagger}$ of the memory mode. We perform a displacement $\mathcal{D}_{\beta}^\dagger$ of amplitude $-\beta$ (sech-shape with $\sigma=$ \SI{13}{ns}) followed by a parity measurement.  $\mathcal{P}=\exp(i\pi m^\dagger m)$ is the photon parity operator. The Wigner functions are measured on a $51$x$51$ square matrix of amplitudes $\beta$ where $\abs{\mathcal{R}e(\beta)},\abs{\mathcal{I}m(\beta)}\leq 2.2$. The measured Wigner functions for mean photon numbers $|\alpha|^2=0,1,1.5$ and $2$ are shown in \cref{fig:Wigner}.a. Each column corresponds to postselected measurements for a given detected photon number $n_2$.

Our numerical model above allows us to compute the predicted Wigner functions for each panel of the figure. The predictions are shown in \cref{fig:Wigner}.b. Note that these figures are obtained by computing the Wigner function directly without modeling the readout of the parity photon number after displacement.

For an arbitrary outcome $n_2$, the photocounter would ideally project the incoming state $\ket{\psi}$ into $\ket{\psi_{n_2}} \propto \sum_j \ket{n_2+4j} \braket{n_2+4j}{\psi}$. We discuss nonidealities in the measurement backaction in the main text. They are mainly due to the finite lifetimes of the qubit and memory for low mean photon numbers $\abs{\alpha}^2$. 

In \cref{fig:Wigner}, some Wigner functions are not invariant by a phase shift as one could expect from mixtures of Fock states. These patterns in the figure indicate coherences between Fock states. Our simulations show that the coherences originate from two main phenomena. First, the photon number measurement is performed modulo $4$, which preserves coherences between different photon numbers modulo $4$ by projection. Second, due to the finite duration of the $\pi/2$ pulses in the pulse sequence that performs question $Q_k$, the encoding of the $k$-th bit of the photon number in the qubit state is imperfect. Therefore, postselecting on the measured binary code $n_2$ preserves some coherence between the Fock states that compose the initial coherent state $\ket{\alpha}$. Finally, the Wigner functions appear distorted due to the memory nonlinear rates $K$ and $K_e$.

\begin{table}
\setlength{\tabcolsep}{6pt}
\setlength\extrarowheight{2pt}
\begin{tabular}{lcccc}
\hline\hline
 $\mathcal{F}(\rho,\rho_{n_2})$   & $n_2=0$ & $n_2=1$ & $n_2=2$ & $n_2=3$  \\[2pt]   \hline
  $\abs{\alpha}^2=0.5$  & $86\%$ & $52\%$  & $32\%$  &  $4.9\%$  \\
  
 $\abs{\alpha}^2=1$  & $ 77\%$ & $ 50\%$  & $ 34\%$  &  $ 11\%$  \\
   
   $\abs{\alpha}^2=1.5$  & $ 58\%$ & $ 48\%$  & $ 38\%$  &  $ 18\%$  \\
    
    $\abs{\alpha}^2=2$  & $ 39\%$ & $ 42\%$  & $ 37\%$  &  $ 22\%$  \\
    \hline\hline
\end{tabular}

 \caption{Fidelities $\mathcal{F}$ between the measured collapsed quantum states $\rho$ and the ideal quantum states $\rho_{n_2}=\ketbra{\psi_{n_2}}{\psi_{n_2}}$ for various outcomes $n_2$ and various mean photon numbers $\abs{\alpha}^2$.}
 \label{tab:Wignerfidelities} 
\end{table}

The deviations from the ideal projected quantum state (fidelities in \cref{tab:Wignerfidelities}) are further investigated in \cref{sec:error_budget}.

\section{Error budget of the photocounter}
\label{sec:error_budget}

In this section we numerically investigate the origin of the errors on the success probabilities $\mathcal{P}_{\ket{n}}(n)$ to find $n$ photons when the incoming wavepacket is in a Fock state $\ket{n}$ and on the QNDness, which is characterized by the fidelities $\mathcal{F}$ above. We study the error budget by sweeping one (or more) parameters independently of the others in our model.
\begin{itemize}

\item  The finite qubit relaxation time $T_{1,\rm q}$ entails different errors depending on the choice of encoding the outcome $n_2$ in the qubit state during questions $Q_k$'s. This choice is done by the sign of the second $\pi/2$ pulse in the sequence of Fig.~3. For each question $Q_k$, the outcome on the $k$-th bit of the photon number corresponding to the qubit excited state will get mixed with the outcome corresponding to the qubit ground state. These errors scale exponentially with $1/T_{1,\rm q}^\mathrm{model}$ (\cref{fig:error_budget}.a and \cref{fig:error_budget_wigner}.a). 

\item The finite memory relaxation time $T_{1,\rm m}$ causes errors except for $\ket{n=0}$ (\cref{fig:error_budget}.b and \cref{fig:error_budget_wigner}.b). The dominant source of error is then the mixing of the outcome $n_2$ with $n_2-1$.

\item The finite lifetimes $T_{1,\rm q}$ and $T_{1,\rm m}$ are our main sources of errors as the counting probabilities $\mathcal{P}_{\ket{n}}(n)$ (\cref{fig:error_budget}.c) and state fidelities $\mathcal{F}$ (\cref{fig:error_budget_wigner}.c) get close to $1$ when both $T_{1,\rm q}$ and $T_{1,\rm m}$ increase. If both $T_{1,\rm q}$ and $T_{1,\rm m}$ increase by an order of magnitude, the success probability will not get below $85\%$ for all outcomes (\cref{fig:error_budget}.c). The QNDness is more demanding and one would need to increase by more than two orders of magnitude the lifetimes in order to get fidelities beyond 80\% (insets of \cref{fig:error_budget_wigner}.a-c). Note that current state of the art in three-dimensional (3D) cavities and new materials demonstrates lifetimes indeed larger than 2 orders of magnitude~\cite{Reagor2016,Place2020}.

\item Our device does not seem to be limited by thermal excitations (\cref{fig:error_budget}.d and \cref{fig:error_budget_wigner}.d). 

\item A more faithful qubit readout would not bring significant improvements in the success probabilities and QNDness (\cref{fig:error_budget}.f and \cref{fig:error_budget_wigner}.f).

\item The memory self-Kerr rate $K$ does not seem to affect the success probabilities and QNDness (not shown). Indeed, the Fock states are eigenstates of the self-Kerr term. However, the additional self-Kerr rate $K_e$ when the qubit is in $\ket{e}$ has an important impact (\cref{fig:error_budget}.e and \cref{fig:error_budget_wigner}.e). 
During the interaction time $T_k$ of question $Q_k$, the qubit acquires an additional parasitic phase $n^2 K_e T_k $ for each Fock state $\ket{n}$. Therefore, for $n \geq 1$ and each question $Q_k$, the qubit phase does not end up in the right value, which undermines the photon number encoding. 
As long as $n^2 K_e 2\pi/(\chi 2^{k}) \ll 1$, this effect can be neglected. For our device, it translates into $n \ll 3.7$. 
This square dependence on the photon number $n$ is the main limitation of this scheme for increasing the maximal number of photons the detector can resolve.

Similar to Ref \cite{Elliott2018}, we compute the rate $K_e$ using perturbation theory to the fourth order in the transverse coupling strength
\[g=\sqrt{\chi \Delta(\Delta -K_q) /(2 K_q)}.\]
It is obtained as a function of the detuning $\Delta=\omega_m-\omega_q$, transmon anharmonicity $-K_q=-E_C/\hbar$ and dispersive shift $\chi$

\begin{equation}
    K_e=\frac{ \chi^2  }{K_q}\frac{(2\Delta^3 - (\Delta-K_q)^3 ) }{2\Delta(\Delta - K_q) (\Delta+K_q)}
\end{equation}

It is then possible to reduce $K_e$ considerably while preserving the behavior of the device for large photon numbers by careful optimization of the device parameters. For example setting the detuning accurately to $\Delta= \frac{K_q}{ (1-\sqrt[3]{2})}$ cancels the rate $K_e$ completely.

\end{itemize}


\begin{thebibliography}{65}%
\makeatletter
\providecommand \@ifxundefined [1]{%
 \@ifx{#1\undefined}
}%
\providecommand \@ifnum [1]{%
 \ifnum #1\expandafter \@firstoftwo
 \else \expandafter \@secondoftwo
 \fi
}%
\providecommand \@ifx [1]{%
 \ifx #1\expandafter \@firstoftwo
 \else \expandafter \@secondoftwo
 \fi
}%
\providecommand \natexlab [1]{#1}%
\providecommand \enquote  [1]{``#1''}%
\providecommand \bibnamefont  [1]{#1}%
\providecommand \bibfnamefont [1]{#1}%
\providecommand \citenamefont [1]{#1}%
\providecommand \href@noop [0]{\@secondoftwo}%
\providecommand \href [0]{\begingroup \@sanitize@url \@href}%
\providecommand \@href[1]{\@@startlink{#1}\@@href}%
\providecommand \@@href[1]{\endgroup#1\@@endlink}%
\providecommand \@sanitize@url [0]{\catcode `\\12\catcode `\$12\catcode
  `\&12\catcode `\#12\catcode `\^12\catcode `\_12\catcode `\%12\relax}%
\providecommand \@@startlink[1]{}%
\providecommand \@@endlink[0]{}%
\providecommand \url  [0]{\begingroup\@sanitize@url \@url }%
\providecommand \@url [1]{\endgroup\@href {#1}{\urlprefix }}%
\providecommand \urlprefix  [0]{URL }%
\providecommand \Eprint [0]{\href }%
\providecommand \doibase [0]{https://doi.org/}%
\providecommand \selectlanguage [0]{\@gobble}%
\providecommand \bibinfo  [0]{\@secondoftwo}%
\providecommand \bibfield  [0]{\@secondoftwo}%
\providecommand \translation [1]{[#1]}%
\providecommand \BibitemOpen [0]{}%
\providecommand \bibitemStop [0]{}%
\providecommand \bibitemNoStop [0]{.\EOS\space}%
\providecommand \EOS [0]{\spacefactor3000\relax}%
\providecommand \BibitemShut  [1]{\csname bibitem#1\endcsname}%
\let\auto@bib@innerbib\@empty
\bibitem [{\citenamefont {Hadfield}(2009)}]{Hadfield2009}%
  \BibitemOpen
  \bibfield  {author} {\bibinfo {author} {\bibfnamefont {R.~H.}\ \bibnamefont
  {Hadfield}},\ }\bibfield  {title} {\bibinfo {title} {{Single-photon detectors
  for optical quantum information applications}},\ }\href
  {https://doi.org/10.1038/nphoton.2009.230} {\bibfield  {journal} {\bibinfo
  {journal} {Nat. Photonics}\ }\textbf {\bibinfo {volume} {3}},\ \bibinfo
  {pages} {696} (\bibinfo {year} {2009})}\BibitemShut {NoStop}%
\bibitem [{\citenamefont {Gleyzes}\ \emph {et~al.}(2007)\citenamefont
  {Gleyzes}, \citenamefont {Kuhr}, \citenamefont {Guerlin}, \citenamefont
  {Bernu}, \citenamefont {Del{\'{e}}glise}, \citenamefont {{Busk Hoff}},
  \citenamefont {Brune}, \citenamefont {Raimond},\ and\ \citenamefont
  {Haroche}}]{Gleyzes2007}%
  \BibitemOpen
  \bibfield  {author} {\bibinfo {author} {\bibfnamefont {S.}~\bibnamefont
  {Gleyzes}}, \bibinfo {author} {\bibfnamefont {S.}~\bibnamefont {Kuhr}},
  \bibinfo {author} {\bibfnamefont {C.}~\bibnamefont {Guerlin}}, \bibinfo
  {author} {\bibfnamefont {J.}~\bibnamefont {Bernu}}, \bibinfo {author}
  {\bibfnamefont {S.}~\bibnamefont {Del{\'{e}}glise}}, \bibinfo {author}
  {\bibfnamefont {U.}~\bibnamefont {{Busk Hoff}}}, \bibinfo {author}
  {\bibfnamefont {M.}~\bibnamefont {Brune}}, \bibinfo {author} {\bibfnamefont
  {J.~M.}\ \bibnamefont {Raimond}},\ and\ \bibinfo {author} {\bibfnamefont
  {S.}~\bibnamefont {Haroche}},\ }\bibfield  {title} {\bibinfo {title}
  {{Quantum jumps of light recording the birth and death of a photon in a
  cavity}},\ }\href {https://doi.org/10.1038/nature05589} {\bibfield  {journal}
  {\bibinfo  {journal} {Nature}\ }\textbf {\bibinfo {volume} {446}},\ \bibinfo
  {pages} {297} (\bibinfo {year} {2007})}\BibitemShut {NoStop}%
\bibitem [{\citenamefont {Guerlin}\ \emph {et~al.}(2007)\citenamefont
  {Guerlin}, \citenamefont {Bernu}, \citenamefont {Del{\'{e}}glise},
  \citenamefont {Sayrin}, \citenamefont {Gleyzes}, \citenamefont {Kuhr},
  \citenamefont {Brune}, \citenamefont {Raimond},\ and\ \citenamefont
  {Haroche}}]{Guerlin2007}%
  \BibitemOpen
  \bibfield  {author} {\bibinfo {author} {\bibfnamefont {C.}~\bibnamefont
  {Guerlin}}, \bibinfo {author} {\bibfnamefont {J.}~\bibnamefont {Bernu}},
  \bibinfo {author} {\bibfnamefont {S.}~\bibnamefont {Del{\'{e}}glise}},
  \bibinfo {author} {\bibfnamefont {C.}~\bibnamefont {Sayrin}}, \bibinfo
  {author} {\bibfnamefont {S.}~\bibnamefont {Gleyzes}}, \bibinfo {author}
  {\bibfnamefont {S.}~\bibnamefont {Kuhr}}, \bibinfo {author} {\bibfnamefont
  {M.}~\bibnamefont {Brune}}, \bibinfo {author} {\bibfnamefont {J.~M.}\
  \bibnamefont {Raimond}},\ and\ \bibinfo {author} {\bibfnamefont
  {S.}~\bibnamefont {Haroche}},\ }\bibfield  {title} {\bibinfo {title}
  {{Progressive field-state collapse and quantum non-demolition photon
  counting}},\ }\href {https://doi.org/10.1038/nature06057} {\bibfield
  {journal} {\bibinfo  {journal} {Nature}\ }\textbf {\bibinfo {volume} {448}},\
  \bibinfo {pages} {889} (\bibinfo {year} {2007})}\BibitemShut {NoStop}%
\bibitem [{\citenamefont {Johnson}\ \emph {et~al.}(2010)\citenamefont
  {Johnson}, \citenamefont {Reed}, \citenamefont {Houck}, \citenamefont
  {Schuster}, \citenamefont {Bishop}, \citenamefont {Ginossar}, \citenamefont
  {Gambetta}, \citenamefont {Dicarlo}, \citenamefont {Frunzio}, \citenamefont
  {Girvin},\ and\ \citenamefont {Schoelkopf}}]{Johnson2010}%
  \BibitemOpen
  \bibfield  {author} {\bibinfo {author} {\bibfnamefont {B.~R.}\ \bibnamefont
  {Johnson}}, \bibinfo {author} {\bibfnamefont {M.~D.}\ \bibnamefont {Reed}},
  \bibinfo {author} {\bibfnamefont {A.~A.}\ \bibnamefont {Houck}}, \bibinfo
  {author} {\bibfnamefont {D.~I.}\ \bibnamefont {Schuster}}, \bibinfo {author}
  {\bibfnamefont {L.~S.}\ \bibnamefont {Bishop}}, \bibinfo {author}
  {\bibfnamefont {E.}~\bibnamefont {Ginossar}}, \bibinfo {author}
  {\bibfnamefont {J.~M.}\ \bibnamefont {Gambetta}}, \bibinfo {author}
  {\bibfnamefont {L.}~\bibnamefont {Dicarlo}}, \bibinfo {author} {\bibfnamefont
  {L.}~\bibnamefont {Frunzio}}, \bibinfo {author} {\bibfnamefont {S.~M.}\
  \bibnamefont {Girvin}},\ and\ \bibinfo {author} {\bibfnamefont {R.~J.}\
  \bibnamefont {Schoelkopf}},\ }\bibfield  {title} {\bibinfo {title} {{Quantum
  non-demolition detection of single microwave photons in a circuit}},\ }\href
  {https://doi.org/10.1038/nphys1710} {\bibfield  {journal} {\bibinfo
  {journal} {Nat. Phys.}\ }\textbf {\bibinfo {volume} {6}},\ \bibinfo {pages}
  {663} (\bibinfo {year} {2010})}\BibitemShut {NoStop}%
\bibitem [{\citenamefont {Leek}\ \emph {et~al.}(2010)\citenamefont {Leek},
  \citenamefont {Baur}, \citenamefont {Fink}, \citenamefont {Bianchetti},
  \citenamefont {Steffen}, \citenamefont {Filipp},\ and\ \citenamefont
  {Wallraff}}]{Leek2010}%
  \BibitemOpen
  \bibfield  {author} {\bibinfo {author} {\bibfnamefont {P.~J.}\ \bibnamefont
  {Leek}}, \bibinfo {author} {\bibfnamefont {M.}~\bibnamefont {Baur}}, \bibinfo
  {author} {\bibfnamefont {J.~M.}\ \bibnamefont {Fink}}, \bibinfo {author}
  {\bibfnamefont {R.}~\bibnamefont {Bianchetti}}, \bibinfo {author}
  {\bibfnamefont {L.}~\bibnamefont {Steffen}}, \bibinfo {author} {\bibfnamefont
  {S.}~\bibnamefont {Filipp}},\ and\ \bibinfo {author} {\bibfnamefont
  {A.}~\bibnamefont {Wallraff}},\ }\bibfield  {title} {\bibinfo {title}
  {{Cavity quantum electrodynamics with separate photon storage and qubit
  readout modes}},\ }\href {https://doi.org/10.1103/PhysRevLett.104.100504}
  {\bibfield  {journal} {\bibinfo  {journal} {Phys. Rev. Lett.}\ }\textbf
  {\bibinfo {volume} {104}},\ \bibinfo {pages} {100504} (\bibinfo {year}
  {2010})}\BibitemShut {NoStop}%
\bibitem [{\citenamefont {Sun}\ \emph {et~al.}(2014)\citenamefont {Sun},
  \citenamefont {Petrenko}, \citenamefont {Leghtas}, \citenamefont {Vlastakis},
  \citenamefont {Kirchmair}, \citenamefont {Sliwa}, \citenamefont {Narla},
  \citenamefont {Hatridge}, \citenamefont {Shankar}, \citenamefont {Blumoff},
  \citenamefont {Frunzio}, \citenamefont {Mirrahimi}, \citenamefont {Devoret},\
  and\ \citenamefont {Schoelkopf}}]{Sun2014}%
  \BibitemOpen
  \bibfield  {author} {\bibinfo {author} {\bibfnamefont {L.}~\bibnamefont
  {Sun}}, \bibinfo {author} {\bibfnamefont {A.}~\bibnamefont {Petrenko}},
  \bibinfo {author} {\bibfnamefont {Z.}~\bibnamefont {Leghtas}}, \bibinfo
  {author} {\bibfnamefont {B.}~\bibnamefont {Vlastakis}}, \bibinfo {author}
  {\bibfnamefont {G.}~\bibnamefont {Kirchmair}}, \bibinfo {author}
  {\bibfnamefont {K.~M.}\ \bibnamefont {Sliwa}}, \bibinfo {author}
  {\bibfnamefont {A.}~\bibnamefont {Narla}}, \bibinfo {author} {\bibfnamefont
  {M.}~\bibnamefont {Hatridge}}, \bibinfo {author} {\bibfnamefont
  {S.}~\bibnamefont {Shankar}}, \bibinfo {author} {\bibfnamefont
  {J.}~\bibnamefont {Blumoff}}, \bibinfo {author} {\bibfnamefont
  {L.}~\bibnamefont {Frunzio}}, \bibinfo {author} {\bibfnamefont
  {M.}~\bibnamefont {Mirrahimi}}, \bibinfo {author} {\bibfnamefont {M.~H.}\
  \bibnamefont {Devoret}},\ and\ \bibinfo {author} {\bibfnamefont {R.~J.}\
  \bibnamefont {Schoelkopf}},\ }\bibfield  {title} {\bibinfo {title} {{Tracking
  photon jumps with repeated quantum non-demolition parity measurements}},\
  }\href {https://doi.org/10.1038/nature13436} {\bibfield  {journal} {\bibinfo
  {journal} {Nature}\ }\textbf {\bibinfo {volume} {511}},\ \bibinfo {pages}
  {444} (\bibinfo {year} {2014})}\BibitemShut {NoStop}%
\bibitem [{\citenamefont {Romero}\ \emph {et~al.}(2009)\citenamefont {Romero},
  \citenamefont {Garc{\'{i}}a-Ripoll},\ and\ \citenamefont
  {Solano}}]{Romero2009}%
  \BibitemOpen
  \bibfield  {author} {\bibinfo {author} {\bibfnamefont {G.}~\bibnamefont
  {Romero}}, \bibinfo {author} {\bibfnamefont {J.~J.}\ \bibnamefont
  {Garc{\'{i}}a-Ripoll}},\ and\ \bibinfo {author} {\bibfnamefont
  {E.}~\bibnamefont {Solano}},\ }\bibfield  {title} {\bibinfo {title}
  {{Microwave photon detector in circuit QED}},\ }\href
  {https://doi.org/10.1103/PhysRevLett.102.173602} {\bibfield  {journal}
  {\bibinfo  {journal} {Phys. Rev. Lett.}\ }\textbf {\bibinfo {volume} {102}},\
  \bibinfo {pages} {173602} (\bibinfo {year} {2009})}\BibitemShut {NoStop}%
\bibitem [{\citenamefont {Helmer}\ \emph {et~al.}(2009)\citenamefont {Helmer},
  \citenamefont {Mariantoni}, \citenamefont {Solano},\ and\ \citenamefont
  {Marquardt}}]{Helmer2009}%
  \BibitemOpen
  \bibfield  {author} {\bibinfo {author} {\bibfnamefont {F.}~\bibnamefont
  {Helmer}}, \bibinfo {author} {\bibfnamefont {M.}~\bibnamefont {Mariantoni}},
  \bibinfo {author} {\bibfnamefont {E.}~\bibnamefont {Solano}},\ and\ \bibinfo
  {author} {\bibfnamefont {F.}~\bibnamefont {Marquardt}},\ }\bibfield  {title}
  {\bibinfo {title} {{Quantum nondemolition photon detection in circuit QED and
  the quantum Zeno effect}},\ }\href
  {https://doi.org/10.1103/PhysRevA.79.052115} {\bibfield  {journal} {\bibinfo
  {journal} {Phys. Rev. A}\ }\textbf {\bibinfo {volume} {79}},\ \bibinfo
  {pages} {052115} (\bibinfo {year} {2009})}\BibitemShut {NoStop}%
\bibitem [{\citenamefont {Koshino}\ \emph {et~al.}(2013)\citenamefont
  {Koshino}, \citenamefont {Inomata}, \citenamefont {Yamamoto},\ and\
  \citenamefont {Nakamura}}]{Koshino2013}%
  \BibitemOpen
  \bibfield  {author} {\bibinfo {author} {\bibfnamefont {K.}~\bibnamefont
  {Koshino}}, \bibinfo {author} {\bibfnamefont {K.}~\bibnamefont {Inomata}},
  \bibinfo {author} {\bibfnamefont {T.}~\bibnamefont {Yamamoto}},\ and\
  \bibinfo {author} {\bibfnamefont {Y.}~\bibnamefont {Nakamura}},\ }\bibfield
  {title} {\bibinfo {title} {Implementation of an impedance-matched {$\Lambda$}
  system by dressed-state engineering},\ }\href
  {https://doi.org/10.1103/PhysRevLett.111.153601} {\bibfield  {journal}
  {\bibinfo  {journal} {Phys. Rev. Lett.}\ }\textbf {\bibinfo {volume} {111}},\
  \bibinfo {pages} {153601} (\bibinfo {year} {2013})}\BibitemShut {NoStop}%
\bibitem [{\citenamefont {Sathyamoorthy}\ \emph {et~al.}(2014)\citenamefont
  {Sathyamoorthy}, \citenamefont {Tornberg}, \citenamefont {Kockum},
  \citenamefont {Baragiola}, \citenamefont {Combes}, \citenamefont {Wilson},
  \citenamefont {Stace},\ and\ \citenamefont {Johansson}}]{Sathyamoorthy2014}%
  \BibitemOpen
  \bibfield  {author} {\bibinfo {author} {\bibfnamefont {S.~R.}\ \bibnamefont
  {Sathyamoorthy}}, \bibinfo {author} {\bibfnamefont {L.}~\bibnamefont
  {Tornberg}}, \bibinfo {author} {\bibfnamefont {A.~F.}\ \bibnamefont
  {Kockum}}, \bibinfo {author} {\bibfnamefont {B.~Q.}\ \bibnamefont
  {Baragiola}}, \bibinfo {author} {\bibfnamefont {J.}~\bibnamefont {Combes}},
  \bibinfo {author} {\bibfnamefont {C.~M.}\ \bibnamefont {Wilson}}, \bibinfo
  {author} {\bibfnamefont {T.~M.}\ \bibnamefont {Stace}},\ and\ \bibinfo
  {author} {\bibfnamefont {G.}~\bibnamefont {Johansson}},\ }\bibfield  {title}
  {\bibinfo {title} {{Quantum nondemolition detection of a propagating
  microwave photon}},\ }\href {https://doi.org/10.1103/PhysRevLett.112.093601}
  {\bibfield  {journal} {\bibinfo  {journal} {Phys. Rev. Lett.}\ }\textbf
  {\bibinfo {volume} {112}},\ \bibinfo {pages} {093601} (\bibinfo {year}
  {2014})}\BibitemShut {NoStop}%
\bibitem [{\citenamefont {Fan}\ \emph {et~al.}(2014)\citenamefont {Fan},
  \citenamefont {Johansson}, \citenamefont {Combes}, \citenamefont {Milburn},\
  and\ \citenamefont {Stace}}]{Fan2014}%
  \BibitemOpen
  \bibfield  {author} {\bibinfo {author} {\bibfnamefont {B.}~\bibnamefont
  {Fan}}, \bibinfo {author} {\bibfnamefont {G.}~\bibnamefont {Johansson}},
  \bibinfo {author} {\bibfnamefont {J.}~\bibnamefont {Combes}}, \bibinfo
  {author} {\bibfnamefont {G.~J.}\ \bibnamefont {Milburn}},\ and\ \bibinfo
  {author} {\bibfnamefont {T.~M.}\ \bibnamefont {Stace}},\ }\bibfield  {title}
  {\bibinfo {title} {{Nonabsorbing high-efficiency counter for itinerant
  microwave photons}},\ }\href {https://doi.org/10.1103/PhysRevB.90.035132}
  {\bibfield  {journal} {\bibinfo  {journal} {Phys. Rev. B}\ }\textbf {\bibinfo
  {volume} {90}},\ \bibinfo {pages} {035132} (\bibinfo {year}
  {2014})}\BibitemShut {NoStop}%
\bibitem [{\citenamefont {Kyriienko}\ and\ \citenamefont
  {S{\o}rensen}(2016)}]{Kyriienko2016}%
  \BibitemOpen
  \bibfield  {author} {\bibinfo {author} {\bibfnamefont {O.}~\bibnamefont
  {Kyriienko}}\ and\ \bibinfo {author} {\bibfnamefont {A.~S.}\ \bibnamefont
  {S{\o}rensen}},\ }\bibfield  {title} {\bibinfo {title} {{Continuous-Wave
  Single-Photon Transistor Based on a Superconducting Circuit}},\ }\href
  {https://doi.org/10.1103/PhysRevLett.117.140503} {\bibfield  {journal}
  {\bibinfo  {journal} {Phys. Rev. Lett.}\ }\textbf {\bibinfo {volume} {117}},\
  \bibinfo {pages} {140503} (\bibinfo {year} {2016})}\BibitemShut {NoStop}%
\bibitem [{\citenamefont {Sathyamoorthy}\ \emph {et~al.}(2016)\citenamefont
  {Sathyamoorthy}, \citenamefont {Stace},\ and\ \citenamefont
  {Johansson}}]{Sathyamoorthy2016}%
  \BibitemOpen
  \bibfield  {author} {\bibinfo {author} {\bibfnamefont {S.~R.}\ \bibnamefont
  {Sathyamoorthy}}, \bibinfo {author} {\bibfnamefont {T.~M.}\ \bibnamefont
  {Stace}},\ and\ \bibinfo {author} {\bibfnamefont {G.}~\bibnamefont
  {Johansson}},\ }\bibfield  {title} {\bibinfo {title} {{Detecting itinerant
  single microwave photons}},\ }\href
  {https://doi.org/10.1016/j.crhy.2016.07.010} {\bibfield  {journal} {\bibinfo
  {journal} {Comptes Rendus Phys.}\ }\textbf {\bibinfo {volume} {17}},\
  \bibinfo {pages} {756} (\bibinfo {year} {2016})}\BibitemShut {NoStop}%
\bibitem [{\citenamefont {Gu}\ \emph {et~al.}(2017)\citenamefont {Gu},
  \citenamefont {Frisk}, \citenamefont {Miranowicz}, \citenamefont {Liu},\ and\
  \citenamefont {Nori}}]{Gu2017}%
  \BibitemOpen
  \bibfield  {author} {\bibinfo {author} {\bibfnamefont {X.}~\bibnamefont
  {Gu}}, \bibinfo {author} {\bibfnamefont {A.}~\bibnamefont {Frisk}}, \bibinfo
  {author} {\bibfnamefont {A.}~\bibnamefont {Miranowicz}}, \bibinfo {author}
  {\bibfnamefont {Y.-x.}\ \bibnamefont {Liu}},\ and\ \bibinfo {author}
  {\bibfnamefont {F.}~\bibnamefont {Nori}},\ }\bibfield  {title} {\bibinfo
  {title} {{Microwave photonics with superconducting quantum circuits}},\
  }\href {https://doi.org/10.1016/j.physrep.2017.10.002} {\bibfield  {journal}
  {\bibinfo  {journal} {Phys. Rep.}\ }\textbf {\bibinfo {volume} {718-719}},\
  \bibinfo {pages} {1} (\bibinfo {year} {2017})}\BibitemShut {NoStop}%
\bibitem [{\citenamefont {Wong}\ and\ \citenamefont
  {Vavilov}(2017)}]{Wong2017}%
  \BibitemOpen
  \bibfield  {author} {\bibinfo {author} {\bibfnamefont {C.~H.}\ \bibnamefont
  {Wong}}\ and\ \bibinfo {author} {\bibfnamefont {M.~G.}\ \bibnamefont
  {Vavilov}},\ }\bibfield  {title} {\bibinfo {title} {Quantum efficiency of a
  single microwave photon detector based on a semiconductor double quantum
  dot},\ }\href {https://doi.org/10.1103/PhysRevA.95.012325} {\bibfield
  {journal} {\bibinfo  {journal} {Phys. Rev. A}\ }\textbf {\bibinfo {volume}
  {95}},\ \bibinfo {pages} {012325} (\bibinfo {year} {2017})}\BibitemShut
  {NoStop}%
\bibitem [{\citenamefont {Lepp{\"{a}}kangas}\ \emph {et~al.}(2018)\citenamefont
  {Lepp{\"{a}}kangas}, \citenamefont {Marthaler}, \citenamefont {Hazra},
  \citenamefont {Jebari}, \citenamefont {Albert}, \citenamefont {Blanchet},
  \citenamefont {Johansson},\ and\ \citenamefont {Hofheinz}}]{Leppakangas2018}%
  \BibitemOpen
  \bibfield  {author} {\bibinfo {author} {\bibfnamefont {J.}~\bibnamefont
  {Lepp{\"{a}}kangas}}, \bibinfo {author} {\bibfnamefont {M.}~\bibnamefont
  {Marthaler}}, \bibinfo {author} {\bibfnamefont {D.}~\bibnamefont {Hazra}},
  \bibinfo {author} {\bibfnamefont {S.}~\bibnamefont {Jebari}}, \bibinfo
  {author} {\bibfnamefont {R.}~\bibnamefont {Albert}}, \bibinfo {author}
  {\bibfnamefont {F.}~\bibnamefont {Blanchet}}, \bibinfo {author}
  {\bibfnamefont {G.}~\bibnamefont {Johansson}},\ and\ \bibinfo {author}
  {\bibfnamefont {M.}~\bibnamefont {Hofheinz}},\ }\bibfield  {title} {\bibinfo
  {title} {{Multiplying and detecting propagating microwave photons using
  inelastic Cooper-pair tunneling}},\ }\href
  {https://doi.org/10.1103/PhysRevA.97.013855} {\bibfield  {journal} {\bibinfo
  {journal} {Phys. Rev. A}\ }\textbf {\bibinfo {volume} {97}},\ \bibinfo
  {pages} {013855} (\bibinfo {year} {2018})}\BibitemShut {NoStop}%
\bibitem [{\citenamefont {Royer}\ \emph {et~al.}(2018)\citenamefont {Royer},
  \citenamefont {Grimsmo}, \citenamefont {Choquette-poitevin},\ and\
  \citenamefont {Blais}}]{Royer2018}%
  \BibitemOpen
  \bibfield  {author} {\bibinfo {author} {\bibfnamefont {B.}~\bibnamefont
  {Royer}}, \bibinfo {author} {\bibfnamefont {A.~L.}\ \bibnamefont {Grimsmo}},
  \bibinfo {author} {\bibfnamefont {A.}~\bibnamefont {Choquette-poitevin}},\
  and\ \bibinfo {author} {\bibfnamefont {A.}~\bibnamefont {Blais}},\ }\bibfield
   {title} {\bibinfo {title} {{Itinerant Microwave Photon Detector}},\ }\href
  {https://doi.org/10.1103/PhysRevLett.120.203602} {\bibfield  {journal}
  {\bibinfo  {journal} {Phys. Rev. Lett.}\ }\textbf {\bibinfo {volume} {120}},\
  \bibinfo {pages} {203602} (\bibinfo {year} {2018})}\BibitemShut {NoStop}%
\bibitem [{\citenamefont {Chen}\ \emph {et~al.}(2011)\citenamefont {Chen},
  \citenamefont {Hover}, \citenamefont {Sendelbach}, \citenamefont {Maurer},
  \citenamefont {Merkel}, \citenamefont {Pritchett}, \citenamefont {Wilhelm},\
  and\ \citenamefont {McDermott}}]{Chen2011}%
  \BibitemOpen
  \bibfield  {author} {\bibinfo {author} {\bibfnamefont {Y.~F.}\ \bibnamefont
  {Chen}}, \bibinfo {author} {\bibfnamefont {D.}~\bibnamefont {Hover}},
  \bibinfo {author} {\bibfnamefont {S.}~\bibnamefont {Sendelbach}}, \bibinfo
  {author} {\bibfnamefont {L.}~\bibnamefont {Maurer}}, \bibinfo {author}
  {\bibfnamefont {S.~T.}\ \bibnamefont {Merkel}}, \bibinfo {author}
  {\bibfnamefont {E.~J.}\ \bibnamefont {Pritchett}}, \bibinfo {author}
  {\bibfnamefont {F.~K.}\ \bibnamefont {Wilhelm}},\ and\ \bibinfo {author}
  {\bibfnamefont {R.}~\bibnamefont {McDermott}},\ }\bibfield  {title} {\bibinfo
  {title} {{Microwave photon counter based on josephson junctions}},\ }\href
  {https://doi.org/10.1103/PhysRevLett.107.217401} {\bibfield  {journal}
  {\bibinfo  {journal} {Phys. Rev. Lett.}\ }\textbf {\bibinfo {volume} {107}},\
  \bibinfo {pages} {217401} (\bibinfo {year} {2011})}\BibitemShut {NoStop}%
\bibitem [{\citenamefont {Inomata}\ \emph {et~al.}(2016)\citenamefont
  {Inomata}, \citenamefont {Lin}, \citenamefont {Koshino}, \citenamefont
  {Oliver}, \citenamefont {Tsai}, \citenamefont {Yamamoto},\ and\ \citenamefont
  {Nakamura}}]{Inomata2016}%
  \BibitemOpen
  \bibfield  {author} {\bibinfo {author} {\bibfnamefont {K.}~\bibnamefont
  {Inomata}}, \bibinfo {author} {\bibfnamefont {Z.}~\bibnamefont {Lin}},
  \bibinfo {author} {\bibfnamefont {K.}~\bibnamefont {Koshino}}, \bibinfo
  {author} {\bibfnamefont {W.~D.}\ \bibnamefont {Oliver}}, \bibinfo {author}
  {\bibfnamefont {J.~S.}\ \bibnamefont {Tsai}}, \bibinfo {author}
  {\bibfnamefont {T.}~\bibnamefont {Yamamoto}},\ and\ \bibinfo {author}
  {\bibfnamefont {Y.}~\bibnamefont {Nakamura}},\ }\bibfield  {title} {\bibinfo
  {title} {{Single microwave-photon detector using an artificial
  {$\Lambda$}-type three-level system}},\ }\href
  {https://doi.org/10.1038/ncomms12303} {\bibfield  {journal} {\bibinfo
  {journal} {Nat. Commun.}\ }\textbf {\bibinfo {volume} {7}},\ \bibinfo {pages}
  {12303} (\bibinfo {year} {2016})}\BibitemShut {NoStop}%
\bibitem [{\citenamefont {Besse}\ \emph {et~al.}(2018)\citenamefont {Besse},
  \citenamefont {Gasparinetti}, \citenamefont {Collodo}, \citenamefont
  {Walter}, \citenamefont {Kurpiers}, \citenamefont {Pechal}, \citenamefont
  {Eichler},\ and\ \citenamefont {Wallraff}}]{Besse2018}%
  \BibitemOpen
  \bibfield  {author} {\bibinfo {author} {\bibfnamefont {J.~C.}\ \bibnamefont
  {Besse}}, \bibinfo {author} {\bibfnamefont {S.}~\bibnamefont {Gasparinetti}},
  \bibinfo {author} {\bibfnamefont {M.~C.}\ \bibnamefont {Collodo}}, \bibinfo
  {author} {\bibfnamefont {T.}~\bibnamefont {Walter}}, \bibinfo {author}
  {\bibfnamefont {P.}~\bibnamefont {Kurpiers}}, \bibinfo {author}
  {\bibfnamefont {M.}~\bibnamefont {Pechal}}, \bibinfo {author} {\bibfnamefont
  {C.}~\bibnamefont {Eichler}},\ and\ \bibinfo {author} {\bibfnamefont
  {A.}~\bibnamefont {Wallraff}},\ }\bibfield  {title} {\bibinfo {title}
  {{Single-Shot Quantum Nondemolition Detection of Individual Itinerant
  Microwave Photons}},\ }\href {https://doi.org/10.1103/PhysRevX.8.021003}
  {\bibfield  {journal} {\bibinfo  {journal} {Phys. Rev. X}\ }\textbf {\bibinfo
  {volume} {8}},\ \bibinfo {pages} {21003} (\bibinfo {year}
  {2018})}\BibitemShut {NoStop}%
\bibitem [{\citenamefont {Kono}\ \emph {et~al.}(2018)\citenamefont {Kono},
  \citenamefont {Koshino}, \citenamefont {Tabuchi}, \citenamefont {Noguchi},\
  and\ \citenamefont {Nakamura}}]{Kono2018}%
  \BibitemOpen
  \bibfield  {author} {\bibinfo {author} {\bibfnamefont {S.}~\bibnamefont
  {Kono}}, \bibinfo {author} {\bibfnamefont {K.}~\bibnamefont {Koshino}},
  \bibinfo {author} {\bibfnamefont {Y.}~\bibnamefont {Tabuchi}}, \bibinfo
  {author} {\bibfnamefont {A.}~\bibnamefont {Noguchi}},\ and\ \bibinfo {author}
  {\bibfnamefont {Y.}~\bibnamefont {Nakamura}},\ }\bibfield  {title} {\bibinfo
  {title} {{Quantum non-demolition detection of an itinerant microwave
  photon}},\ }\href {https://doi.org/10.1038/s41567-018-0066-3} {\bibfield
  {journal} {\bibinfo  {journal} {Nat. Phys.}\ }\textbf {\bibinfo {volume}
  {14}},\ \bibinfo {pages} {546} (\bibinfo {year} {2018})}\BibitemShut
  {NoStop}%
\bibitem [{\citenamefont {Narla}\ \emph {et~al.}(2016)\citenamefont {Narla},
  \citenamefont {Shankar}, \citenamefont {Hatridge}, \citenamefont {Leghtas},
  \citenamefont {Sliwa}, \citenamefont {Zalys-Geller}, \citenamefont
  {Mundhada}, \citenamefont {Pfaff}, \citenamefont {Frunzio}, \citenamefont
  {Schoelkopf},\ and\ \citenamefont {Devoret}}]{Narla2016}%
  \BibitemOpen
  \bibfield  {author} {\bibinfo {author} {\bibfnamefont {A.}~\bibnamefont
  {Narla}}, \bibinfo {author} {\bibfnamefont {S.}~\bibnamefont {Shankar}},
  \bibinfo {author} {\bibfnamefont {M.}~\bibnamefont {Hatridge}}, \bibinfo
  {author} {\bibfnamefont {Z.}~\bibnamefont {Leghtas}}, \bibinfo {author}
  {\bibfnamefont {K.~M.}\ \bibnamefont {Sliwa}}, \bibinfo {author}
  {\bibfnamefont {E.}~\bibnamefont {Zalys-Geller}}, \bibinfo {author}
  {\bibfnamefont {S.~O.}\ \bibnamefont {Mundhada}}, \bibinfo {author}
  {\bibfnamefont {W.}~\bibnamefont {Pfaff}}, \bibinfo {author} {\bibfnamefont
  {L.}~\bibnamefont {Frunzio}}, \bibinfo {author} {\bibfnamefont {R.~J.}\
  \bibnamefont {Schoelkopf}},\ and\ \bibinfo {author} {\bibfnamefont {M.~H.}\
  \bibnamefont {Devoret}},\ }\bibfield  {title} {\bibinfo {title} {{Robust
  concurrent remote entanglement between two superconducting qubits}},\ }\href
  {https://link.aps.org/doi/10.1103/PhysRevX.6.031036} {\bibfield  {journal}
  {\bibinfo  {journal} {Phys. Rev. X}\ }\textbf {\bibinfo {volume} {6}},\
  \bibinfo {pages} {031036} (\bibinfo {year} {2016})}\BibitemShut {NoStop}%
\bibitem [{\citenamefont {Lescanne}\ \emph {et~al.}(2020)\citenamefont
  {Lescanne}, \citenamefont {Del\'eglise}, \citenamefont {Albertinale},
  \citenamefont {R\'eglade}, \citenamefont {Capelle}, \citenamefont {Ivanov},
  \citenamefont {Jacqmin}, \citenamefont {Leghtas},\ and\ \citenamefont
  {Flurin}}]{Lescanne2020}%
  \BibitemOpen
  \bibfield  {author} {\bibinfo {author} {\bibfnamefont {R.}~\bibnamefont
  {Lescanne}}, \bibinfo {author} {\bibfnamefont {S.}~\bibnamefont
  {Del\'eglise}}, \bibinfo {author} {\bibfnamefont {E.}~\bibnamefont
  {Albertinale}}, \bibinfo {author} {\bibfnamefont {U.}~\bibnamefont
  {R\'eglade}}, \bibinfo {author} {\bibfnamefont {T.}~\bibnamefont {Capelle}},
  \bibinfo {author} {\bibfnamefont {E.}~\bibnamefont {Ivanov}}, \bibinfo
  {author} {\bibfnamefont {T.}~\bibnamefont {Jacqmin}}, \bibinfo {author}
  {\bibfnamefont {Z.}~\bibnamefont {Leghtas}},\ and\ \bibinfo {author}
  {\bibfnamefont {E.}~\bibnamefont {Flurin}},\ }\bibfield  {title} {\bibinfo
  {title} {Irreversible qubit-photon coupling for the detection of itinerant
  microwave photons},\ }\href {https://doi.org/10.1103/PhysRevX.10.021038}
  {\bibfield  {journal} {\bibinfo  {journal} {Phys. Rev. X}\ }\textbf {\bibinfo
  {volume} {10}},\ \bibinfo {pages} {021038} (\bibinfo {year}
  {2020})}\BibitemShut {NoStop}%
\bibitem [{\citenamefont {Sokolov}\ and\ \citenamefont
  {Wilhelm}(2020)}]{Sokolov2020}%
  \BibitemOpen
  \bibfield  {author} {\bibinfo {author} {\bibfnamefont {A.~M.}\ \bibnamefont
  {Sokolov}}\ and\ \bibinfo {author} {\bibfnamefont {F.~K.}\ \bibnamefont
  {Wilhelm}},\ }\href@noop {} {\bibinfo {title} {A superconducting detector
  that counts microwave photons up to two}} (\bibinfo {year} {2020}),\ \Eprint
  {https://arxiv.org/abs/2003.04625} {arXiv:2003.04625 [quant-ph]} \BibitemShut
  {NoStop}%
\bibitem [{\citenamefont {Grimsmo}\ \emph {et~al.}(2020)\citenamefont
  {Grimsmo}, \citenamefont {Royer}, \citenamefont {Kreikebaum}, \citenamefont
  {Ye}, \citenamefont {O'Brien}, \citenamefont {Siddiqi},\ and\ \citenamefont
  {Blais}}]{grimsmo2020quantum}%
  \BibitemOpen
  \bibfield  {author} {\bibinfo {author} {\bibfnamefont {A.~L.}\ \bibnamefont
  {Grimsmo}}, \bibinfo {author} {\bibfnamefont {B.}~\bibnamefont {Royer}},
  \bibinfo {author} {\bibfnamefont {J.~M.}\ \bibnamefont {Kreikebaum}},
  \bibinfo {author} {\bibfnamefont {Y.}~\bibnamefont {Ye}}, \bibinfo {author}
  {\bibfnamefont {K.}~\bibnamefont {O'Brien}}, \bibinfo {author} {\bibfnamefont
  {I.}~\bibnamefont {Siddiqi}},\ and\ \bibinfo {author} {\bibfnamefont
  {A.}~\bibnamefont {Blais}},\ }\href@noop {} {\bibinfo {title} {Quantum
  metamaterial for nondestructive microwave photon counting}} (\bibinfo {year}
  {2020}),\ \Eprint {https://arxiv.org/abs/2005.06483} {arXiv:2005.06483
  [quant-ph]} \BibitemShut {NoStop}%
\bibitem [{\citenamefont {Bergeal}\ \emph {et~al.}(2010)\citenamefont
  {Bergeal}, \citenamefont {Schackert}, \citenamefont {Metcalfe}, \citenamefont
  {Vijay}, \citenamefont {Manucharyan}, \citenamefont {Frunzio}, \citenamefont
  {Prober}, \citenamefont {Schoelkopf}, \citenamefont {Girvin},\ and\
  \citenamefont {Devoret}}]{Bergeal2010}%
  \BibitemOpen
  \bibfield  {author} {\bibinfo {author} {\bibfnamefont {N.}~\bibnamefont
  {Bergeal}}, \bibinfo {author} {\bibfnamefont {F.}~\bibnamefont {Schackert}},
  \bibinfo {author} {\bibfnamefont {M.}~\bibnamefont {Metcalfe}}, \bibinfo
  {author} {\bibfnamefont {R.}~\bibnamefont {Vijay}}, \bibinfo {author}
  {\bibfnamefont {V.~E.}\ \bibnamefont {Manucharyan}}, \bibinfo {author}
  {\bibfnamefont {L.}~\bibnamefont {Frunzio}}, \bibinfo {author} {\bibfnamefont
  {D.~E.}\ \bibnamefont {Prober}}, \bibinfo {author} {\bibfnamefont {R.~J.}\
  \bibnamefont {Schoelkopf}}, \bibinfo {author} {\bibfnamefont {S.~M.}\
  \bibnamefont {Girvin}},\ and\ \bibinfo {author} {\bibfnamefont {M.~H.}\
  \bibnamefont {Devoret}},\ }\bibfield  {title} {\bibinfo {title}
  {{Phase-preserving amplification near the quantum limit with a Josephson ring
  modulator}},\ }\href {https://doi.org/10.1038/nature09035} {\bibfield
  {journal} {\bibinfo  {journal} {Nature}\ }\textbf {\bibinfo {volume} {465}},\
  \bibinfo {pages} {64} (\bibinfo {year} {2010})}\BibitemShut {NoStop}%
\bibitem [{\citenamefont {Roch}\ \emph {et~al.}(2012)\citenamefont {Roch},
  \citenamefont {Flurin}, \citenamefont {Nguyen}, \citenamefont {Morfin},
  \citenamefont {Campagne-Ibarcq}, \citenamefont {Devoret},\ and\ \citenamefont
  {Huard}}]{Roch2012}%
  \BibitemOpen
  \bibfield  {author} {\bibinfo {author} {\bibfnamefont {N.}~\bibnamefont
  {Roch}}, \bibinfo {author} {\bibfnamefont {E.}~\bibnamefont {Flurin}},
  \bibinfo {author} {\bibfnamefont {F.}~\bibnamefont {Nguyen}}, \bibinfo
  {author} {\bibfnamefont {P.}~\bibnamefont {Morfin}}, \bibinfo {author}
  {\bibfnamefont {P.}~\bibnamefont {Campagne-Ibarcq}}, \bibinfo {author}
  {\bibfnamefont {M.~H.}\ \bibnamefont {Devoret}},\ and\ \bibinfo {author}
  {\bibfnamefont {B.}~\bibnamefont {Huard}},\ }\bibfield  {title} {\bibinfo
  {title} {{Widely Tunable, Nondegenerate Three-Wave Mixing Microwave Device
  Operating near the Quantum Limit}},\ }\href
  {https://doi.org/10.1103/PhysRevLett.108.147701} {\bibfield  {journal}
  {\bibinfo  {journal} {Phys. Rev. Lett.}\ }\textbf {\bibinfo {volume} {108}},\
  \bibinfo {pages} {147701} (\bibinfo {year} {2012})}\BibitemShut {NoStop}%
\bibitem [{\citenamefont {Peronnin}\ \emph {et~al.}(2020)\citenamefont
  {Peronnin}, \citenamefont {Markovi\ifmmode~\acute{c}\else \'{c}\fi{}},
  \citenamefont {Ficheux},\ and\ \citenamefont {Huard}}]{Peronnin2019}%
  \BibitemOpen
  \bibfield  {author} {\bibinfo {author} {\bibfnamefont {T.}~\bibnamefont
  {Peronnin}}, \bibinfo {author} {\bibfnamefont {D.}~\bibnamefont
  {Markovi\ifmmode~\acute{c}\else \'{c}\fi{}}}, \bibinfo {author}
  {\bibfnamefont {Q.}~\bibnamefont {Ficheux}},\ and\ \bibinfo {author}
  {\bibfnamefont {B.}~\bibnamefont {Huard}},\ }\bibfield  {title} {\bibinfo
  {title} {Sequential dispersive measurement of a superconducting qubit},\
  }\href {https://doi.org/10.1103/PhysRevLett.124.180502} {\bibfield  {journal}
  {\bibinfo  {journal} {Phys. Rev. Lett.}\ }\textbf {\bibinfo {volume} {124}},\
  \bibinfo {pages} {180502} (\bibinfo {year} {2020})}\BibitemShut {NoStop}%
\bibitem [{\citenamefont {Yin}\ \emph {et~al.}(2013)\citenamefont {Yin},
  \citenamefont {Chen}, \citenamefont {Sank}, \citenamefont {O'Malley},
  \citenamefont {White}, \citenamefont {Barends}, \citenamefont {Kelly},
  \citenamefont {Lucero}, \citenamefont {Mariantoni}, \citenamefont {Megrant},
  \citenamefont {Neill}, \citenamefont {Vainsencher}, \citenamefont {Wenner},
  \citenamefont {Korotkov}, \citenamefont {Cleland},\ and\ \citenamefont
  {Martinis}}]{Yin2013}%
  \BibitemOpen
  \bibfield  {author} {\bibinfo {author} {\bibfnamefont {Y.}~\bibnamefont
  {Yin}}, \bibinfo {author} {\bibfnamefont {Y.}~\bibnamefont {Chen}}, \bibinfo
  {author} {\bibfnamefont {D.}~\bibnamefont {Sank}}, \bibinfo {author}
  {\bibfnamefont {P.~J.~J.}\ \bibnamefont {O'Malley}}, \bibinfo {author}
  {\bibfnamefont {T.~C.}\ \bibnamefont {White}}, \bibinfo {author}
  {\bibfnamefont {R.}~\bibnamefont {Barends}}, \bibinfo {author} {\bibfnamefont
  {J.}~\bibnamefont {Kelly}}, \bibinfo {author} {\bibfnamefont
  {E.}~\bibnamefont {Lucero}}, \bibinfo {author} {\bibfnamefont
  {M.}~\bibnamefont {Mariantoni}}, \bibinfo {author} {\bibfnamefont
  {A.}~\bibnamefont {Megrant}}, \bibinfo {author} {\bibfnamefont
  {C.}~\bibnamefont {Neill}}, \bibinfo {author} {\bibfnamefont
  {A.}~\bibnamefont {Vainsencher}}, \bibinfo {author} {\bibfnamefont
  {J.}~\bibnamefont {Wenner}}, \bibinfo {author} {\bibfnamefont {A.~N.}\
  \bibnamefont {Korotkov}}, \bibinfo {author} {\bibfnamefont {A.~N.}\
  \bibnamefont {Cleland}},\ and\ \bibinfo {author} {\bibfnamefont {J.~M.}\
  \bibnamefont {Martinis}},\ }\bibfield  {title} {\bibinfo {title} {{Catch and
  release of microwave photon states}},\ }\href
  {https://doi.org/10.1103/PhysRevLett.110.107001} {\bibfield  {journal}
  {\bibinfo  {journal} {Phys. Rev. Lett.}\ }\textbf {\bibinfo {volume} {110}},\
  \bibinfo {pages} {107001} (\bibinfo {year} {2013})}\BibitemShut {NoStop}%
\bibitem [{\citenamefont {Wenner}\ \emph {et~al.}(2014)\citenamefont {Wenner},
  \citenamefont {Yin}, \citenamefont {Chen}, \citenamefont {Barends},
  \citenamefont {Chiaro}, \citenamefont {Jeffrey}, \citenamefont {Kelly},
  \citenamefont {Megrant}, \citenamefont {Mutus}, \citenamefont {Neill},
  \citenamefont {O'Malley}, \citenamefont {Roushan}, \citenamefont {Sank},
  \citenamefont {Vainsencher}, \citenamefont {White}, \citenamefont {Korotkov},
  \citenamefont {Cleland},\ and\ \citenamefont {Martinis}}]{Wenner2014}%
  \BibitemOpen
  \bibfield  {author} {\bibinfo {author} {\bibfnamefont {J.}~\bibnamefont
  {Wenner}}, \bibinfo {author} {\bibfnamefont {Y.}~\bibnamefont {Yin}},
  \bibinfo {author} {\bibfnamefont {Y.}~\bibnamefont {Chen}}, \bibinfo {author}
  {\bibfnamefont {R.}~\bibnamefont {Barends}}, \bibinfo {author} {\bibfnamefont
  {B.}~\bibnamefont {Chiaro}}, \bibinfo {author} {\bibfnamefont
  {E.}~\bibnamefont {Jeffrey}}, \bibinfo {author} {\bibfnamefont
  {J.}~\bibnamefont {Kelly}}, \bibinfo {author} {\bibfnamefont
  {A.}~\bibnamefont {Megrant}}, \bibinfo {author} {\bibfnamefont {J.~Y.}\
  \bibnamefont {Mutus}}, \bibinfo {author} {\bibfnamefont {C.}~\bibnamefont
  {Neill}}, \bibinfo {author} {\bibfnamefont {P.~J.~J.}\ \bibnamefont
  {O'Malley}}, \bibinfo {author} {\bibfnamefont {P.}~\bibnamefont {Roushan}},
  \bibinfo {author} {\bibfnamefont {D.}~\bibnamefont {Sank}}, \bibinfo {author}
  {\bibfnamefont {A.}~\bibnamefont {Vainsencher}}, \bibinfo {author}
  {\bibfnamefont {T.~C.}\ \bibnamefont {White}}, \bibinfo {author}
  {\bibfnamefont {A.~N.}\ \bibnamefont {Korotkov}}, \bibinfo {author}
  {\bibfnamefont {A.~N.}\ \bibnamefont {Cleland}},\ and\ \bibinfo {author}
  {\bibfnamefont {J.~M.}\ \bibnamefont {Martinis}},\ }\bibfield  {title}
  {\bibinfo {title} {{Catching time-reversed microwave coherent state photons
  with 99.4{\%} absorption efficiency}},\ }\href
  {https://doi.org/10.1103/PhysRevLett.112.210501} {\bibfield  {journal}
  {\bibinfo  {journal} {Phys. Rev. Lett.}\ }\textbf {\bibinfo {volume} {112}},\
  \bibinfo {pages} {210501} (\bibinfo {year} {2014})}\BibitemShut {NoStop}%
\bibitem [{\citenamefont {Flurin}(2014)}]{Flurin2014}%
  \BibitemOpen
  \bibfield  {author} {\bibinfo {author} {\bibfnamefont {E.}~\bibnamefont
  {Flurin}},\ }\emph {\bibinfo {title} {{The Josephson Mixer, a Swiss army
  knife for microwave quantum optics}}},\ \href
  {https://tel.archives-ouvertes.fr/tel-01241123} {Ph.D. thesis},\ \bibinfo
  {school} {{\'{E}}cole Normale Sup{\'{e}}rieure} (\bibinfo {year}
  {2014})\BibitemShut {NoStop}%
\bibitem [{\citenamefont {Axline}\ \emph {et~al.}(2018)\citenamefont {Axline},
  \citenamefont {Burkhart}, \citenamefont {Pfaff}, \citenamefont {Zhang},
  \citenamefont {Chou}, \citenamefont {Campagne-Ibarcq}, \citenamefont
  {Reinhold}, \citenamefont {Frunzio}, \citenamefont {Girvin}, \citenamefont
  {Jiang}, \citenamefont {Devoret},\ and\ \citenamefont
  {Schoelkopf}}]{Axline2018}%
  \BibitemOpen
  \bibfield  {author} {\bibinfo {author} {\bibfnamefont {C.~J.}\ \bibnamefont
  {Axline}}, \bibinfo {author} {\bibfnamefont {L.~D.}\ \bibnamefont
  {Burkhart}}, \bibinfo {author} {\bibfnamefont {W.}~\bibnamefont {Pfaff}},
  \bibinfo {author} {\bibfnamefont {M.}~\bibnamefont {Zhang}}, \bibinfo
  {author} {\bibfnamefont {K.}~\bibnamefont {Chou}}, \bibinfo {author}
  {\bibfnamefont {P.}~\bibnamefont {Campagne-Ibarcq}}, \bibinfo {author}
  {\bibfnamefont {P.}~\bibnamefont {Reinhold}}, \bibinfo {author}
  {\bibfnamefont {L.}~\bibnamefont {Frunzio}}, \bibinfo {author} {\bibfnamefont
  {S.~M.}\ \bibnamefont {Girvin}}, \bibinfo {author} {\bibfnamefont
  {L.}~\bibnamefont {Jiang}}, \bibinfo {author} {\bibfnamefont {M.~H.}\
  \bibnamefont {Devoret}},\ and\ \bibinfo {author} {\bibfnamefont {R.~J.}\
  \bibnamefont {Schoelkopf}},\ }\bibfield  {title} {\bibinfo {title}
  {{On-demand quantum state transfer and entanglement between remote microwave
  cavity memories}},\ }\href {https://doi.org/10.1038/s41567-018-0115-y}
  {\bibfield  {journal} {\bibinfo  {journal} {Nat. Phys.}\ }\textbf {\bibinfo
  {volume} {14}},\ \bibinfo {pages} {705} (\bibinfo {year} {2018})}\BibitemShut
  {NoStop}%
\bibitem [{\citenamefont {Zhong}\ \emph {et~al.}(2019)\citenamefont {Zhong},
  \citenamefont {Chang}, \citenamefont {Satzinger}, \citenamefont {Chou},
  \citenamefont {Bienfait}, \citenamefont {Conner}, \citenamefont {Dumur},
  \citenamefont {Grebel}, \citenamefont {Peairs}, \citenamefont {Povey},
  \citenamefont {Schuster},\ and\ \citenamefont {Cleland}}]{Zhong2019}%
  \BibitemOpen
  \bibfield  {author} {\bibinfo {author} {\bibfnamefont {Y.~P.}\ \bibnamefont
  {Zhong}}, \bibinfo {author} {\bibfnamefont {H.~S.}\ \bibnamefont {Chang}},
  \bibinfo {author} {\bibfnamefont {K.~J.}\ \bibnamefont {Satzinger}}, \bibinfo
  {author} {\bibfnamefont {M.~H.}\ \bibnamefont {Chou}}, \bibinfo {author}
  {\bibfnamefont {A.}~\bibnamefont {Bienfait}}, \bibinfo {author}
  {\bibfnamefont {C.~R.}\ \bibnamefont {Conner}}, \bibinfo {author}
  {\bibnamefont {Dumur}}, \bibinfo {author} {\bibfnamefont {J.}~\bibnamefont
  {Grebel}}, \bibinfo {author} {\bibfnamefont {G.~A.}\ \bibnamefont {Peairs}},
  \bibinfo {author} {\bibfnamefont {R.~G.}\ \bibnamefont {Povey}}, \bibinfo
  {author} {\bibfnamefont {D.~I.}\ \bibnamefont {Schuster}},\ and\ \bibinfo
  {author} {\bibfnamefont {A.~N.}\ \bibnamefont {Cleland}},\ }\bibfield
  {title} {\bibinfo {title} {{Violating Bell's inequality with remotely
  connected superconducting qubits}},\ }\href
  {https://doi.org/10.1038/s41567-019-0507-7} {\bibfield  {journal} {\bibinfo
  {journal} {Nat. Phys.}\ }\textbf {\bibinfo {volume} {15}},\ \bibinfo {pages}
  {741–744} (\bibinfo {year} {2019})}\BibitemShut {NoStop}%
\bibitem [{\citenamefont {Campagne-Ibarcq}\ \emph {et~al.}(2018)\citenamefont
  {Campagne-Ibarcq}, \citenamefont {Zalys-Geller}, \citenamefont {Narla},
  \citenamefont {Shankar}, \citenamefont {Reinhold}, \citenamefont {Burkhart},
  \citenamefont {Axline}, \citenamefont {Pfaff}, \citenamefont {Frunzio},
  \citenamefont {Schoelkopf},\ and\ \citenamefont
  {Devoret}}]{Campagne-Ibarcq2018}%
  \BibitemOpen
  \bibfield  {author} {\bibinfo {author} {\bibfnamefont {P.}~\bibnamefont
  {Campagne-Ibarcq}}, \bibinfo {author} {\bibfnamefont {E.}~\bibnamefont
  {Zalys-Geller}}, \bibinfo {author} {\bibfnamefont {A.}~\bibnamefont {Narla}},
  \bibinfo {author} {\bibfnamefont {S.}~\bibnamefont {Shankar}}, \bibinfo
  {author} {\bibfnamefont {P.}~\bibnamefont {Reinhold}}, \bibinfo {author}
  {\bibfnamefont {L.}~\bibnamefont {Burkhart}}, \bibinfo {author}
  {\bibfnamefont {C.}~\bibnamefont {Axline}}, \bibinfo {author} {\bibfnamefont
  {W.}~\bibnamefont {Pfaff}}, \bibinfo {author} {\bibfnamefont
  {L.}~\bibnamefont {Frunzio}}, \bibinfo {author} {\bibfnamefont {R.~J.}\
  \bibnamefont {Schoelkopf}},\ and\ \bibinfo {author} {\bibfnamefont {M.~H.}\
  \bibnamefont {Devoret}},\ }\bibfield  {title} {\bibinfo {title}
  {{Deterministic Remote Entanglement of Superconducting Circuits through
  Microwave Two-Photon Transitions}},\ }\href
  {https://doi.org/10.1103/PhysRevLett.120.200501} {\bibfield  {journal}
  {\bibinfo  {journal} {Phys. Rev. Lett.}\ }\textbf {\bibinfo {volume} {120}},\
  \bibinfo {pages} {200501} (\bibinfo {year} {2018})}\BibitemShut {NoStop}%
\bibitem [{\citenamefont {Kurpiers}\ \emph {et~al.}(2018)\citenamefont
  {Kurpiers}, \citenamefont {Magnard}, \citenamefont {Walter}, \citenamefont
  {Royer}, \citenamefont {Pechal}, \citenamefont {Heinsoo}, \citenamefont
  {Salath{\'{e}}}, \citenamefont {Akin}, \citenamefont {Storz}, \citenamefont
  {Besse}, \citenamefont {Gasparinetti}, \citenamefont {Blais},\ and\
  \citenamefont {Wallraff}}]{Kurpiers2018}%
  \BibitemOpen
  \bibfield  {author} {\bibinfo {author} {\bibfnamefont {P.}~\bibnamefont
  {Kurpiers}}, \bibinfo {author} {\bibfnamefont {P.}~\bibnamefont {Magnard}},
  \bibinfo {author} {\bibfnamefont {T.}~\bibnamefont {Walter}}, \bibinfo
  {author} {\bibfnamefont {B.}~\bibnamefont {Royer}}, \bibinfo {author}
  {\bibfnamefont {M.}~\bibnamefont {Pechal}}, \bibinfo {author} {\bibfnamefont
  {J.}~\bibnamefont {Heinsoo}}, \bibinfo {author} {\bibfnamefont
  {Y.}~\bibnamefont {Salath{\'{e}}}}, \bibinfo {author} {\bibfnamefont
  {A.}~\bibnamefont {Akin}}, \bibinfo {author} {\bibfnamefont {S.}~\bibnamefont
  {Storz}}, \bibinfo {author} {\bibfnamefont {J.~C.}\ \bibnamefont {Besse}},
  \bibinfo {author} {\bibfnamefont {S.}~\bibnamefont {Gasparinetti}}, \bibinfo
  {author} {\bibfnamefont {A.}~\bibnamefont {Blais}},\ and\ \bibinfo {author}
  {\bibfnamefont {A.}~\bibnamefont {Wallraff}},\ }\bibfield  {title} {\bibinfo
  {title} {{Deterministic quantum state transfer and remote entanglement using
  microwave photons}},\ }\href {https://doi.org/10.1038/s41586-018-0195-y}
  {\bibfield  {journal} {\bibinfo  {journal} {Nature}\ }\textbf {\bibinfo
  {volume} {558}},\ \bibinfo {pages} {264} (\bibinfo {year}
  {2018})}\BibitemShut {NoStop}%
\bibitem [{\citenamefont {Korotkov}(2011)}]{Korotkov2011}%
  \BibitemOpen
  \bibfield  {author} {\bibinfo {author} {\bibfnamefont {A.~N.}\ \bibnamefont
  {Korotkov}},\ }\bibfield  {title} {\bibinfo {title} {{Flying microwave qubits
  with nearly perfect transfer efficiency}},\ }\href
  {https://doi.org/10.1103/PhysRevB.84.014510} {\bibfield  {journal} {\bibinfo
  {journal} {Phys. Rev. B}\ }\textbf {\bibinfo {volume} {84}},\ \bibinfo
  {pages} {014510} (\bibinfo {year} {2011})}\BibitemShut {NoStop}%
\bibitem [{\citenamefont {Flurin}\ \emph {et~al.}(2015)\citenamefont {Flurin},
  \citenamefont {Roch}, \citenamefont {Pillet}, \citenamefont {Mallet},\ and\
  \citenamefont {Huard}}]{Flurin2015}%
  \BibitemOpen
  \bibfield  {author} {\bibinfo {author} {\bibfnamefont {E.}~\bibnamefont
  {Flurin}}, \bibinfo {author} {\bibfnamefont {N.}~\bibnamefont {Roch}},
  \bibinfo {author} {\bibfnamefont {J.~D.}\ \bibnamefont {Pillet}}, \bibinfo
  {author} {\bibfnamefont {F.}~\bibnamefont {Mallet}},\ and\ \bibinfo {author}
  {\bibfnamefont {B.}~\bibnamefont {Huard}},\ }\bibfield  {title} {\bibinfo
  {title} {{Superconducting quantum node for entanglement and storage of
  microwave radiation}},\ }\href
  {https://doi.org/10.1103/PhysRevLett.114.090503} {\bibfield  {journal}
  {\bibinfo  {journal} {Phys. Rev. Lett.}\ }\textbf {\bibinfo {volume} {114}},\
  \bibinfo {pages} {1} (\bibinfo {year} {2015})}\BibitemShut {NoStop}%
\bibitem [{\citenamefont {Schuster}\ \emph {et~al.}(2007)\citenamefont
  {Schuster}, \citenamefont {Houck}, \citenamefont {Schreier}, \citenamefont
  {Wallraff}, \citenamefont {Gambetta}, \citenamefont {Blais}, \citenamefont
  {Frunzio}, \citenamefont {Majer}, \citenamefont {Johnson}, \citenamefont
  {Devoret}, \citenamefont {Girvin},\ and\ \citenamefont
  {Schoelkopf}}]{Schuster2007}%
  \BibitemOpen
  \bibfield  {author} {\bibinfo {author} {\bibfnamefont {D.~I.}\ \bibnamefont
  {Schuster}}, \bibinfo {author} {\bibfnamefont {A.~A.}\ \bibnamefont {Houck}},
  \bibinfo {author} {\bibfnamefont {J.~A.}\ \bibnamefont {Schreier}}, \bibinfo
  {author} {\bibfnamefont {A.}~\bibnamefont {Wallraff}}, \bibinfo {author}
  {\bibfnamefont {J.~M.}\ \bibnamefont {Gambetta}}, \bibinfo {author}
  {\bibfnamefont {A.}~\bibnamefont {Blais}}, \bibinfo {author} {\bibfnamefont
  {L.}~\bibnamefont {Frunzio}}, \bibinfo {author} {\bibfnamefont
  {J.}~\bibnamefont {Majer}}, \bibinfo {author} {\bibfnamefont
  {B.}~\bibnamefont {Johnson}}, \bibinfo {author} {\bibfnamefont {M.~H.}\
  \bibnamefont {Devoret}}, \bibinfo {author} {\bibfnamefont {S.~M.}\
  \bibnamefont {Girvin}},\ and\ \bibinfo {author} {\bibfnamefont {R.~J.}\
  \bibnamefont {Schoelkopf}},\ }\bibfield  {title} {\bibinfo {title}
  {{Resolving photon number states in a superconducting circuit}},\ }\href
  {https://doi.org/10.1038/nature05461} {\bibfield  {journal} {\bibinfo
  {journal} {Nature}\ }\textbf {\bibinfo {volume} {445}},\ \bibinfo {pages}
  {515} (\bibinfo {year} {2007})}\BibitemShut {NoStop}%
\bibitem [{\citenamefont {McClure}\ \emph {et~al.}(2016)\citenamefont
  {McClure}, \citenamefont {Paik}, \citenamefont {Bishop}, \citenamefont
  {Steffen}, \citenamefont {Chow},\ and\ \citenamefont
  {Gambetta}}]{McClure2016}%
  \BibitemOpen
  \bibfield  {author} {\bibinfo {author} {\bibfnamefont {D.~T.}\ \bibnamefont
  {McClure}}, \bibinfo {author} {\bibfnamefont {H.}~\bibnamefont {Paik}},
  \bibinfo {author} {\bibfnamefont {L.~S.}\ \bibnamefont {Bishop}}, \bibinfo
  {author} {\bibfnamefont {M.}~\bibnamefont {Steffen}}, \bibinfo {author}
  {\bibfnamefont {J.~M.}\ \bibnamefont {Chow}},\ and\ \bibinfo {author}
  {\bibfnamefont {J.~M.}\ \bibnamefont {Gambetta}},\ }\bibfield  {title}
  {\bibinfo {title} {{Rapid Driven Reset of a Qubit Readout Resonator}},\
  }\href {https://doi.org/10.1103/PhysRevApplied.5.011001} {\bibfield
  {journal} {\bibinfo  {journal} {Phys. Rev. Applied}\ }\textbf {\bibinfo
  {volume} {5}},\ \bibinfo {pages} {11001} (\bibinfo {year}
  {2016})}\BibitemShut {NoStop}%
\bibitem [{\citenamefont {Haroche}\ \emph {et~al.}(1992)\citenamefont
  {Haroche}, \citenamefont {Brune},\ and\ \citenamefont
  {Raimond}}]{Haroche1992}%
  \BibitemOpen
  \bibfield  {author} {\bibinfo {author} {\bibfnamefont {S.}~\bibnamefont
  {Haroche}}, \bibinfo {author} {\bibfnamefont {M.}~\bibnamefont {Brune}},\
  and\ \bibinfo {author} {\bibfnamefont {J.}~\bibnamefont {Raimond}},\
  }\bibfield  {title} {\bibinfo {title} {{Measuring photon numbers in a cavity
  by atomic interferometry: optimizing the convergence procedure}},\ }\href
  {https://doi.org/10.1051/jp2:1992157} {\bibfield  {journal} {\bibinfo
  {journal} {{Journal de Physique II}}\ }\textbf {\bibinfo {volume} {2}},\
  \bibinfo {pages} {659} (\bibinfo {year} {1992})}\BibitemShut {NoStop}%
\bibitem [{\citenamefont {Heeres}\ \emph {et~al.}(2016)\citenamefont {Heeres},
  \citenamefont {Reinhold},\ and\ \citenamefont {Schoelkopf}}]{Heeres2016}%
  \BibitemOpen
  \bibfield  {author} {\bibinfo {author} {\bibfnamefont {R.}~\bibnamefont
  {Heeres}}, \bibinfo {author} {\bibfnamefont {P.}~\bibnamefont {Reinhold}},\
  and\ \bibinfo {author} {\bibfnamefont {R.}~\bibnamefont {Schoelkopf}},\
  }\href@noop {} {\bibinfo {title} {{Private communication}}} (\bibinfo {year}
  {2016})\BibitemShut {NoStop}%
\bibitem [{\citenamefont {Wang}\ \emph {et~al.}(2020)\citenamefont {Wang},
  \citenamefont {Curtis}, \citenamefont {Lester}, \citenamefont {Zhang},
  \citenamefont {Gao}, \citenamefont {Freeze}, \citenamefont {Batista},
  \citenamefont {Vaccaro}, \citenamefont {Chuang}, \citenamefont {Frunzio},
  \citenamefont {Jiang}, \citenamefont {Girvin},\ and\ \citenamefont
  {Schoelkopf}}]{Wang2020}%
  \BibitemOpen
  \bibfield  {author} {\bibinfo {author} {\bibfnamefont {C.~S.}\ \bibnamefont
  {Wang}}, \bibinfo {author} {\bibfnamefont {J.~C.}\ \bibnamefont {Curtis}},
  \bibinfo {author} {\bibfnamefont {B.~J.}\ \bibnamefont {Lester}}, \bibinfo
  {author} {\bibfnamefont {Y.}~\bibnamefont {Zhang}}, \bibinfo {author}
  {\bibfnamefont {Y.~Y.}\ \bibnamefont {Gao}}, \bibinfo {author} {\bibfnamefont
  {J.}~\bibnamefont {Freeze}}, \bibinfo {author} {\bibfnamefont {V.~S.}\
  \bibnamefont {Batista}}, \bibinfo {author} {\bibfnamefont {P.~H.}\
  \bibnamefont {Vaccaro}}, \bibinfo {author} {\bibfnamefont {I.~L.}\
  \bibnamefont {Chuang}}, \bibinfo {author} {\bibfnamefont {L.}~\bibnamefont
  {Frunzio}}, \bibinfo {author} {\bibfnamefont {L.}~\bibnamefont {Jiang}},
  \bibinfo {author} {\bibfnamefont {S.~M.}\ \bibnamefont {Girvin}},\ and\
  \bibinfo {author} {\bibfnamefont {R.~J.}\ \bibnamefont {Schoelkopf}},\
  }\bibfield  {title} {\bibinfo {title} {Efficient multiphoton sampling of
  molecular vibronic spectra on a superconducting bosonic processor},\ }\href
  {https://doi.org/10.1103/PhysRevX.10.021060} {\bibfield  {journal} {\bibinfo
  {journal} {Phys. Rev. X}\ }\textbf {\bibinfo {volume} {10}},\ \bibinfo
  {pages} {021060} (\bibinfo {year} {2020})}\BibitemShut {NoStop}%
\bibitem [{\citenamefont {Motzoi}\ \emph {et~al.}(2009)\citenamefont {Motzoi},
  \citenamefont {Gambetta}, \citenamefont {Rebentrost},\ and\ \citenamefont
  {Wilhelm}}]{Motzoi2009}%
  \BibitemOpen
  \bibfield  {author} {\bibinfo {author} {\bibfnamefont {F.}~\bibnamefont
  {Motzoi}}, \bibinfo {author} {\bibfnamefont {J.~M.}\ \bibnamefont
  {Gambetta}}, \bibinfo {author} {\bibfnamefont {P.}~\bibnamefont
  {Rebentrost}},\ and\ \bibinfo {author} {\bibfnamefont {F.~K.}\ \bibnamefont
  {Wilhelm}},\ }\bibfield  {title} {\bibinfo {title} {{Simple Pulses for
  Elimination of Leakage in Weakly Nonlinear Qubits}},\ }\href
  {https://doi.org/10.1103/PhysRevLett.103.110501} {\bibfield  {journal}
  {\bibinfo  {journal} {Phys. Rev. Lett.}\ }\textbf {\bibinfo {volume} {103}},\
  \bibinfo {pages} {110501} (\bibinfo {year} {2009})}\BibitemShut {NoStop}%
\bibitem [{\citenamefont {Khezri}\ \emph {et~al.}(2016)\citenamefont {Khezri},
  \citenamefont {Mlinar}, \citenamefont {Dressel},\ and\ \citenamefont
  {Korotkov}}]{Khezri2016}%
  \BibitemOpen
  \bibfield  {author} {\bibinfo {author} {\bibfnamefont {M.}~\bibnamefont
  {Khezri}}, \bibinfo {author} {\bibfnamefont {E.}~\bibnamefont {Mlinar}},
  \bibinfo {author} {\bibfnamefont {J.}~\bibnamefont {Dressel}},\ and\ \bibinfo
  {author} {\bibfnamefont {A.~N.}\ \bibnamefont {Korotkov}},\ }\bibfield
  {title} {\bibinfo {title} {{Measuring a transmon qubit in circuit QED:
  Dressed squeezed states}},\ }\href
  {https://doi.org/10.1103/PhysRevA.94.012347} {\bibfield  {journal} {\bibinfo
  {journal} {Phys. Rev. A}\ }\textbf {\bibinfo {volume} {94}},\ \bibinfo
  {pages} {12347} (\bibinfo {year} {2016})}\BibitemShut {NoStop}%
\bibitem [{\citenamefont {Lutterbach}\ and\ \citenamefont
  {Davidovich}(1997)}]{Lutterbach1997}%
  \BibitemOpen
  \bibfield  {author} {\bibinfo {author} {\bibfnamefont {L.~G.}\ \bibnamefont
  {Lutterbach}}\ and\ \bibinfo {author} {\bibfnamefont {L.}~\bibnamefont
  {Davidovich}},\ }\bibfield  {title} {\bibinfo {title} {{Method for Direct
  Measurement of the Wigner Function in Cavity QED and Ion Traps}},\ }\href
  {https://doi.org/10.1103/PhysRevLett.78.2547} {\bibfield  {journal} {\bibinfo
   {journal} {Phys. Rev. Lett.}\ }\textbf {\bibinfo {volume} {78}},\ \bibinfo
  {pages} {2547} (\bibinfo {year} {1997})}\BibitemShut {NoStop}%
\bibitem [{\citenamefont {Bertet}\ \emph {et~al.}(2002)\citenamefont {Bertet},
  \citenamefont {Auffeves}, \citenamefont {Maioli}, \citenamefont {Osnaghi},
  \citenamefont {Meunier}, \citenamefont {Brune}, \citenamefont {Raimond},\
  and\ \citenamefont {Haroche}}]{Bertet2002}%
  \BibitemOpen
  \bibfield  {author} {\bibinfo {author} {\bibfnamefont {P.}~\bibnamefont
  {Bertet}}, \bibinfo {author} {\bibfnamefont {A.}~\bibnamefont {Auffeves}},
  \bibinfo {author} {\bibfnamefont {P.}~\bibnamefont {Maioli}}, \bibinfo
  {author} {\bibfnamefont {S.}~\bibnamefont {Osnaghi}}, \bibinfo {author}
  {\bibfnamefont {T.}~\bibnamefont {Meunier}}, \bibinfo {author} {\bibfnamefont
  {M.}~\bibnamefont {Brune}}, \bibinfo {author} {\bibfnamefont {J.~M.}\
  \bibnamefont {Raimond}},\ and\ \bibinfo {author} {\bibfnamefont
  {S.}~\bibnamefont {Haroche}},\ }\bibfield  {title} {\bibinfo {title} {{Direct
  Measurement of the Wigner Function of a One-Photon Fock State in a Cavity}},\
  }\href {https://doi.org/10.1103/PhysRevLett.89.200402} {\bibfield  {journal}
  {\bibinfo  {journal} {Phys. Rev. Lett.}\ }\textbf {\bibinfo {volume} {89}},\
  \bibinfo {pages} {200402} (\bibinfo {year} {2002})}\BibitemShut {NoStop}%
\bibitem [{\citenamefont {Vlastakis}\ \emph {et~al.}(2013)\citenamefont
  {Vlastakis}, \citenamefont {Kirchmair}, \citenamefont {Leghtas},
  \citenamefont {Nigg}, \citenamefont {Frunzio}, \citenamefont {Girvin},
  \citenamefont {Mirrahimi}, \citenamefont {Devoret},\ and\ \citenamefont
  {Schoelkopf}}]{Vlastakis2013}%
  \BibitemOpen
  \bibfield  {author} {\bibinfo {author} {\bibfnamefont {B.}~\bibnamefont
  {Vlastakis}}, \bibinfo {author} {\bibfnamefont {G.}~\bibnamefont
  {Kirchmair}}, \bibinfo {author} {\bibfnamefont {Z.}~\bibnamefont {Leghtas}},
  \bibinfo {author} {\bibfnamefont {S.~E.}\ \bibnamefont {Nigg}}, \bibinfo
  {author} {\bibfnamefont {L.}~\bibnamefont {Frunzio}}, \bibinfo {author}
  {\bibfnamefont {S.~M.}\ \bibnamefont {Girvin}}, \bibinfo {author}
  {\bibfnamefont {M.}~\bibnamefont {Mirrahimi}}, \bibinfo {author}
  {\bibfnamefont {M.~H.}\ \bibnamefont {Devoret}},\ and\ \bibinfo {author}
  {\bibfnamefont {R.~J.}\ \bibnamefont {Schoelkopf}},\ }\bibfield  {title}
  {\bibinfo {title} {Deterministically encoding quantum information using
  100-photon {S}chr{\"o}dinger cat states},\ }\href
  {https://doi.org/10.1126/science.1243289} {\bibfield  {journal} {\bibinfo
  {journal} {Science}\ }\textbf {\bibinfo {volume} {342}},\ \bibinfo {pages}
  {607} (\bibinfo {year} {2013})}\BibitemShut {NoStop}%
\bibitem [{\citenamefont {Mendon\ifmmode~\mbox{\c{c}}\else \c{c}\fi{}a}\ \emph
  {et~al.}(2008)\citenamefont {Mendon\ifmmode~\mbox{\c{c}}\else \c{c}\fi{}a},
  \citenamefont {Napolitano}, \citenamefont {Marchiolli}, \citenamefont
  {Foster},\ and\ \citenamefont {Liang}}]{Paulo2008}%
  \BibitemOpen
  \bibfield  {author} {\bibinfo {author} {\bibfnamefont {P.~E. M.~F.}\
  \bibnamefont {Mendon\ifmmode~\mbox{\c{c}}\else \c{c}\fi{}a}}, \bibinfo
  {author} {\bibfnamefont {R.~d.~J.}\ \bibnamefont {Napolitano}}, \bibinfo
  {author} {\bibfnamefont {M.~A.}\ \bibnamefont {Marchiolli}}, \bibinfo
  {author} {\bibfnamefont {C.~J.}\ \bibnamefont {Foster}},\ and\ \bibinfo
  {author} {\bibfnamefont {Y.-C.}\ \bibnamefont {Liang}},\ }\bibfield  {title}
  {\bibinfo {title} {Alternative fidelity measure between quantum states},\
  }\href {https://doi.org/10.1103/PhysRevA.78.052330} {\bibfield  {journal}
  {\bibinfo  {journal} {Phys. Rev. A}\ }\textbf {\bibinfo {volume} {78}},\
  \bibinfo {pages} {052330} (\bibinfo {year} {2008})}\BibitemShut {NoStop}%
\bibitem [{\citenamefont {Miszczak}\ \emph {et~al.}(2009)\citenamefont
  {Miszczak}, \citenamefont {Pucha\l{}a}, \citenamefont {Horodecki},
  \citenamefont {Uhlmann},\ and\ \citenamefont {Zyczkowski}}]{Miszczak2009}%
  \BibitemOpen
  \bibfield  {author} {\bibinfo {author} {\bibfnamefont {J.~A.}\ \bibnamefont
  {Miszczak}}, \bibinfo {author} {\bibfnamefont {Z.}~\bibnamefont
  {Pucha\l{}a}}, \bibinfo {author} {\bibfnamefont {P.}~\bibnamefont
  {Horodecki}}, \bibinfo {author} {\bibfnamefont {A.}~\bibnamefont {Uhlmann}},\
  and\ \bibinfo {author} {\bibfnamefont {K.}~\bibnamefont {Zyczkowski}},\
  }\bibfield  {title} {\bibinfo {title} {Sub- and super-fidelity as bounds for
  quantum fidelity},\ }\href@noop {} {\bibfield  {journal} {\bibinfo  {journal}
  {Quantum Info. Comput.}\ }\textbf {\bibinfo {volume} {9}},\ \bibinfo {pages}
  {103–130} (\bibinfo {year} {2009})}\BibitemShut {NoStop}%
\bibitem [{\citenamefont {Besse}\ \emph {et~al.}(2019)\citenamefont {Besse},
  \citenamefont {Gasparinetti}, \citenamefont {Collodo}, \citenamefont
  {Walter}, \citenamefont {Remm}, \citenamefont {Krause}, \citenamefont
  {Eichler},\ and\ \citenamefont {Wallraff}}]{Besse2019}%
  \BibitemOpen
  \bibfield  {author} {\bibinfo {author} {\bibfnamefont {J.-c.}\ \bibnamefont
  {Besse}}, \bibinfo {author} {\bibfnamefont {S.}~\bibnamefont {Gasparinetti}},
  \bibinfo {author} {\bibfnamefont {M.~C.}\ \bibnamefont {Collodo}}, \bibinfo
  {author} {\bibfnamefont {T.}~\bibnamefont {Walter}}, \bibinfo {author}
  {\bibfnamefont {A.}~\bibnamefont {Remm}}, \bibinfo {author} {\bibfnamefont
  {J.}~\bibnamefont {Krause}}, \bibinfo {author} {\bibfnamefont
  {C.}~\bibnamefont {Eichler}},\ and\ \bibinfo {author} {\bibfnamefont
  {A.}~\bibnamefont {Wallraff}},\ }\bibfield  {title} {\bibinfo {title}
  {{Parity Detection of Propagating Microwave Fields}},\ }\href
  {https://doi.org/10.1103/PhysRevX.10.011046} {\bibfield  {journal} {\bibinfo
  {journal} {Phys. Rev. X}\ }\textbf {\bibinfo {volume} {10}},\ \bibinfo
  {pages} {11046} (\bibinfo {year} {2019})}\BibitemShut {NoStop}%
\bibitem [{\citenamefont {Dolinar}(1973)}]{Dolinar1973}%
  \BibitemOpen
  \bibfield  {author} {\bibinfo {author} {\bibfnamefont {S.~J.}\ \bibnamefont
  {Dolinar}},\ }\bibfield  {title} {\bibinfo {title} {{An optimum receiver for
  the binary coherent state quantum channel}},\ }\href
  {https://dspace.mit.edu/bitstream/handle/1721.1/56414/RLE_QPR_111_VII.pdf}
  {\bibfield  {journal} {\bibinfo  {journal} {MIT Research Laboratory of
  Electronics Quarterly Progress Report}\ }\textbf {\bibinfo {volume} {111}},\
  \bibinfo {pages} {115} (\bibinfo {year} {1973})}\BibitemShut {NoStop}%
\bibitem [{Git()}]{Github}%
  \BibitemOpen
  \href {https://github.com/Quantum-Circuit-Group/photocounting-OPX} {\bibinfo
  {title}
  {{https://github.com/Quantum-Circuit-Group/photocounting-OPX}}}\BibitemShut
  {NoStop}%
\bibitem [{\citenamefont {Macklin}\ \emph {et~al.}(2015)\citenamefont
  {Macklin}, \citenamefont {O{\textquoteright}Brien}, \citenamefont {Hover},
  \citenamefont {Schwartz}, \citenamefont {Bolkhovsky}, \citenamefont {Zhang},
  \citenamefont {Oliver},\ and\ \citenamefont {Siddiqi}}]{Macklin2015}%
  \BibitemOpen
  \bibfield  {author} {\bibinfo {author} {\bibfnamefont {C.}~\bibnamefont
  {Macklin}}, \bibinfo {author} {\bibfnamefont {K.}~\bibnamefont
  {O{\textquoteright}Brien}}, \bibinfo {author} {\bibfnamefont
  {D.}~\bibnamefont {Hover}}, \bibinfo {author} {\bibfnamefont {M.~E.}\
  \bibnamefont {Schwartz}}, \bibinfo {author} {\bibfnamefont {V.}~\bibnamefont
  {Bolkhovsky}}, \bibinfo {author} {\bibfnamefont {X.}~\bibnamefont {Zhang}},
  \bibinfo {author} {\bibfnamefont {W.~D.}\ \bibnamefont {Oliver}},\ and\
  \bibinfo {author} {\bibfnamefont {I.}~\bibnamefont {Siddiqi}},\ }\bibfield
  {title} {\bibinfo {title} {A near{\textendash}quantum-limited josephson
  traveling-wave parametric amplifier},\ }\href
  {https://doi.org/10.1126/science.aaa8525} {\bibfield  {journal} {\bibinfo
  {journal} {Science}\ }\textbf {\bibinfo {volume} {350}},\ \bibinfo {pages}
  {307} (\bibinfo {year} {2015})}\BibitemShut {NoStop}%
\bibitem [{\citenamefont {Sank}\ \emph {et~al.}(2016)\citenamefont {Sank},
  \citenamefont {Chen}, \citenamefont {Khezri}, \citenamefont {Kelly},
  \citenamefont {Barends}, \citenamefont {Campbell}, \citenamefont {Chen},
  \citenamefont {Chiaro}, \citenamefont {Dunsworth}, \citenamefont {Fowler},
  \citenamefont {Jeffrey}, \citenamefont {Lucero}, \citenamefont {Megrant},
  \citenamefont {Mutus}, \citenamefont {Neeley}, \citenamefont {Neill},
  \citenamefont {O'Malley}, \citenamefont {Quintana}, \citenamefont {Roushan},
  \citenamefont {Vainsencher}, \citenamefont {White}, \citenamefont {Wenner},
  \citenamefont {Korotkov},\ and\ \citenamefont {Martinis}}]{Sank2016}%
  \BibitemOpen
  \bibfield  {author} {\bibinfo {author} {\bibfnamefont {D.}~\bibnamefont
  {Sank}}, \bibinfo {author} {\bibfnamefont {Z.}~\bibnamefont {Chen}}, \bibinfo
  {author} {\bibfnamefont {M.}~\bibnamefont {Khezri}}, \bibinfo {author}
  {\bibfnamefont {J.}~\bibnamefont {Kelly}}, \bibinfo {author} {\bibfnamefont
  {R.}~\bibnamefont {Barends}}, \bibinfo {author} {\bibfnamefont
  {B.}~\bibnamefont {Campbell}}, \bibinfo {author} {\bibfnamefont
  {Y.}~\bibnamefont {Chen}}, \bibinfo {author} {\bibfnamefont {B.}~\bibnamefont
  {Chiaro}}, \bibinfo {author} {\bibfnamefont {A.}~\bibnamefont {Dunsworth}},
  \bibinfo {author} {\bibfnamefont {A.}~\bibnamefont {Fowler}}, \bibinfo
  {author} {\bibfnamefont {E.}~\bibnamefont {Jeffrey}}, \bibinfo {author}
  {\bibfnamefont {E.}~\bibnamefont {Lucero}}, \bibinfo {author} {\bibfnamefont
  {A.}~\bibnamefont {Megrant}}, \bibinfo {author} {\bibfnamefont
  {J.}~\bibnamefont {Mutus}}, \bibinfo {author} {\bibfnamefont
  {M.}~\bibnamefont {Neeley}}, \bibinfo {author} {\bibfnamefont
  {C.}~\bibnamefont {Neill}}, \bibinfo {author} {\bibfnamefont {P.~J.~J.}\
  \bibnamefont {O'Malley}}, \bibinfo {author} {\bibfnamefont {C.}~\bibnamefont
  {Quintana}}, \bibinfo {author} {\bibfnamefont {P.}~\bibnamefont {Roushan}},
  \bibinfo {author} {\bibfnamefont {A.}~\bibnamefont {Vainsencher}}, \bibinfo
  {author} {\bibfnamefont {T.}~\bibnamefont {White}}, \bibinfo {author}
  {\bibfnamefont {J.}~\bibnamefont {Wenner}}, \bibinfo {author} {\bibfnamefont
  {A.~N.}\ \bibnamefont {Korotkov}},\ and\ \bibinfo {author} {\bibfnamefont
  {J.~M.}\ \bibnamefont {Martinis}},\ }\bibfield  {title} {\bibinfo {title}
  {{Measurement-Induced State Transitions in a Superconducting Qubit: Beyond
  the Rotating Wave Approximation}},\ }\href
  {https://doi.org/10.1103/PhysRevLett.117.190503} {\bibfield  {journal}
  {\bibinfo  {journal} {Phys. Rev. Lett.}\ }\textbf {\bibinfo {volume} {117}},\
  \bibinfo {pages} {190503} (\bibinfo {year} {2016})}\BibitemShut {NoStop}%
\bibitem [{\citenamefont {Touzard}\ \emph {et~al.}(2019)\citenamefont
  {Touzard}, \citenamefont {Kou}, \citenamefont {Frattini}, \citenamefont
  {Sivak}, \citenamefont {Puri}, \citenamefont {Grimm}, \citenamefont
  {Frunzio}, \citenamefont {Shankar},\ and\ \citenamefont
  {Devoret}}]{Touzard2019}%
  \BibitemOpen
  \bibfield  {author} {\bibinfo {author} {\bibfnamefont {S.}~\bibnamefont
  {Touzard}}, \bibinfo {author} {\bibfnamefont {A.}~\bibnamefont {Kou}},
  \bibinfo {author} {\bibfnamefont {N.~E.}\ \bibnamefont {Frattini}}, \bibinfo
  {author} {\bibfnamefont {V.~V.}\ \bibnamefont {Sivak}}, \bibinfo {author}
  {\bibfnamefont {S.}~\bibnamefont {Puri}}, \bibinfo {author} {\bibfnamefont
  {A.}~\bibnamefont {Grimm}}, \bibinfo {author} {\bibfnamefont
  {L.}~\bibnamefont {Frunzio}}, \bibinfo {author} {\bibfnamefont
  {S.}~\bibnamefont {Shankar}},\ and\ \bibinfo {author} {\bibfnamefont {M.~H.}\
  \bibnamefont {Devoret}},\ }\bibfield  {title} {\bibinfo {title} {{Gated
  Conditional Displacement Readout of Superconducting Qubits}},\ }\href
  {https://doi.org/10.1103/PhysRevLett.122.080502} {\bibfield  {journal}
  {\bibinfo  {journal} {Phys. Rev. Lett.}\ }\textbf {\bibinfo {volume} {122}},\
  \bibinfo {pages} {80502} (\bibinfo {year} {2019})}\BibitemShut {NoStop}%
\bibitem [{\citenamefont {Ikonen}\ \emph {et~al.}(2019)\citenamefont {Ikonen},
  \citenamefont {Goetz}, \citenamefont {Ilves}, \citenamefont {Ker{\"{a}}nen},
  \citenamefont {Gunyho}, \citenamefont {Partanen}, \citenamefont {Tan},
  \citenamefont {Hazra}, \citenamefont {Gr{\"{o}}nberg}, \citenamefont
  {Vesterinen}, \citenamefont {Simbierowicz}, \citenamefont {Hassel},\ and\
  \citenamefont {M{\"{o}}tt{\"{o}}nen}}]{Ikonen2019}%
  \BibitemOpen
  \bibfield  {author} {\bibinfo {author} {\bibfnamefont {J.}~\bibnamefont
  {Ikonen}}, \bibinfo {author} {\bibfnamefont {J.}~\bibnamefont {Goetz}},
  \bibinfo {author} {\bibfnamefont {J.}~\bibnamefont {Ilves}}, \bibinfo
  {author} {\bibfnamefont {A.}~\bibnamefont {Ker{\"{a}}nen}}, \bibinfo {author}
  {\bibfnamefont {A.~M.}\ \bibnamefont {Gunyho}}, \bibinfo {author}
  {\bibfnamefont {M.}~\bibnamefont {Partanen}}, \bibinfo {author}
  {\bibfnamefont {K.~Y.}\ \bibnamefont {Tan}}, \bibinfo {author} {\bibfnamefont
  {D.}~\bibnamefont {Hazra}}, \bibinfo {author} {\bibfnamefont
  {L.}~\bibnamefont {Gr{\"{o}}nberg}}, \bibinfo {author} {\bibfnamefont
  {V.}~\bibnamefont {Vesterinen}}, \bibinfo {author} {\bibfnamefont
  {S.}~\bibnamefont {Simbierowicz}}, \bibinfo {author} {\bibfnamefont
  {J.}~\bibnamefont {Hassel}},\ and\ \bibinfo {author} {\bibfnamefont
  {M.}~\bibnamefont {M{\"{o}}tt{\"{o}}nen}},\ }\bibfield  {title} {\bibinfo
  {title} {{Qubit Measurement by Multichannel Driving}},\ }\href
  {https://doi.org/10.1103/PhysRevLett.122.080503} {\bibfield  {journal}
  {\bibinfo  {journal} {Phys. Rev. Lett.}\ }\textbf {\bibinfo {volume} {122}},\
  \bibinfo {pages} {80503} (\bibinfo {year} {2019})}\BibitemShut {NoStop}%
\bibitem [{\citenamefont {Dassonneville}\ \emph {et~al.}(2020)\citenamefont
  {Dassonneville}, \citenamefont {Ramos}, \citenamefont {Milchakov},
  \citenamefont {Planat}, \citenamefont {Dumur}, \citenamefont {Foroughi},
  \citenamefont {Puertas}, \citenamefont {Leger}, \citenamefont {Bharadwaj},
  \citenamefont {Delaforce}, \citenamefont {Naud}, \citenamefont
  {Hasch-Guichard}, \citenamefont {Garc\'{\i}a-Ripoll}, \citenamefont {Roch},\
  and\ \citenamefont {Buisson}}]{dassonneville2020}%
  \BibitemOpen
  \bibfield  {author} {\bibinfo {author} {\bibfnamefont {R.}~\bibnamefont
  {Dassonneville}}, \bibinfo {author} {\bibfnamefont {T.}~\bibnamefont
  {Ramos}}, \bibinfo {author} {\bibfnamefont {V.}~\bibnamefont {Milchakov}},
  \bibinfo {author} {\bibfnamefont {L.}~\bibnamefont {Planat}}, \bibinfo
  {author} {\bibfnamefont {{\'{E}}.}~\bibnamefont {Dumur}}, \bibinfo {author}
  {\bibfnamefont {F.}~\bibnamefont {Foroughi}}, \bibinfo {author}
  {\bibfnamefont {J.}~\bibnamefont {Puertas}}, \bibinfo {author} {\bibfnamefont
  {S.}~\bibnamefont {Leger}}, \bibinfo {author} {\bibfnamefont
  {K.}~\bibnamefont {Bharadwaj}}, \bibinfo {author} {\bibfnamefont
  {J.}~\bibnamefont {Delaforce}}, \bibinfo {author} {\bibfnamefont
  {C.}~\bibnamefont {Naud}}, \bibinfo {author} {\bibfnamefont {W.}~\bibnamefont
  {Hasch-Guichard}}, \bibinfo {author} {\bibfnamefont {J.~J.}\ \bibnamefont
  {Garc\'{\i}a-Ripoll}}, \bibinfo {author} {\bibfnamefont {N.}~\bibnamefont
  {Roch}},\ and\ \bibinfo {author} {\bibfnamefont {O.}~\bibnamefont
  {Buisson}},\ }\bibfield  {title} {\bibinfo {title} {{Fast High-Fidelity
  Quantum Nondemolition Qubit Readout via a Nonperturbative Cross-Kerr
  Coupling}},\ }\href {https://doi.org/10.1103/PhysRevX.10.011045} {\bibfield
  {journal} {\bibinfo  {journal} {Phys. Rev. X}\ }\textbf {\bibinfo {volume}
  {10}},\ \bibinfo {pages} {11045} (\bibinfo {year} {2020})}\BibitemShut
  {NoStop}%
\bibitem [{\citenamefont {Ryan}\ \emph {et~al.}(2015)\citenamefont {Ryan},
  \citenamefont {Johnson}, \citenamefont {Gambetta}, \citenamefont {Chow},
  \citenamefont {da~Silva}, \citenamefont {Dial},\ and\ \citenamefont
  {Ohki}}]{Ryan2015}%
  \BibitemOpen
  \bibfield  {author} {\bibinfo {author} {\bibfnamefont {C.~A.}\ \bibnamefont
  {Ryan}}, \bibinfo {author} {\bibfnamefont {B.~R.}\ \bibnamefont {Johnson}},
  \bibinfo {author} {\bibfnamefont {J.~M.}\ \bibnamefont {Gambetta}}, \bibinfo
  {author} {\bibfnamefont {J.~M.}\ \bibnamefont {Chow}}, \bibinfo {author}
  {\bibfnamefont {M.~P.}\ \bibnamefont {da~Silva}}, \bibinfo {author}
  {\bibfnamefont {O.~E.}\ \bibnamefont {Dial}},\ and\ \bibinfo {author}
  {\bibfnamefont {T.~A.}\ \bibnamefont {Ohki}},\ }\bibfield  {title} {\bibinfo
  {title} {{Tomography via correlation of noisy measurement records}},\ }\href
  {https://doi.org/10.1103/PhysRevA.91.022118} {\bibfield  {journal} {\bibinfo
  {journal} {Phys. Rev. A}\ }\textbf {\bibinfo {volume} {91}},\ \bibinfo
  {pages} {22118} (\bibinfo {year} {2015})}\BibitemShut {NoStop}%
\bibitem [{\citenamefont {Walter}\ \emph {et~al.}(2017)\citenamefont {Walter},
  \citenamefont {Kurpiers}, \citenamefont {Gasparinetti}, \citenamefont
  {Magnard}, \citenamefont {Poto\ifmmode~\check{c}\else \v{c}\fi{}nik},
  \citenamefont {Salath\'e}, \citenamefont {Pechal}, \citenamefont {Mondal},
  \citenamefont {Oppliger}, \citenamefont {Eichler},\ and\ \citenamefont
  {Wallraff}}]{Walter2017}%
  \BibitemOpen
  \bibfield  {author} {\bibinfo {author} {\bibfnamefont {T.}~\bibnamefont
  {Walter}}, \bibinfo {author} {\bibfnamefont {P.}~\bibnamefont {Kurpiers}},
  \bibinfo {author} {\bibfnamefont {S.}~\bibnamefont {Gasparinetti}}, \bibinfo
  {author} {\bibfnamefont {P.}~\bibnamefont {Magnard}}, \bibinfo {author}
  {\bibfnamefont {A.}~\bibnamefont {Poto\ifmmode~\check{c}\else
  \v{c}\fi{}nik}}, \bibinfo {author} {\bibfnamefont {Y.}~\bibnamefont
  {Salath\'e}}, \bibinfo {author} {\bibfnamefont {M.}~\bibnamefont {Pechal}},
  \bibinfo {author} {\bibfnamefont {M.}~\bibnamefont {Mondal}}, \bibinfo
  {author} {\bibfnamefont {M.}~\bibnamefont {Oppliger}}, \bibinfo {author}
  {\bibfnamefont {C.}~\bibnamefont {Eichler}},\ and\ \bibinfo {author}
  {\bibfnamefont {A.}~\bibnamefont {Wallraff}},\ }\bibfield  {title} {\bibinfo
  {title} {Rapid high-fidelity single-shot dispersive readout of
  superconducting qubits},\ }\href
  {https://doi.org/10.1103/PhysRevApplied.7.054020} {\bibfield  {journal}
  {\bibinfo  {journal} {Phys. Rev. Applied}\ }\textbf {\bibinfo {volume} {7}},\
  \bibinfo {pages} {054020} (\bibinfo {year} {2017})}\BibitemShut {NoStop}%
\bibitem [{\citenamefont {Fliess}\ \emph {et~al.}(1995)\citenamefont {Fliess},
  \citenamefont {Lévine}, \citenamefont {Martin},\ and\ \citenamefont
  {Rouchon}}]{Fliess1995}%
  \BibitemOpen
  \bibfield  {author} {\bibinfo {author} {\bibfnamefont {M.}~\bibnamefont
  {Fliess}}, \bibinfo {author} {\bibfnamefont {J.}~\bibnamefont {Lévine}},
  \bibinfo {author} {\bibfnamefont {P.}~\bibnamefont {Martin}},\ and\ \bibinfo
  {author} {\bibfnamefont {P.}~\bibnamefont {Rouchon}},\ }\bibfield  {title}
  {\bibinfo {title} {Flatness and defect of non-linear systems: introductory
  theory and examples},\ }\href {https://doi.org/10.1080/00207179508921959}
  {\bibfield  {journal} {\bibinfo  {journal} {International Journal of
  Control}\ }\textbf {\bibinfo {volume} {61}},\ \bibinfo {pages} {1327}
  (\bibinfo {year} {1995})}\BibitemShut {NoStop}%
\bibitem [{\citenamefont {Campagne-Ibarcq}(2015)}]{Campagne-ibarcq2015}%
  \BibitemOpen
  \bibfield  {author} {\bibinfo {author} {\bibfnamefont {P.}~\bibnamefont
  {Campagne-Ibarcq}},\ }\emph {\bibinfo {title} {{Measurement back action and
  feedback in superconducting circuits}}},\ \href
  {https://hal.archives-ouvertes.fr/tel-01248789} {Ph.D. thesis},\ \bibinfo
  {school} {{{\'E}cole Normale Sup{\'e}rieure (ENS)}} (\bibinfo {year}
  {2015})\BibitemShut {NoStop}%
\bibitem [{\citenamefont {Kr{\"{a}}mer}\ \emph {et~al.}(2018)\citenamefont
  {Kr{\"{a}}mer}, \citenamefont {Plankensteiner}, \citenamefont {Ostermann},\
  and\ \citenamefont {Ritsch}}]{KRAMER2018109}%
  \BibitemOpen
  \bibfield  {author} {\bibinfo {author} {\bibfnamefont {S.}~\bibnamefont
  {Kr{\"{a}}mer}}, \bibinfo {author} {\bibfnamefont {D.}~\bibnamefont
  {Plankensteiner}}, \bibinfo {author} {\bibfnamefont {L.}~\bibnamefont
  {Ostermann}},\ and\ \bibinfo {author} {\bibfnamefont {H.}~\bibnamefont
  {Ritsch}},\ }\bibfield  {title} {\bibinfo {title} {{QuantumOptics.jl: A Julia
  framework for simulating open quantum systems}},\ }\href
  {https://doi.org/10.1016/j.cpc.2018.02.004} {\bibfield  {journal} {\bibinfo
  {journal} {Computer Physics Communications}\ }\textbf {\bibinfo {volume}
  {227}},\ \bibinfo {pages} {109} (\bibinfo {year} {2018})}\BibitemShut
  {NoStop}%
\bibitem [{\citenamefont {Reagor}\ \emph {et~al.}(2016)\citenamefont {Reagor},
  \citenamefont {Pfaff}, \citenamefont {Axline}, \citenamefont {Heeres},
  \citenamefont {Ofek}, \citenamefont {Sliwa}, \citenamefont {Holland},
  \citenamefont {Wang}, \citenamefont {Blumoff}, \citenamefont {Chou},
  \citenamefont {Hatridge}, \citenamefont {Frunzio}, \citenamefont {Devoret},
  \citenamefont {Jiang},\ and\ \citenamefont {Schoelkopf}}]{Reagor2016}%
  \BibitemOpen
  \bibfield  {author} {\bibinfo {author} {\bibfnamefont {M.}~\bibnamefont
  {Reagor}}, \bibinfo {author} {\bibfnamefont {W.}~\bibnamefont {Pfaff}},
  \bibinfo {author} {\bibfnamefont {C.}~\bibnamefont {Axline}}, \bibinfo
  {author} {\bibfnamefont {R.~W.}\ \bibnamefont {Heeres}}, \bibinfo {author}
  {\bibfnamefont {N.}~\bibnamefont {Ofek}}, \bibinfo {author} {\bibfnamefont
  {K.}~\bibnamefont {Sliwa}}, \bibinfo {author} {\bibfnamefont
  {E.}~\bibnamefont {Holland}}, \bibinfo {author} {\bibfnamefont
  {C.}~\bibnamefont {Wang}}, \bibinfo {author} {\bibfnamefont {J.}~\bibnamefont
  {Blumoff}}, \bibinfo {author} {\bibfnamefont {K.}~\bibnamefont {Chou}},
  \bibinfo {author} {\bibfnamefont {M.~J.}\ \bibnamefont {Hatridge}}, \bibinfo
  {author} {\bibfnamefont {L.}~\bibnamefont {Frunzio}}, \bibinfo {author}
  {\bibfnamefont {M.~H.}\ \bibnamefont {Devoret}}, \bibinfo {author}
  {\bibfnamefont {L.}~\bibnamefont {Jiang}},\ and\ \bibinfo {author}
  {\bibfnamefont {R.~J.}\ \bibnamefont {Schoelkopf}},\ }\bibfield  {title}
  {\bibinfo {title} {Quantum memory with millisecond coherence in circuit
  qed},\ }\href {https://doi.org/10.1103/PhysRevB.94.014506} {\bibfield
  {journal} {\bibinfo  {journal} {Phys. Rev. B}\ }\textbf {\bibinfo {volume}
  {94}},\ \bibinfo {pages} {014506} (\bibinfo {year} {2016})}\BibitemShut
  {NoStop}%
\bibitem [{\citenamefont {Place}\ \emph {et~al.}(2020)\citenamefont {Place},
  \citenamefont {Rodgers}, \citenamefont {Mundada}, \citenamefont {Smitham},
  \citenamefont {Fitzpatrick}, \citenamefont {Leng}, \citenamefont {Premkumar},
  \citenamefont {Bryon}, \citenamefont {Sussman}, \citenamefont {Cheng},
  \citenamefont {Madhavan}, \citenamefont {Babla}, \citenamefont {Jaeck},
  \citenamefont {Gyenis}, \citenamefont {Yao}, \citenamefont {Cava},
  \citenamefont {de~Leon},\ and\ \citenamefont {Houck}}]{Place2020}%
  \BibitemOpen
  \bibfield  {author} {\bibinfo {author} {\bibfnamefont {A.~P.~M.}\
  \bibnamefont {Place}}, \bibinfo {author} {\bibfnamefont {L.~V.~H.}\
  \bibnamefont {Rodgers}}, \bibinfo {author} {\bibfnamefont {P.}~\bibnamefont
  {Mundada}}, \bibinfo {author} {\bibfnamefont {B.~M.}\ \bibnamefont
  {Smitham}}, \bibinfo {author} {\bibfnamefont {M.}~\bibnamefont
  {Fitzpatrick}}, \bibinfo {author} {\bibfnamefont {Z.}~\bibnamefont {Leng}},
  \bibinfo {author} {\bibfnamefont {A.}~\bibnamefont {Premkumar}}, \bibinfo
  {author} {\bibfnamefont {J.}~\bibnamefont {Bryon}}, \bibinfo {author}
  {\bibfnamefont {S.}~\bibnamefont {Sussman}}, \bibinfo {author} {\bibfnamefont
  {G.}~\bibnamefont {Cheng}}, \bibinfo {author} {\bibfnamefont
  {T.}~\bibnamefont {Madhavan}}, \bibinfo {author} {\bibfnamefont {H.~K.}\
  \bibnamefont {Babla}}, \bibinfo {author} {\bibfnamefont {B.}~\bibnamefont
  {Jaeck}}, \bibinfo {author} {\bibfnamefont {A.}~\bibnamefont {Gyenis}},
  \bibinfo {author} {\bibfnamefont {N.}~\bibnamefont {Yao}}, \bibinfo {author}
  {\bibfnamefont {R.~J.}\ \bibnamefont {Cava}}, \bibinfo {author}
  {\bibfnamefont {N.~P.}\ \bibnamefont {de~Leon}},\ and\ \bibinfo {author}
  {\bibfnamefont {A.~A.}\ \bibnamefont {Houck}},\ }\href@noop {} {\bibinfo
  {title} {New material platform for superconducting transmon qubits with
  coherence times exceeding 0.3 milliseconds}} (\bibinfo {year} {2020}),\
  \Eprint {https://arxiv.org/abs/2003.00024} {arXiv:2003.00024 [quant-ph]}
  \BibitemShut {NoStop}%
\bibitem [{\citenamefont {Elliott}\ \emph {et~al.}(2018)\citenamefont
  {Elliott}, \citenamefont {Joo},\ and\ \citenamefont
  {Ginossar}}]{Elliott2018}%
  \BibitemOpen
  \bibfield  {author} {\bibinfo {author} {\bibfnamefont {M.}~\bibnamefont
  {Elliott}}, \bibinfo {author} {\bibfnamefont {J.}~\bibnamefont {Joo}},\ and\
  \bibinfo {author} {\bibfnamefont {E.}~\bibnamefont {Ginossar}},\ }\bibfield
  {title} {\bibinfo {title} {Designing {K}err interactions using multiple
  superconducting qubit types in a single circuit},\ }\href
  {https://doi.org/10.1088/1367-2630/aa9243} {\bibfield  {journal} {\bibinfo
  {journal} {New Journal of Physics}\ }\textbf {\bibinfo {volume} {20}},\
  \bibinfo {pages} {023037} (\bibinfo {year} {2018})}\BibitemShut {NoStop}%
\end{thebibliography}
%

\end{document}